\documentclass[12pt]{JHEP3}

\usepackage[all]{xy}

\usepackage{amsmath,amsthm,amsfonts,amssymb,amscd,mathrsfs,graphicx,fontenc,Glebdefs,
}

\usepackage{enumitem}

\usepackage{relsize}

\usepackage{color}
\let\normalcolor\relax
\usepackage{colordefs}

\def\tu{\tilde{u}}
\def\cu{\check{u}}
\def\bee{\begin{enumerate}}
\def\eee{\end{enumerate}}
\def\bei{\begin{itemize}}
\def\eei{\end{itemize}}

\author{Gleb Arutyunov$^a$\footnote{Email: G.E.Arutyunov@uu.nl, frolovs@maths.tcd.ie, s.j.vantongeren@uu.nl} {}\footnote{Correspondent fellow at Steklov
Mathematical Institute, Moscow.}\,,  Sergey Frolov$^{b\,\dagger}$  and\, Stijn J. van Tongeren$^a$
 \\ $^{a}$ {\it Institute for Theoretical
Physics and Spinoza Institute,\\ ~~Utrecht University, 3508 TD
Utrecht, The Netherlands} \\ $^b$ {\it Hamilton Mathematics Institute and School of Mathematics, \\
~~Trinity College, Dublin 2, Ireland} }

\abstract{ The spectrum of the light-cone $\AdS$ superstring
contains states composed of particles with
complex momenta including in particular those which turn into bound states in the decompactification limit.
We propose the mirror TBA description for these states.  We focus on a three-particle state
which is a finite-size representative of a scattering state
of a fundamental particle and a two-particle bound state and
dual to an operator from the $\su(2)$ sector of ${\cal N}=4$ SYM.
We find that the analytic behavior
of Y-functions  differs drastically from the case of states with real momenta. Most importantly,
$Y_Q$-functions exhibit poles in the analyticity strip which leads to the appearance of new terms in the formula
for the energy of this state.  In addition, the TBA equations are supplied by quantization conditions which involve
$Y_2$. Considering yet another example of a three-particle state, we find that the corresponding
quantization conditions do not even involve $Y_1$.   Our treatment can be generalized to a wide class of states
with complex momenta.

}

\title{Bound States in the Mirror TBA}

\preprint{
\scriptsize{ITP-UU-11-39}\\[-.5ex]
\scriptsize{SPIN-11-30}\\[-.5ex]
\scriptsize{TCDMATH 11-12}\\[-.5ex]
\scriptsize{HMI-11-04}
}

\begin{document}
\renewcommand{\thefootnote}{\arabic{footnote}}
\setcounter{footnote}{0}

\section{Introduction}
In this paper we continue our studies of the mirror Thermodynamic Bethe Ansatz (TBA) as a tool to determine the spectrum of the $\AdS$ superstring, and through the gauge-string correspondence \cite{M} the spectrum of conformal dimensions of composite primary operators in planar  ${\cal N}=4$ super Yang-Mills theory. We will show by way of example how to construct TBA equations describing string excitations with complex momenta.

\smallskip

The main idea of the TBA approach originally developed for two-dimensional relativistic theories \cite{Zamolodchikov90}
is to reformulate the finite-size spectral problem for an integrable model in terms of thermodynamics of the accompanying mirror model. Integrability of the mirror model allows one to compute the necessary thermodynamic quantities and,
as a result, to determine the spectrum of the original model\footnote{See, {\it e.g.} \cite{Kuniba2,Bajnok:2010ke}
for  recent reviews of the TBA techniques.
Concerning string integrability, the reader may consult  the reviews \cite{Arutyunov:2009ga,Brev}.  }. As a necessary step towards realization of this idea, one needs to classify solutions of the mirror Bethe-Yang (BY) equations contributing in the
thermodynamic limit which is known under the name of string hypothesis \cite{Takahashi72}.  For the $\AdS$ mirror the  BY equations were obtained in \cite{AF07} and the corresponding string hypothesis was formulated in
\cite{AF09a}. This led to the construction of the ground state \cite{AF09b}-\cite{GKKV09} and excited state TBA equations
for string states with real momenta \cite{GKKV09}-\cite{Sfondrini:2011rr}.
The TBA equations can be formulated in a variety of different forms: canonical \cite{AF09b}-\cite{GKKV09},
simplified \cite{AF09b,AF09d}, hybrid \cite{AFS09}, and quasi-local \cite{Balog:2011cx}; each of these forms is best suited
for studying particular analytic or numerical aspects of the corresponding solution.   Also,
the TBA equations have been investigated  for particular states in different regimes.
Numerically, for intermediate values of the
coupling  the TBA equations for a state dual to the Konishi operator in the gauge theory were solved in
 \cite{GKV09b,Frolov:2010wt} and the results obtained agree  with various string theory computations
  \cite {Gromov:2011de}-\cite{Beccaria:2011uz}. Next, a relation between the TBA equations and the
  semi-classical description of string states
has been elucidated in \cite{Gromov09a}.
Finally, at weak coupling the TBA equations for the Konishi operator were shown
\cite{AFS10,BH10a,BH10b} to agree with L\"uscher's perturbative treatment \cite{BJ08}-\cite{Janik:2010kd}
and at four loops with explicit field-theoretic computations \cite{Sieg,Velizhanin:2008jd}.

\smallskip

To expose new features of the TBA approach, in this work
we turn our attention to the $\su(2)$ sector  where
particles may have complex momenta, and in particular there are bound states arising in the large $J$ limit
due to poles of the world-sheet scattering matrix. Here $J$ is the angular momentum of a string
rotating around the equator of ${\rm S}^5$ which is related to
the length $L$ of a gauge theory operator from the $\su(2)$ sector with
$M$ excitations (magnons) as $L=J+M$. Furthermore, $J$ is related to the length parameter
$L_{\rm TBA}$ entering the TBA equations
 as $L_{\rm TBA}=J+2$, which is the maximum $J$-charge in a typical
multiplet of $\psu(2,2|4)$ algebra \cite{Arutyunov:2011uz}.
\smallskip

The construction of TBA equations for generic states based on the contour deformation trick,
a procedure inspired by the work  \cite{DT96,BLZe}, has been elaborated upon in
\cite{AFS09,Arutyunov:2011uz}.
It assumes that {\it for finite $J$ and small coupling $g$},  states
are described as solutions of the BY equations \cite{BS}.
Here $g=\frac{\sqrt{\lambda}}{2\pi}$, where $\lambda$ is the 't Hooft coupling.
Picking up a state, {\it i.e.} a concrete solution
of the BY equations, we then construct
the corresponding asymptotic Y-functions and determine their analytic properties, in particular, the location of zeros and poles.
This analytic structure is then used to find proper integration contours and engineer the TBA equations of interest,
such that they are solved by the asymptotic Y-functions upon omitting contributions which
vanish in the limit $g\to 0$, {\it e.g.} terms such as $\log(1+Y_Q)$. Furthermore, {\it quantization conditions}
which fix the location of singularities of the {\it exact} Y-functions must be imposed. In particular,
the exact rapidities $u_k$ of fundamental particles are found from the exact Bethe equations $Y_{1*}(u_k)=-1$,
which themselves are obtained by analytically continuing the TBA equation for $Y_1$ to the string region. These are the quantization conditions for $u_k$; the finite-size analogues of the BY equations.

\smallskip

The procedure of constructing TBA equations explained above relies on
the assumption that the analytic properties of the asymptotic and exact  Y-functions are similar, {\it i.e.} that the locations of zeroes and poles of $Y$ and $1+Y$ in their analyticity strip are smoothly deformed in passing from the asymptotic to the exact solution.
In particular, this means that no new singularities can be formed.
In this paper we show that the same strategy of constructing TBA equations also applies to states with complex momenta, at least for the cases
we considered explicitly.

\smallskip

We start by considering the simplest three-particle, {\it i.e.} $M=3$, state in the $\su(2)$ sector which involves complex momenta --
a configuration where the first particle has real (positive) momentum and the other two have complex conjugate momenta, such that the level-matching condition is satisfied.  Search for solutions of the BY equations in the limit
$g\to 0$ reveals that such configurations exist; the first one shows up for $J=4$, that is for $L=7$.\footnote{For $L=6$ there is a singular solution composed of a particle with momentum $\pi$ and a two-particle bound state with momentum $-\pi$ \cite{BKS,BDS}. It is unknown how to handle such a state in the TBA framework. }
This solution shows several remarkable related features which we list below
\begin{enumerate}[label=\it{\arabic{*}})~~]
\item In the limit $g\to 0$, the complex rapidities $u_2$ and $u_3$ of the second and third particle respectively,
lie {\it outside} the analyticity strip, which is in between two lines running parallel to the real axis
at $\frac{i}{g}$ and $-\frac{i}{g}$,
\item As $g$ is increased, $u_2$ and $u_3$ move towards the analyticity strip, more precisely,
to the points $-2-\frac{i}{g}$ and $-2+\frac{i}{g}$. Further increasing $g$ leads to a breakdown of the asymptotic
Bethe Ansatz, as the energy of the corresponding configuration becomes complex.   This breakdown happens before
$u_2$ and $u_3$ reach the boundaries of the analyticity strip,
\item The first three $Y_Q$-functions, $Y_1$, $Y_2$ and $Y_3$, computed for the asymptotic solution, exhibit poles
located {\it inside} the analyticity strip; the poles of $Y_2$ being closest to the real line are
at\footnote{Throughout the paper the superscript
${\pm}$ means the shift of the function argument by $\pm \frac{i}{g}$ with obvious generalization to many $\pm\ldots\pm$.} $u_2^+$
and $u_3^-$.

\end{enumerate}

\noindent
Concerning the first point, we made a wide numerical search for solutions of the one-loop BY equations for three-particle configurations of the type described above, and could not find any solution with
$u_2$ and $u_3$ falling inside the analyticity strip.
There are however many three-particle  solutions with complex roots being in any of the $k$th strips $(k-1)/g < |{\rm Im}(u)| < k/g$, $k=2,3,\ldots$
A configuration with complex roots within the analyticity
strip  can be  found for  a four-particle configuration and we will come back to its discussion later.

\smallskip

\noindent
Concerning the second point, we expect that while the
asymptotic roots
$u_2$ and $u_3$ move towards the boundaries of the analyticity strip, they cannot cross them because
the S-matrix entering the  BY equations develops a singularity as $u_3-u_2\to\frac{2i}{g}$. Also, the breakdown of the BY equations simply reflects their asymptotic nature in comparison to the exact TBA equations.
Nevertheless, in the weak coupling expansion
the exact Bethe equations must coincide with the asymptotic Bethe Ansatz
up to the first order of wrapping, which for an operator of length $L$ from the $\su(2)$ sector means up to order $g^{2L}$.

\smallskip

\noindent
Concerning the third point, occurrence of poles  for some of $Y_Q$ inside the analyticity strip
is a new phenomenon in comparison to the analytic structure of states from the $\sl(2)$ sector
and it will have important implications for construction of the corresponding TBA equations.
We also point out that  as $g$ increases
the poles of $Y_2$ move towards the real line; nevertheless  for $g$ sufficiently small $Y_2$ remains small in the vicinity of the real line, {\it i.e.} for these values of $g$ we can trust the asymptotic solution.

\smallskip

The main observation which allows us to construct consistent TBA equations is
as follows.  If a $Y_Q$-function has a pole
at  a point $u_{\infty}$ inside the analyticity strip,
then, as we will show, it must be equal to $-1$ at a point $u_{-1}$ which is located close to the pole. Both $u_{\infty}$
and $u_{-1}$ can in general depend on $Q$. In the limit $g\to 0$ we can estimate their difference from
the asymptotic expression for $Y_1$, obtaining
$$
\delta u=u_{\infty}-u_{-1}\sim g^{2L}\, .
$$
Indeed, as we see the roots start to differ from each other precisely at the $L$-loop order! As we will
explain, this guarantees that in the weak coupling expansion the asymptotic Bethe Ansatz
agrees with the TBA up to $g^{2L}$.  It is interesting to point out that an analytic structure similar to the one we encounter here
is realized in  the relativistic ${\rm SU}(N)$ principal model
for states describing fundamental particles with complex momenta \cite{Kazakov:2010kf,Balog}.

\smallskip

Having understood the analytic structure of the exact solution, we then proceed with the construction of the TBA
equations by means of the contour deformation trick. We begin with the canonical TBA equations
because there the choice of integration contours can be made most transparent.
In particular, in this case the poles of the auxiliary $Y$-functions play no role, {\it i.e.} only the contributions of zeroes should be taken into account. Most importantly, we find that the contours
must enclose all real zeroes of $1+Y_Q$ which are in the string region, and all zeroes  and poles related to the complex Bethe roots which are below the real line of the mirror region.\footnote{This means, for instance, that contours never
enclose the Bethe root $u_3$ which is in the intersection of the string  and anti-mirror regions.}
Finally, we use the canonical equations to derive the corresponding simplified
and hybrid equations.

\smallskip

The driving terms in the resulting TBA equations have quite an intricate structure. They appear to depend  on
$u_{2,3}^{(1,2)} $ related to singularities of $Y_1$ and $Y_2$,
the real root $u_1$, and additional roots $r^M$
related to auxiliary functions $Y_-$ and $Y_{M|w}$.
The exact values of these roots are fixed by the corresponding exact Bethe equations. It is worthwhile to point out that for the state we consider, several apparently different quantization conditions for the Bethe roots arise. For instance for $u_3^{(1)}$ we find
$$
Y_1(u_3^{(1)})=-1~~ \Leftrightarrow ~~Y_1(u_3^{(1)--}) =-1  ~~\Leftrightarrow ~~Y_{1*}(u_3^{(1)})=-1\, .
$$
The first two conditions follow from our assumptions on the analytic structure and the last one,
which involves $Y_{1*}$, the analytic continuation of $Y_1$ to the string region, is the quantization condition
we expect as a finite-size analogue of the BY equation. We show that
the exact Bethe equations representing these quantization conditions are compatible in a rather non-trivial manner
which involves, in particular, crossing symmetry. This is a strong consistency check of our construction.
There are similar quantization conditions involving $Y_2$. For instance, the location of $u_3^{(2)}$ is determined by the
following compatible exact Bethe equations
$$
Y_2(u_3^{(2)-})=-1~~~ \Leftrightarrow ~~~Y_2(u_3^{(2)---})=-1\, .
$$

Our next interesting observation concerns the energy formula. The fact that $1+Y_1$ and $1+Y_2$ functions have zeroes and poles in the analyticity strip in conjunction with our choice for the integration contours leads to
the following energy formula
 \begin{align}
E=&\
\sum_{i=1}^3\E(u_i^{(1)})
-{1\ov 2\pi}\sum_{Q=1}^{\infty}\int_{-\infty}^\infty\, du {d\tilde{p}_Q\ov du}\log(1+Y_Q)
\nonumber\\
&\quad -i\tilde{p}_2(u_2^{(1)+})+i\tilde{p}_2(u_2^{(2)+})
-i\tilde{p}_2(u_3^{(2)-})+i\tilde{p}_2(u_3^{(1)-})\,,
\nonumber
\end{align}
where $u_1^{(1)}\equiv u_1$ and $\E(u)$ is the dispersion relation of a fundamental particle with rapidity variable $u$, while $\tilde{p}_Q$ is the momentum of a mirror $Q$-particle.

\smallskip

The expression for $E$ is exact and it can be used to compute corrections to the Bethe Ansatz energy
in the limit $g\to 0$ and $J$ finite, and in the limit $J\to \infty$ and $g$ finite. The first limit provides
the leading wrapping correction which is given by
\begin{align}
\Delta E^{\rm wrap}=&
-\frac{1}{2\pi}\sum_{Q=1}^{\infty}\int_{-\infty}^\infty\, du{d\tilde{p}_Q\ov du} \, Y_Q
\nonumber
\\
& -i\Big[\, {\rm Res}\Big(\frac{d\tilde{p}_2}{du}(u_2^+)Y_2(u_2^+)\Big)
- {\rm Res}\Big(\frac{d\tilde{p}_2}{du}(u_3^-)Y_2(u_3^-)\Big)\Big]\, .
\nonumber
\end{align}
The last line  in the above formula is nothing else but the residue of the integrand for $Y_2$, the function which in comparison to the other $Y_Q$-functions has poles closest to the real line.  The residue terms are of the same order
as the integral term.

\smallskip

In the second limit corrections are expected to be exponentially small in $J$ which for simple models or states are
given by the generalized L\"uscher's formula \cite{BJ08}. In particular, in this limit the $Y_Q$-functions are exponentially small
and the integral term takes the same form as in the expression for $\Delta E^{\rm wrap}$. This term is usually interpreted as the F-term. However, in our case the situation is much more complicated because in the limit $J\to \infty$
the function $Y_2$ develops a double pole on the real line so that we cannot replace $\log(1+Y_2)$ by $Y_2$.
Therefore, the large $J$-correction coming from the integral term is not given by the F-term. To our knowledge,
the $\tilde{p}$-dependent terms in the expression for $E$ are new and, as far as we can see, they cannot be interpreted as
L\"uscher's $\mu$-terms. It would be interesting to find the large $J$ expansion of the energy formula.

\smallskip

Finally, to check universality of our approach we studied another three-particle state. This state has
$L=40$ with complex rapidities $u_2$ and $u_3$ falling inside the third strip. The analytic structure
of asymptotic and exact Y-functions is very similar to the one previously considered with an exception that
now the first four $Y_Q$-functions  have poles inside the analyticity strip; $Y_1$ and $Y_3$ have poles
closest to the real line.  We obtain the canonical TBA equations by picking up the same contours as before.
This time the driving terms depend on $u^{(2,3)}_{2,3}$ which are related to singularities of $Y_2$ and $Y_3$.
The rapidities $u_{2,3}^{(2,3)}$ should be found from the corresponding exact Bethe equations.
It is pretty surprising that the ``standard" Bethe equations $Y_1(u_{2,3}^{(1)})=-1$ do not play any role
for the description of this state, because the TBA equations do not explicitly involve these roots at all!

\smallskip

With two examples at hand, a generalization of our construction to a three-particle state
with $u_2$ and $u_3$ lying in the $k$th strip seems to be straightforward.
Four functions $Y_{k-2},\ldots, Y_{k+1}$
will have poles in the analyticity strip, with the poles of $Y_{k-2}$ and $Y_k$ being closest to the real line. The driving terms in the corresponding TBA equations will depend on $u_{2,3}^{(k-1)}$ and $u_{2,3}^{(k)}$ whose locations
are determined by the corresponding exact Bethe equations for $Y_{k-1}$ and $Y_k$.
The energy formula is  then given by
 \begin{align}
E=&\
\sum_{i=1}^3\E(u_i^{(1)})
-{1\ov 2\pi}\sum_{Q=1}^{\infty}\int_{-\infty}^\infty\, du {d\tilde{p}_Q\ov du}\log(1+Y_Q)
\nonumber\\
&\quad\quad  -i\tilde{p}_k\Big(u_2^{(k-1)}+{\textstyle{(k-1)}}\tfrac{i}{g}\Big)
+i\tilde{p}_k\Big(u_2^{(k)}+(k-1)\tfrac{i}{g}\Big) \nonumber \\
&\hskip 2.5cm
-i\tilde{p}_k\Big(u_3^{(k)}-(k-1)\tfrac{i}{g}\Big)+i\tilde{p}_k\Big(u_3^{(k-1)}-(k-1)\tfrac{i}{g}\Big)\, .
\nonumber
\end{align}
This completes our discussion of the TBA approach for the three-particle states with complex momenta.

\smallskip
Let us now come back to the four-particle solution mentioned earlier. The type of solution we considered is given by a  symmetric configuration of particles with momenta $\{p_i\} = \{p, p^*, -p, -p^*\}$. For  $L\geq 10$ such configurations exist with rapidities \emph{inside} the analyticity strip.  As the coupling is increased the rapidities tend to the boundaries of the strip from the inside.
For numerical reasons we  explicitly study a state with $L=16$. It appears that for this case only $Y_2$ has poles inside the analyticity strip. Nevertheless, the fact that all rapidities are inside
the analyticity strip clearly distinguishes this state from the three-particle case discussed above.
In short, in choosing the integration contours we found no reason
to pick  up contributions of the poles  and zeros of $1+Y_2$.
The TBA and exact Bethe equations are constructed in essentially the same fashion as for states with real momenta.
It would be important to further clarify what precisely makes complex configurations with rapidities inside and
outside the analyticity strip so different in the TBA treatment. Certainly, this must be related to the fact that the corresponding rapidities do or do not lie in the overlap of the string and mirror regions respectively.

\smallskip

The paper is organized as follows. In section \ref{su2states} we consider three-particle states in the $\su(2)$ sector
and in section \ref{sect:L7} we discuss the relevant analytic properties of the
asymptotic and exact solution for our main state of interest.
Section \ref{sect:canL7} is devoted to the derivation of  the canonical TBA equations via the contour deformation trick.
We also present expressions for the energy and momentum.
In section \ref{sect:simpTBA} the canonical equations are cast into the simplified and hybrid forms.
In section \ref{sect:EBandQ} the exact Bethe equations are presented and various consistency conditions
are verified. We also discuss the relation of the exact Bethe equations to the asymptotic Bethe Ansatz.
In the conclusions we indicate some interesting questions and discuss a potential fate of three-particle bound states when $g$ becomes large.
Finally, in appendices \ref{L40} and \ref{4pt} we study in some detail the $L=40$ three-particle state with roots in the third strip and the four-particle state with roots in the first strip. We present the corresponding TBA and exact Bethe equations.
Various technical details  are relegated to other appendices.

\section{Three-particle states in the $\su(2)$ sector}\la{su2states}
We consider three-particle  $\AdS$ superstring excited states with vanishing total momentum which carry two $\su(4)$ charges $J_1=J$ and $J_2=3$.
They are dual to operators of length $L=J+3$ from the $\su(2)$ sector of ${\cal N}=4$ SYM.
Such states can be  composed of either three fundamental particles carrying real momenta or of one particle with a real momentum and two particles with complex momenta which are conjugate to each other at any $L$ for small enough values of the coupling constant $g$. The TBA and exact Bethe equations for states with real momenta are similar to the ones for the $\sl(2)$ states, and in this paper we will discuss only states with complex momenta.

\smallskip
We denote the real momentum of the fundamental particle as $p_1\equiv p$ and assume that it is positive. Then, the complex momenta of two other particles are $p_2=-{p\ov2}+iq$ and $p_3=-{p\ov2}-iq$, where the parameter $q$ has a positive real part Re$(q)>0$.
It is worth mentioning that for infinite $L$ such a state is a scattering state of a fundamental particle and a two-particle bound state, and that $q$ becomes complex for $g$ exceeding a special value depending on $p$. For these values of $g$ and $p$ the exponentially suppressed corrections to the energy of the string state computed by using the BY equations are complex as well, indicating a breakdown of the BY equations \cite{AF07}.
\smallskip

The two independent BY equations in the $\su(2)$-sector \cite{AFS} for the state under consideration can be written in the form
\bea
\begin{aligned}
e^{ip_1 L}\frac{u_1-u_2-{2i\ov g}}{u_1-u_2+{2i\ov g}}\,\frac{u_1-u_3-{2i\ov g}}{u_1-u_3+{2i\ov g}}\,{1\ov \sigma(p_1,p_2)^2\sigma(p_1,p_3)^2}=1
\, ,\la{BYE}\\
e^{ip_2 L}\frac{u_2-u_1-{2i\ov g}}{u_2-u_1+{2i\ov g}}\,\frac{u_2-u_3-{2i\ov g}}{u_2-u_3+{2i\ov g}}\,{1\ov \sigma(p_2,p_1)^2\sigma(p_2,p_3)^2}=1
\, ,
\end{aligned}
\eea
where $\s$ is the BES dressing factor \cite{BES}, and $u_k$ are the $u$-plane rapidity variables related to $p_k$ as \cite{BDS}
\bea\la{pvsu}
u = {1\ov g} \cot{p\ov2}\,\sqrt{1+4g^2\sin^2{p\ov2}}\,.
\eea
Taking the logarithm  of the BY equations, we get
\bea\la{BYe}
\log( \, {\rm l.h.s.}_1\, )=2\pi i\, n_1\,,\quad \log(\, {\rm l.h.s.}_2\, )=-2\pi i\, n_2\,,
\eea
where $n_1$ and $n_2$ are positive integers because  $p_1$ is positive. Due to the level matching condition they should satisfy the relation
$n_2\equiv n=2n_1$. As was shown in  \cite{AFS}, at large values of $g$ the integer $n$ is equal to the string level of the state.

\smallskip
Analyzing solutions of the BY equations, we find that for small values of $g$
there is {\it no} solution with complex roots $u_2$ and $u_3$ lying in the analyticity strip
$-1/g < {\rm Im}(u) < 1/g$.
The fact that the complex roots are outside the analyticity strip leads to dramatic changes in the analytic properties of the Y-functions in comparison to the case with real momenta.

\smallskip
Changing the values of $L$ and $n$, it is possible to find solutions with complex rapidities lying in any of the strips $(k-1)/g < |{\rm Im}(u)| < k/g$, $k=2,3,\ldots$. Thus, such states can be characterized not only by $L$ and $n$ but also by the positive integer $k$ which indicates the strips the complex roots $u_2$ and $u_3$ are located in for small values of $g$. Solving the BY equations \eqref{BYE} for increasing values of $g$, we observe that for all solutions the complex roots move towards the boundaries of the analyticity strip, {\it i.e.} the lines $|{\rm Im}(u)|=1/g$. They cannot however cross them because the S-matrix has a pole if Im$(u_3)=-$Im$(u_2)=1/g$. As a result, as soon as the coupling constant exceeds a critical value, $u_1$ becomes complex and $u_2$ and $u_3$ are repelled from the lines Im$(u)=\mp 1/g$.
In addition the asymptotic energy of such a state becomes complex clearly demonstrating a breakdown of the BY equations.

\smallskip

In the next sections  we discuss one example of the states of this type with $L=7$, $n=2$ and $k=2$ in full detail, and we present the necessary results for the  $L=40$, $n=2$, $k=3$ case in appendix \ref{L40}. Most of our considerations can be generalized to any $L$, $n$ and $k$.

\section{The $L=7$, $n=2$, $k=2$ state and Y-functions}\label{sect:L7}

\subsection*{The $L=7$, $n=2$, $k=2$ state}\la{su2state}
An $\AdS$ superstring excited state with complex roots located in the second strip $1/g < |{\rm Im}(u)| < 2/g$
can be thought of as a finite-size analog of a scattering state of a fundamental particle and a two-particle bound state, because complex roots of such a state approximately satisfy the bound state condition $u_3-u_2=2i/g$. We will only consider the simplest state of this type with $n=2$ and $L=7$ but our consideration can be applied to any state with $k=2$.\footnote{ For $L=7$ we found only one such state with $n=2$ and no state with $n\ge 4$. For large values of $L$, $n$ should be increased to find solutions with $k=2$.}
\smallskip

We solved the BY equations \eqref{BYe} numerically\footnote{The  equations can be  solved only numerically even at $g=0$.} for $0\le g\le 0.5$ with step size $0.1$, for $0.5< g\le 0.53$ with step size $0.01$, and finally for $g= 0.5301$ and $g= 0.5302$. In table \ref{pqasym} we show the results for $p$ and $q$.

\begin{table}
 {\small
\bea\nonumber
\hspace{12pt}\begin{array}{|c||c|c||c|c|c|}
\hline
g& p&q&g& p&q\\ \hline
 0. & 2.3129&  0.926075&0.5 & 2.24919 &1.23789 \\
 0.1 & 2.3098&  0.933177&0.51 & 2.24704 & 1.27083\\
 0.2 & 2.30088& 0.955744& 0.52 & 2.2449 & 1.31517\\
 0.3 & 2.28709& 0.99838 & 0.53 & 2.24302 & 1.40691 \\
 0.4 & 2.26953 &  1.0737 &
 0.5301 & 2.24303 & 1.41083  \\  \hline
 ~~~~~~&~~~~~~~~~~~~~~~~~~~&~~~~~~~~~~~~~~~~~~~~&~0.5302 ~& ~2.2431-0.00001 i~&~1.41983-0.001 i ~\\  \hline
\end{array}~~~~~
\eea
}
\caption{Numerical solution of the BY equations for the $L=7$ state.}
\label{pqasym}
\end{table}

\begin{figure}[t]
\begin{center}
\includegraphics[width=7cm]{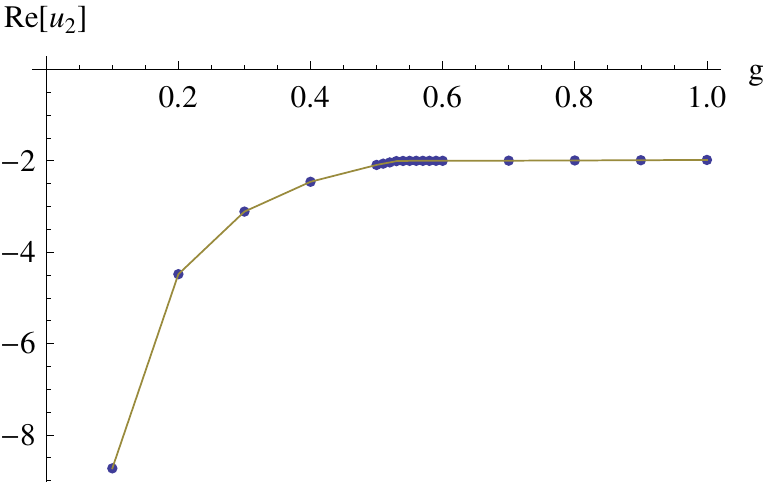}\quad \includegraphics[width=7cm]{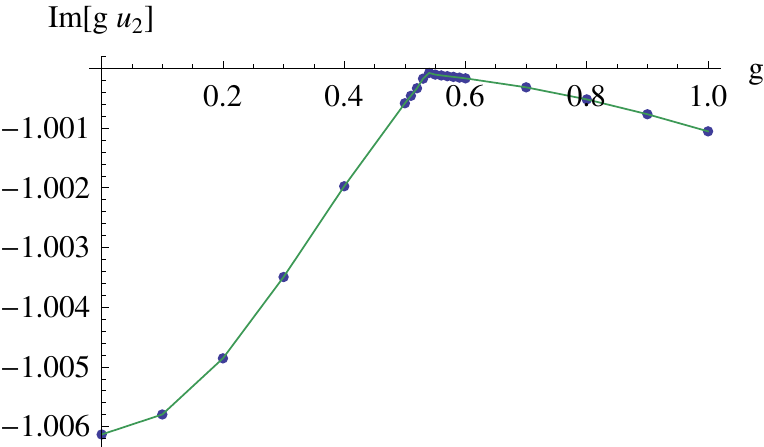}
\caption{The solution to the Bethe Yang equation for $u_2$ at $L=7$. For the imaginary part, the rapidity has been rescaled by a factor of $g$. Note that the rapidity asymptotes to $2-i/g$ before breakdown of the BY equations.}
\label{ReImu2}
\end{center}
\end{figure}
\smallskip
We see from the table that  $p$ and $q$ become complex at $g=0.5302$, and the BY equations cannot be used anymore. In fact the BY equations can probably not even be trusted at $g=0.5301$ because the momentum at this coupling is greater than its value at $g=0.53$, while the momentum has been steadily decreasing up to $g=0.53$. To understand a reason for the breakdown of the BY equations it is convenient to analyze the corresponding values of the $u$-plane rapidity variables $u_k$ and also their rescaled values $u_k^{rescaled} = g u_k$ which are more convenient for small values of $g$. The results are shown in table \ref{uuasym}.
\begin{table}
{\small
\bea
\nonumber
\hspace{12pt}\begin{array}{|c||c|c||c|c|}
\hline
g& u_1&u_2&u_1^{rescaled}&u_2^{rescaled}\\\hline
 0. & {0.439807\ov 0}&{-0.865401 - 1.00613 i\ov 0} & 0.439807&-0.865401 - 1.00613 i\\
 0.1 &4.48989&-8.73211 - 10.058 i&0.448989&-0.873211 - 1.0058 i \\
 0.2 &2.37935 &-4.48112 - 5.02428 i & 0.47587& -0.896224 - 1.00486 i \\
 0.3 &1.72919 &-3.11126 - 3.34498i &0.518756 & -0.933377 - 1.00349 i \\
 0.4 & 1.43888&-2.45839 - 2.50493 i &0.575551 & -0.983356 - 1.00197i \\
 0.5 & 1.28853&-2.0896 - 2.00117i  & 0.644265& -1.0448 - 1.00058 i \\
 0.51 &1.27788 & -2.06169 - 1.96168 i&0.651717 & -1.05146 - 1.00046 i \\
 0.52 & 1.26779&-2.03478 - 1.92372i & 0.659252&  -1.05809 - 1.00033 i \\
 0.53 &1.25786 &-2.006 - 1.88712i &0.666668 & -1.06318 - 1.00017i \\
 0.5301 & 1.25772&-2.00538 - 1.88675 i &0.666719 & -1.06305 - 1.00017i \\
 0.5302 &1.26 + 0.00002 i & -2.0041 - 1.88652 i&0.67 + 0.00001i &  -1.06257 - 1.00024 i
 \\\hline
\end{array}~~~~~
\eea
}
\caption{Numerical solution of the BY equation for the $L=7$ state in terms of (rescaled) rapidities. Note that at
$g=0.5302$ the rapidity $u_1$ becomes complex.}
\label{uuasym}
\end{table}

\smallskip
Figure \ref{ReImu2} and table  \ref{uuasym} show that as $g$ increases $u_2$ approaches $-2-i/g$ which is a branch point of $x(u+i/g)$. It cannot however cross the cut Im$(u)=-1/g$ because the S-matrix has a pole if Im$(u_3)=-$Im$(u_2)=1/g$. As a result, as soon as $g\gtrsim 0.5301$, $u_1$ becomes complex, and $u_2$ and $u_3$ are repelled from the cuts Im$(u)=\mp 1/g$. Let us finally mention that the asymptotic energy of the state at $g=0.5302$ is complex which makes inapplicability of the BY equations for  $g\gtrsim 0.5301$ obvious.

\begin{figure}[t]
\begin{center}
\includegraphics[width=1.\textwidth]{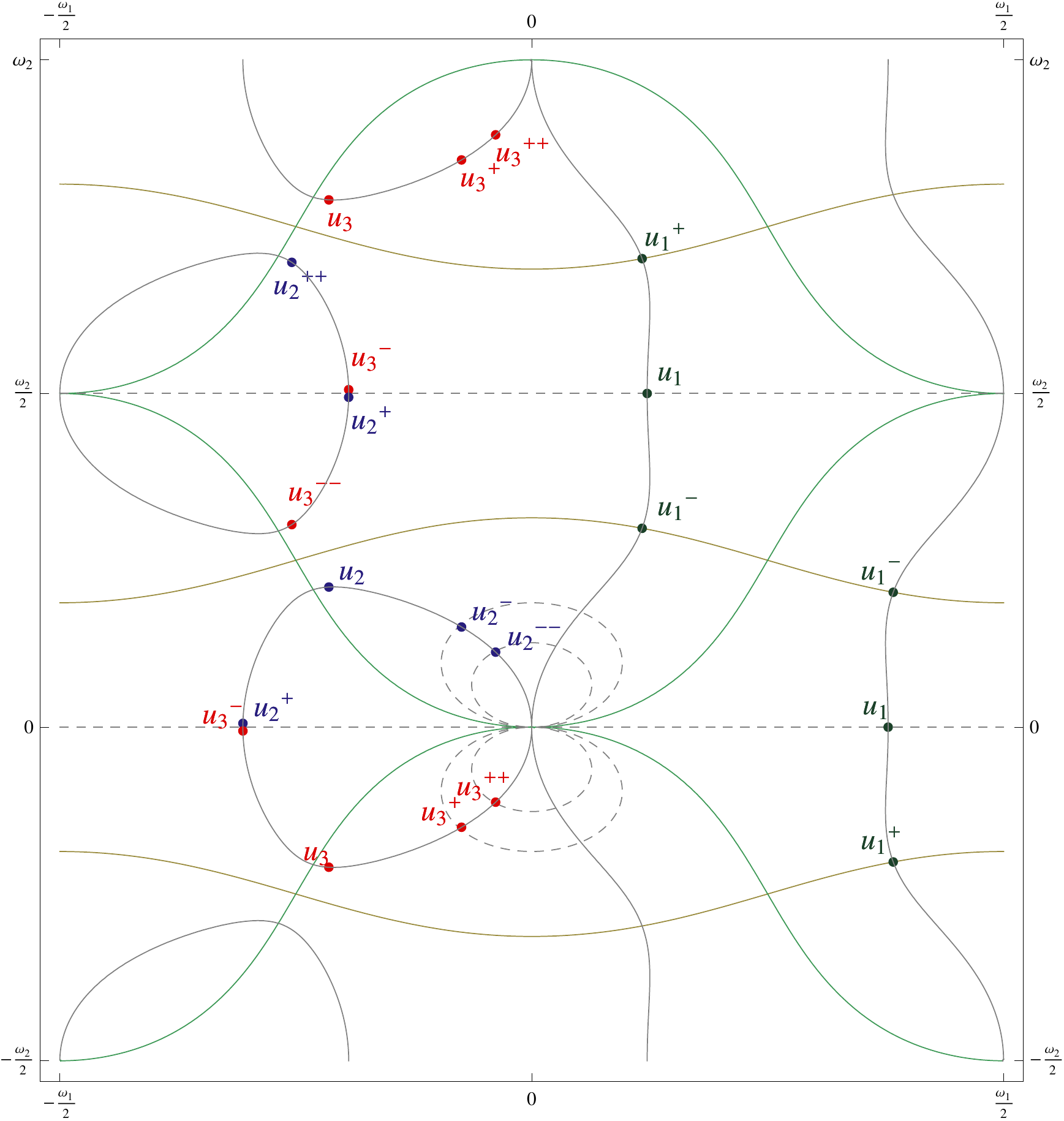}
\caption{The location of the (shifted) rapidities on the $z$-torus, at $g=1/2$. The green and yellow lines outline the mirror and string regions respectively, while the gray lines are the contours $\mbox{Re}(u(z)) = \mbox{Re}(u_i)$. The curved dashed gray lines correspond to the lines at $-2 i/g$ and $-3 i/g$ in the mirror $u$-plane. The straight dashed lines are the real mirror and real string line respectively.}
\label{fig:contour}
\end{center}
\end{figure}

\smallskip
To apply the contour deformation trick, it is convenient to know the location of the Bethe roots on the $z$-torus, as we have indicated in figure \ref{fig:contour}. We see, in particular, that the root $u_2$ is in the intersection of the string and mirror regions. We also see that as $g$ increases  the root $u_2$ approaches the point of intersection of the boundaries of the mirror and string regions.

\subsection*{Analytic properties of asymptotic Y-functions}

The numerical solution for $u_k(g)$ can be used to analyze the analytic properties of asymptotic Y-functions considered as functions of $g$.
According to the contour deformation trick, all driving terms in the TBA equations should come from zeroes and poles of $Y$--  and $1+Y$--functions.
In table \ref{zepo} we only list zeroes and poles relevant for constructing the TBA equations for the state, omitting those which do not appear in the equations.
For $Q\ge 4$ the poles of $Y_Q^o$ at $u_3-{i\ov g}(Q-1)$ lie below the analyticity strip
and are located on the grey curves associated to the complex rapidities, in the mirror region of figure \ref{fig:contour}. They are close to the points $u_2-{i\ov g}(Q-3)$ but lie on the other side of the line $-{i\ov g}(Q-2)$.


\smallskip

\begin{table}
\begin{center}
\begin{tabular}{|c|c|c|}
\hline Y${}^o$-function  & Zeroes   & Poles\\
\hline $Y_{M|w}$   & $r_{M\pm1}$ &  \\
\hline $1+Y_{M|w}$ & $r_M^-\,,\ r_M^+$ & $u_2-(M+1)i/g\,,\ u_3+(M+1)i/g$  \\
\hline $Y_{1|vw}$ &$u_1\,,\ r_0$ & \\
\hline $1+Y_{M|vw}$ && $u_2+(M+1)i/g\,,\ u_3-(M+1)i/g$ \\
\hline $Y_-$      & $u_2^-\,,\ u_3^+$ & $r_1\,,\ u_2^+\,,\ u_3^-$ \\
\hline $Y_+$       & & $r_1\,,\ u_1^-$\\
\hline $1-Y_-$   & $r_0^-\,,\ r_0^+$ & \\
\hline $1-Y_+$   &&\\
\hline $Y_1$   & $r_0$&$u_2^{++}\,,\ u_3^{--}$ \\
\hline $Y_2$   & &$u_2^{+}\,,\ u_3^{-}$ \\
\hline $Y_Q\,, Q\ge 3$   & &$u_2+{i\ov g}(Q-1)\,,\ u_3-{i\ov g}(Q-1)$ \\
\hline
\end{tabular}
\end{center}
\caption{Relevant roots and poles of asymptotic Y-functions within the mirror region.
}
\label{zepo}
\end{table}

For $g=1/2$  the rapidities and the first five roots take the following values
\bea\nonumber
&&\{ u_1,u_2,u_3\}=\{1.28853, -2.0896 - 2.00117 i, -2.0896 + 2.00117 i\}\,,\\\nonumber
&&\{ r_0,r_1,r_2,r_3,r_4\}=\{-1.28046, -1.18687, -1.16032, -1.14978, -1.14463\}\,.
\eea
We find that all the roots $r_M$ are real and they approach a limiting value at $M\to\infty$.  Next, note that $u_2^+$, $u_2^{++}$ and $u_3^-$, $u_3^{--}$ are within the analyticity strip $-1/g< {\rm Im}\, u<1/g$.
Thus, the first three $Y_Q$-functions have poles there. This is a drastically different situation compared to all previously studied states.
The function $Y_2$ in particular has two complex-conjugate poles located very close to the real line.


\subsection*{Analytic properties of exact  $Y_Q$-functions}\la{exactYQ}

In the last subsection we pointed out that the asymptotic functions $Y_1$, $Y_2$ and $Y_3$ have poles which lie within the analyticity strip. This leads to a dramatic change in the analyticity structure of the exact $Y_Q$-functions. In particular, we will show that this immediately implies that for small values of $g$ these functions must satisfy the exact Bethe equations $Y_Q(u^{(Q)})=-1$, where $u^{(Q)}$ is located close to a pole of $Y_Q$. The consideration is general and  works either for finite $J$ and small $g$ (which is the case we are interested in) or for finite $g$ and large $J$. To simplify the notations, we drop the index $Q$ and represent a Y-function in the form
\be
Y(u) = {y(u)\ov u-u_\infty}\,,
\ee
where  $y(u)$ is regular and  does not vanish at $u_\infty$ but it may have poles and zeroes elsewhere. Moreover, for any $u$ within the analyticity strip which is not its pole, $y(u)$ is of order $g^{2L-1}$  while $u_\infty$ scales as $1/g$ for small values of $g$.

We want to find $u_{-1}$ close to $u_\infty$ such that $Y(u_{-1})=-1$. We get immediately
\be
u_{-1}-u_\infty+y(u_{-1}) = 0
\ee
and expanding $y(u_{-1})$ around $u_\infty$  we obtain
\be\la{u1u0}
u_{-1} \approx u_\infty - y(u_\infty) = u_\infty - {\rm Res}\, Y(u_\infty)\,.
\ee
Since $y(u_\infty)$ is small $u_{-1}$ is close to $u_\infty$.

\medskip

Let us denote by $\tilde{u}_2^{(Q)}\approx u_2$ and $\tilde{u}_3^{(Q)}\approx u_3$ the points which are related to the exact locations of the poles of $Y_Q$ functions in the analyticity strip. The poles can be (and in general are) slightly shifted from their asymptotic positions for small but finite $g$. We assume that all Y-functions are real analytic in the mirror plane, that is $Y(u)^*=Y(u^*)$. Therefore $\tilde{u}_2^{(Q)}$ and $\tilde{u}_3^{(Q)}$ are complex conjugate to each other. Then from table \ref{zepo} we have
\bea
&&Y_1(\tu_2^{(1)++})=\infty\,,\quad Y_2(\tu_2^{(2)+})=\infty\,,\quad Y_3(\tu_2^{(3)++})=\infty\,,
\eea
where for definiteness we discuss the pole locations related to $\tilde{u}_2^{(Q)}$ only.

According to the discussion above, there are complex conjugate points $u_2^{(Q)}$ and $u_3^{(Q)}$ which are close to $\tilde{u}_2^{(Q)}$ and $\tilde{u}_3^{(Q)}$ (and to the asymptotic points $u_2\,, u_3$) such that
\bea
&&1+Y_1(u_2^{(1)++})=0\,,\quad 1+Y_2(u_2^{(2)+})=0\,,\quad 1+Y_3(u_2^{(3)++})=0\,.
\eea

We now show that the pole locations are determined by the zeroes of the functions $1+Y_Q$.
To this end we assume that
 for finite $g$ the exact $Y_Q$ functions  have the following representation  \cite{Kuniba1,Kuniba2}
\bea\la{YQexact}
 Y_Q&=&{\Upsilon}_Q\,{T_{Q,-1}\,T_{Q,1}\ov T_{Q-1,0}T_{Q+1,0}}\,,\quad
 {\Upsilon}_Q(v)=e^{-J\widetilde{\cal E}_Q(v)}\,  \prod_{i=1}^{N} S_{\sl(2)}^{Q1_*}(v,u_i)\,, ~~~ \eea
where in the $g\to 0$ limit the T-functions $T_{Q,\pm1}$ reduce to the asymptotic transfer matrices, $T_{0,0}=1$ and $T_{Q,0}$ reduce to 1.

\smallskip

The poles of the asymptotic $Y_Q^o$ appear due to poles in ${\Upsilon}_Q$.
As was mentioned above, the poles of the exact $Y_Q$-functions
are shifted from their asymptotic positions for finite $g$. This means that the T-function $T_{Q,0}$ must have a pole at $u_2+{i\ov g}Q$ and a zero at $\check{u}_2+{i\ov g}Q$ closed to the pole. Thus $T_{Q,0}$
satisfies
\bea\la{Tn0exact}
T_{Q,0}(u_2+{i\ov g}Q)=\infty\,,\quad T_{Q,0}(\check{u}_2^{(Q)}+{i\ov g}Q)=0\, ,~~~~
\eea
with similar properties for $u_3\,, \check{u}_3^{(Q)}$. In what follows we will assume  that \eqref{Tn0exact} hold for any $Q$ and that $\check{u}_2^{(Q)}\neq \check{u}_2^{(Q')}$ for any $Q\neq Q'$.

\smallskip

 The zeroes of $T_{Q,0}$ are obviously related to poles of $Y_Q$.
 In addition they are also related to the zeroes of $1+Y_Q$ as follows from the second representation for $Y_Q$
\bea\la{YQexact2}
1+Y_Q =  {T_{Q,0}^+\,T_{Q,0}^-\ov T_{Q-1,0}T_{Q+1,0}}\,,
\eea
which is valid if the T-functions satisfy the Hirota equations \cite{Hirota}.
Indeed
\bea
\cu_2^{(1)} = \tu_2^{(2)} =u_2^{(1)}\,,\quad \cu_2^{(2)} = \tu_2^{(1)} = \tu_2^{(3)} =u_2^{(2)}\,,\quad \cu_2^{(3)} =u_2^{(3)}\,,
\eea
and in general $\cu_2^{(Q)} =u_2^{(Q)}$.
Moreover, the conditions
\bea\la{Tn0exactb}
&&T_{Q,0}(u_2^{(Q)}+{i\ov g}Q)=0\,,\quad T_{Q,0}(u_3^{(Q)}-{i\ov g}Q)=0\,~~~~
\eea
imply that in the mirror $u$-plane the function $1+Y_Q$ for $Q\ge 2$ has  zeroes at
$$
 u_2^{(Q)}+{i\ov g}(Q-1)\,, \ u_2^{(Q)}+{i\ov g}(Q+1)\,, \ u_3^{(Q)}-{i\ov g}(Q-1)\,, \ u_3^{(Q)}-{i\ov g}(Q+1)\,,
 $$
 and poles at
$$u_2^{(Q-1)}+{i\ov g}(Q-1)\,,\ u_2^{(Q+1)}+{i\ov g}(Q+1)\,,\ u_3^{(Q-1)}-{i\ov g}(Q-1)\,,\ u_3^{(Q+1)}-{i\ov g}(Q+1)\,.$$
Since $1+Y_1$ has just $T_{2,0}$ in its denominator it only has poles at $u_2^{(2)++}$ and $u_3^{(2)--}$ while it has zeroes at
$u_2^{(1)++}$, $u_3^{(1)--}$, $u_2^{(1)}$ and $u_3^{(1)}$. In addition in the string $u$-plane it should have an extra zero at $u_1$ so that $Y_1$ satisfies the exact Bethe equation there. It is worth mentioning that the $Y_Q$-functions have additional poles related to the real Bethe root $u_1$, {\it e.g.} $Y_2$ has a pole at $u_1^-$. These additional poles however always lie outside integration contours and  therefore are irrelevant for constructing the TBA equations.

\subsection*{Analytic properties of auxiliary Y-functions}

The analytic properties of exact auxiliary Y-functions are similar to those of the asymptotic ones.  Basically, all zeroes and poles which depended on $u_2$ and  $u_3$ would now depend on $u_2^{(Q)}$ and $u_3^{(Q)}$.
In fact, all information about $u_2\,,\ u_3$ goes away and all Y-functions can only have singularities related to $u_2^{(Q)}$ and $u_3^{(Q)}$ as can be seen by performing a redefinition of T-functions which removes $\Upsilon_Q$ from $Y_Q$, see {\it e.g.} \cite{Suzuki:2011dj}.

\smallskip

The exact Y-functions can be expressed in terms of T-functions satisfying the Hirota equations in the standard form
except for $Y_{Q,0}=Y_Q$ for which we keep the conventional $\Upsilon_Q$ factor
\bea\la{YvT}
Y_{a,s}={T_{a,s-1}T_{a,s+1}\ov T_{a-1,s}T_{a+1,s}}\,,\quad 1+Y_{a,s} = {T_{a,s}^+T_{a,s}^-\ov T_{a-1,s}T_{a+1,s}}\,,\quad 1+{1\ov Y_{a,s}} = {T_{a,s}^+T_{a,s}^-\ov T_{a,s-1}T_{a,s+1}}\,,~~~
\eea
where $Y_{a,s}$ are related to our Y-functions as
\bea
&&Y_{1,-1}=-\frac{1}{Y_-^{(-)}}\,, \quad Y_{1,1}=-\frac{1}{Y_-^{(+)}}\,,
\qquad Y_{2,-2}=-Y_+^{(-)}\, ,\quad Y_{2,2}=-Y_+^{(+)}\, ,\\\nonumber
&&Y_{Q+1,-1}=\frac{1}{Y_{Q|vw}^{(-)}}\,,\quad
Y_{Q+1,1}=\frac{1}{Y_{Q|vw}^{(+)}}\, ,\qquad
Y_{1,-Q-1}=Y_{Q|w}^{(-)}\,,\quad
Y_{1,Q+1}=Y_{Q|w}^{(+)}\,.~~~~~
\eea
For states from the $\su(2)$ sector the auxiliary Y-functions from the left and right wings are equal, and we can drop the superscripts ${}^{(\pm)}$
and consider only the right wing Y-functions.
We want to know how the singularities of Y-functions related to the complex Bethe roots $u_2$ and $u_3$ are shifted due to the presence of $T_{a,0}$-functions in  \eqref{YvT}. Thus, we
discuss the $s=1$ case which includes $Y_-$ and $Y_{M|vw}$; the singularities of  $Y_{M|w}$ are shifted as well but they lie outside the analyticity strip and appear to be  irrelevant for the construction of the TBA equations. Concretely, we have

\bigskip
 \noindent
$\bullet$\ $Y_{M|vw}=1/Y_{M+1,1}$, $\ 1+Y_{M|vw} =  {T_{M+1,1}^+T_{M+1,1}^-\ov T_{M+1,0}T_{M+1,2}}$

\medskip

As we know the asymptotic $Y_{M|vw}$ function has poles at $u_2+(M+1)i/g$ and $u_3-(M+1)i/g$.  These poles disappear because $T_{M+1,0}$ has poles there. However new poles at
$u_2^{(M+1)}+(M+1)i/g$ and $u_3^{(M+1)}-(M+1)i/g$ appear because $T_{M+1,0}$  has zeroes there.

\bigskip
 \noindent
$\bullet$\ $Y_{-}=-1/Y_{1,1}$, $\
1-Y_{-} =  {T_{1,1}^+T_{1,1}^-\ov T_{1,0}T_{1,2}}$

\medskip

The  poles at $u_2^{+}$ and $u_3^{-}$ are shifted to
$u_2^{(1)+}$ and $u_3^{(1)-}$ because $T_{1,0}$  has zeroes there. Asymptotically $Y_-$ has zeroes at $u_2^-$ and $u_3^+$. The location of these zeroes is shifted too but we do not need them to write the TBA equations for $Y_{M|w}$ because
they lie outside the analyticity strip.

\section{Canonical TBA equations}\label{sect:canL7}

We begin our discussion of the TBA equations for the $L=7$, $n=2$ state with their canonical form even though the simplified TBA equations for $Y_{M|w}$, $Y_{M|vw}$ and $Y_{Q}, Q\ge2$ are completely fixed by the zeroes and poles of these functions in the analyticity strip. The main reason for this choice is that in the canonical TBA equations the auxiliary functions $Y_\pm$, $Y_{M|w}$ and $Y_{M|vw}$ appear in the form $1+1/Y$ while the $Y_Q$-functions appear in the form $1+Y_Q$, and therefore the poles of the auxiliary Y-functions and the zeroes of the $Y_Q$-functions do not produce any driving terms, meaning they play no role in the choice of the integration contours. In addition, the kernels appearing in the canonical TBA equations for $Y_\pm$ and $Y_1$ functions have a simpler analytic structure than those in the simplified and hybrid TBA equations which makes the analysis clearer.

\subsection*{Integration contour}

There is a choice of integration contours for $Y_Q$-functions which we believe is universal for the type of states under consideration.
We suggest that  for any state
the integration contours for $Y_Q$ are chosen such that
they enclose all the real zeroes of $1+Y_Q$ which are in the string region, and all the zeroes  and poles related to the complex Bethe roots which are below the real line of the mirror region, see figure \ref{fig:contour}. In particular,
the contours never go to the anti-mirror region of the $z$-torus. For the $L=7$, $n=2$ state this means that we take into account the poles of $Y_1$ at $u_3^{(2)--}$, of $Y_2$ at $u_2^{(1)+}$ and $u_3^{(3)}-{3i\ov g}$, and of $Y_Q, Q\ge3$ at $u_3^{(Q-1)}-{i\ov g}(Q-1)$ and $u_3^{(Q+1)}-{i\ov g}(Q+1)$,
and then the zeroes of $1+Y_1$ at $u_3^{(1)--}$ and $u_2^{(1)}$, of
$1+Y_2$ at $u_2^{(2)+}$ and $u_3^{(2)}-{3i\ov g}$, and of $1+Y_Q, Q\ge3$ at $u_3^{(Q)}-{i\ov g}(Q-1)$ and $u_3^{(Q)}-{i\ov g}(Q+1)$
 in the mirror $u$-plane, and finally the zero of $1+Y_1$ at $u_1$ in the string $u$-plane. The net result of these contributions is discussed in appendix \ref{sec:contcontrib}.  Let us stress that we do {\it not} take into account the complex Bethe root $u_3$ which is in the intersection of the string  and anti-mirror regions.
The choice of integration contours is not unique, and we will see that for
 the $L=7$, $n=2$ state we can make a simpler choice where we only take the contributions of the real zero of $1+Y_1$ in the string $u$-plane, the zeroes $u_2$ and $u_3$  of $1+Y_1$ in the mirror $u$-plane,  and all zeroes and poles of $1+Y_2$ in the analyticity strip of the mirror $u$-plane into account. With this choice the integration contours avoid all other zeroes and poles of $1+Y_Q$, even those which are inside the analyticity strip of the mirror $u$-plane.

\smallskip

The integration contours for all auxiliary Y-functions,  collectively denoted $Y_{\rm aux}$, run along the real line of the mirror region, lie  above the zeroes of Y-functions at real Bethe roots  and below all other real zeroes, and enclose all zeroes of
$Y_{\rm aux}$ and $1+ Y_{\rm aux}$ which are inside the analyticity strip of the mirror $u$-plane (including its boundary) but below the real line.

\smallskip

It is worth stressing that the integration contours discussed above are for the canonical TBA equations, and they are different from the contours for the simplified equations.
In particular, in the simplified TBA equations
 the integration contour for $Y_+$ should enclose the points
$u_k^-$ in the mirror $u$-plane for {\it real} Bethe roots $u_k$.

\smallskip

Let us now use the integration contours to derive the energy and momentum formulae, and the canonical TBA equations for the $L=7$, $n=2$ state.
We use the kernels and S-matrices  defined in \cite{AFS09}.

\subsection*{ Energy  formula}

According to the contour deformation trick the energy of an excited state is given by the formula
\bea
E=-{1\ov 2\pi}\int_{C_Q}\, du\, {d\tilde{p}_Q\ov du}\log(1+Y_Q)\,,
\eea
where $C_{Q}$  are the integration contours for  $Y_Q$ functions.

Formula \eqref{YQKQ} can be  used to take the integration contours back to the real line of the mirror $u$-plane.
We can think of ${1\ov 2\pi}{d\tilde{p}_Q\ov du}$ as a kernel with
$i\tilde{p}_Q$ being identified with $\log {\cal S}_Q$ in \eqref{YQKQ}.
It satisfies the discrete Laplace equation, and therefore the energy is
 \begin{align}
E=&\
i\tilde{p}_{1_*}(u_1)+i\tilde{p}_{1}(u_2^{(1)})
-i\tilde{p}_1(u_3^{(1)})-{1\ov 2\pi}\int_{-\infty}^\infty \, du\, {d\tilde{p}_Q\ov du}\log(1+Y_Q)\nonumber\\
&\quad -i\tilde{p}_2(u_2^{(1)+})+i\tilde{p}_2(u_2^{(2)+})
-i\tilde{p}_2(u_3^{(2)-})+i\tilde{p}_2(u_3^{(1)-})
\,.
\la{Ener}
\end{align}
Taking into account that
\bea
i\tilde{p}_{1_*}(u_1, v)=\E(u_1)
 \,,\quad i\tilde{p}_1(u_2^{(1)})=\E(u_2^{(1)}) \,,\quad -i\tilde{p}_{1}(u_3^{(1)}) = \E(u_3^{(1)})\,,
\eea
we get
 \begin{align}
E=&\
\sum_{i}\E(u_i^{(1)})
-{1\ov 2\pi}\int_{-\infty}^\infty\, du {d\tilde{p}_Q\ov du}\log(1+Y_Q)
\nonumber\\
&\quad -i\tilde{p}_2(u_2^{(1)+})+i\tilde{p}_2(u_2^{(2)+})
-i\tilde{p}_2(u_3^{(2)-})+i\tilde{p}_2(u_3^{(1)-})\,,
\la{Ener2}
\end{align}
where $u_1^{(1)}\equiv u_1$ and $\E(u)$ is the dispersion relation of a fundamental particle with rapidity variable $u$.

The energy of the state depends only on the singularities of $Y_1$ and $Y_2$. The contributions coming from the other $Y_Q$-functions cancel out, and the rapidity dependent terms
can also be thought of as purely originating from the zeroes of $1+Y_1$ in the string region, and the zeroes and poles of $Y_2$ in the analyticity strip of the mirror $u$-plane.

Let us also mention that the terms on the second line can be written as energies of two-particle bound states analytically continued to the mirror region.

\subsection*{Momentum formula}

Similar consideration can be applied to
 the formula for the total momentum (which should vanish for our state)
 given by
 \bea
P=-{1\ov 2\pi}\int_{C_Q}\, du\, {d\tilde{\E}_Q\ov du}\log(1+Y_Q)\,.
\eea
Since also ${1\ov 2\pi}{d\tilde{\E}_Q\ov du}$ satisfies  the discrete Laplace equation,
identifying
$i\tilde{\E}_Q$ with $\log {\cal S}_Q$ in \eqref{YQKQ},  we obtain
\begin{align}
P=&
i\tilde{\E}_{1_*}(u_1)+i\tilde{\E}_{1}(u_2^{(1)})
-i\tilde{\E}_1(u_3^{(1)})-{1\ov 2\pi}\int_{-\infty}^\infty\, du\, {d\tilde{\E}_Q\ov du}\log(1+Y_Q)
\nonumber\\
&-i\tilde{\E}_2(u_2^{(1)+})+i\tilde{\E}_2(u_2^{(2)+})
-i\tilde{\E}_2(u_3^{(2)-})+i\tilde{\E}_2(u_3^{(1)-})\,.
\la{momen}
\end{align}
Taking into account that
 \begin{align}\nonumber
i\tilde{\E}_{1_*}(u_1, v)=p(u_1)\equiv p_1\,,\quad i\tilde{\E}_1(u_2^{(1)})=p(u_2^{(1)})\equiv p_2 \,,\quad -i\tilde{\E}_{1}(u_3^{(1)}) = p(u_3^{(1)})\equiv p_3 \,,
\end{align}
we get the following formula for the total momentum
 \begin{align}
P=&
\sum_i\, p_i-{1\ov 2\pi}\int_{-\infty}^\infty\, du\, {d\tilde{\E}_Q\ov du}\log(1+Y_Q)
\nonumber\\
&-i\tilde{\E}_2(u_2^{(1)+})+i\tilde{\E}_2(u_2^{(2)+})
-i\tilde{\E}_2(u_3^{(2)-})+i\tilde{\E}_2(u_3^{(1)-})\,.
\la{momen2}
\end{align}
It was noticed in \cite{CFT10} that the TBA equations imply a quantization condition for the total momentum. Thus, since the total momentum vanishes as $g\to 0$ and it changes continuously with $g$ the total momentum should vanish for any $g$.

\subsection*{Canonical equations for $w$-strings}

The excited state canonical TBA equations for $w$ strings are given by
\begin{align}\nonumber
\log Y_{M|w} = &\log( 1+\frac{1}{Y_{N|w}})\star_{C_{N|w}} K_{NM} + \log \frac{1-\frac{1}{Y_-}}{1-\frac{1}{Y_+}} \star_{C_{\pm}} K_M \, ,
\end{align}
where $C_{N|w}$, $C_{-}$ and $C_{+}$  are the integration contours for $Y_{N|w}$, $Y_-$ and $Y_+$ functions.
Taking the integration contours back to real line of the mirror $u$-plane, $Y_+$ does not produce any driving term, the zero of $1-Y_-$ at $r_0^-$ produces
$-\log S_{M} (r_0^--v)$, and finally the zeroes of $Y_{N|w}$ at $r_{N-1}$ and $r_{N+1}$, and the zeroes of
$1+Y_{N|w}$ at $r_{N}^-$ give
$$
+\tfrac{1}{2}\sum_{N=1}^{\infty} \log S_{NM} (r_{N-1}-v)S_{NM} (r_{N+1}-v) - \sum_{N=1}^{\infty} \log S_{NM} (r_N^--v)\,,
$$
where $+1/2$ in the first term appears due to the principal value prescription in (\ref{YwcanTBA}).

Taking into account that $S_{NM} (u-v)$ satisfies the discrete Laplace equation
$$
S_{N-1,M} (u-v)S_{N+1,M} (u-v)=S_{NM} (u^--v)S_{NM} (u^+-v) \,,
$$
we can write the canonical TBA equations for $w$ strings
in the form
\begin{align}
\nonumber
\log Y_{M|w} = & \log( 1+\frac{1}{Y_{N|w}})\star_{p.v.} K_{NM} + \log \frac{1-\frac{1}{Y_-}}{1-\frac{1}{Y_+}}\hstar K_M \\
\la{YwcanTBA}
& +\tfrac{1}{2}\sum_{N=1}^{\infty} \log \frac{S_{NM} (r_{N}^+-v)}{S_{NM} (r_{N}^--v)}
+ \tfrac{1}{2}\log S_{1M} (r_0-v)-\log S_{M} (r_0^--v) \, ,
\end{align}
where $\log S_{1M} (r_0-v)$ should be understood as $\log S_{M-1} (r_0-v) +\log S_{M+1} (r_0-v)$.

\subsection*{Canonical  equations for $vw$-strings}

The excited state  canonical TBA equations for $vw$ strings are given by
\begin{align}\nonumber
\log Y_{M|vw} = & \log( 1+\frac{1}{Y_{N|vw}})\star_{C_{N|vw}} K_{NM} + \log \frac{1-\frac{1}{Y_-}}{1-\frac{1}{Y_+}}\star_{C_{\pm}}  K_M- \log (1+Y_Q)\star_{C_{Q}}  K^{QM}_{xv}  \, ,
\end{align}
where $C_{N|vw}$ and $C_{Q}$  are the integration contours for $Y_{N|vw}$, and $Y_Q$ functions.
Taking the integration contours back to real line of the mirror $u$-plane and using
formula \eqref{YQKQ}, we can bring the canonical TBA equations for $vw$ strings to the form
\begin{align}
\log Y_{M|vw} &=   \log( 1+\frac{1}{Y_{N|vw}})\star_{p.v.} K_{NM} + \log \frac{1-\frac{1}{Y_-}}{1-\frac{1}{Y_+}}\hstar K_M - \log (1+Y_Q)\star K^{QM}_{xv} \nonumber \\
&+ \frac{1}{2} \log \frac{S_{1M} (r_0-v)}{S_{1M} (u_1-v)} - \log S_{M} (r_0^--v)-\log \frac{S^{1M}_{xv} (u_3^{(1)},v)}{S^{1M}_{xv} (u_2^{(1)},v)}\nonumber\\
& \quad \qquad  + \log S^{1_*M}_{xv} (u_1,v) - \log \frac{S^{2M}_{xv} (u_3^{(2)-} ,v)}{S^{2M}_{xv} (u_3^{(1)-},v)}\frac{S^{2M}_{xv} (u_2^{(1)+},v)}{S^{2M}_{xv} (u_2^{(2)+},v)} \, ,
\end{align}
where the first term on the second line appears due to the zeroes of $Y_{1|vw}$ at $u_1$ and $r_0$, the second term arises because of the zero of $1+Y_{-}$  at $r_0^-$.

\subsection*{Canonical  equations for $Y_+/Y_-$}

Formula \eqref{YQKQ}  can be used to write the TBA equation for
$Y_+/Y_-$
\begin{align}\la{ypovym}
\log {Y_+\ov Y_-} = \,   \log(1 +  Y_{Q})\star K_{Qy} -& \sum_i \log S_{1_*y}(u_i^{(1)},v)  +\log {S_{2y}(u_2^{(1)+}, v)\ov S_{2y}(u_2^{(2)+}, v)}{S_{2y}(u_3^{(2)-}, v)\ov S_{2y}(u_3^{(1)-}, v)}\,,
\end{align}
where  we have used that
 \bea\nonumber
 S_{1_*y}(u_3^{(1)},v) = 1/S_{1y}(u_3^{(1)},v)\,,\quad S_{1_*y}(u_2^{(1)},v) = S_{1y}(u_2^{(1)},v)
 \,.~~~~~
\eea
The driving terms can be also explained by contours which enclose only the zeroes of $1+Y_1$ in the string region, and all zeroes and poles of $Y_2$ in the analyticity strip of the mirror $u$-plane, while avoiding other zeroes and poles of $1+Y_Q$.

\subsection*{Canonical equation for $Y_+Y_-$}

Let us now analyze the canonical TBA equation for $Y_+Y_-$ given by
 \begin{align}
\la{Yfory2}
 &\log {Y_+ Y_-} =  - \log\left(1+Y_Q \right)\star_{C_{Q}}  K_Q
\\\nonumber
&+2\log \big(1+{1\ov Y_{M|vw}}
\big)\star_{C_{M|vw}}  K_M - 2\log \big(1+{1\ov Y_{M|w}}\big)\star_{C_{M|w}}  K_M\,.
\end{align}
$\bullet$ The term $- \log(1 +  Y_{Q} )\star_{C_{Q}}  K_Q$ produces
 \begin{align}
&
\log {S}_{1}(u_1- v)+\log {{S}_1(u_2^{(1)}- v)\ov {S}_{1}(u_3^{(1)}- v)}-\log {{S}_2(u_2^{(1)+}- v)\ov {S}_2(u_2^{(2)+}-v)}{{S}_2(u_3^{(2)-}- v)\ov {S}_2(u_3^{(1)-}- v)}
\end{align}

 \noindent $\bullet$ The term $2\log (1+{1\ov Y_{M|vw}}) \star_{C_{M|vw}}  K_M$ produces
 \begin{align}
&2\log (1+{1\ov Y_{M|vw}}) \star_{p.v.} K_M
-\log {S}_{1}(u_1- v)+\log {S}_{1}(r_0- v)\,,
\end{align}
where we take into account that the contour runs above $u_1$ but below $r_M$.

 \noindent $\bullet$ The term $2\log (1+{1\ov Y_{M|w}}) \star_{C_{M|w}}  K_M$ produces
 \begin{align}
&2\log (1+{1\ov Y_{M|w}}) \star_{p.v.} K_M +
\log {S}_{M}(r_{M-1}- v){S}_{M}(r_{M+1}- v) -2\log {S}_{M}(r_{M}^- - v)\,,
\end{align}
where we sum over $M$ from 1 to $\infty$. Computing the sum we get
\begin{align}
&2\log (1+{1\ov Y_{M|w}}) \star_{p.v.} K_M +\log {S}_{1}(r_{0}- v)-
\sum_{M=1}^\infty\log {{S}_{M}(r_{M}^-- v)\ov {S}_{M}(r_{M}^+- v) }\,.
\end{align}
Thus, the canonical TBA equation for $Y_+Y_-$ is
\begin{align}
\la{TBAcanypym} \log {Y_+ Y_-} &=  - \log\left(1+Y_Q
\right)\star K_Q+2\log {1+{1\ov Y_{M|vw}}
 \ov 1+{1\ov Y_{M|w}}}\star_{p.v.} K_M
 \\\nonumber
&+\log {{S}_1(u_2^{(1)}- v)\ov {S}_{1}(u_3^{(1)}- v)}-\log {{S}_2(u_2^{(1)+}- v)\ov {S}_2(u_2^{(2)+}-v)}{{S}_2(u_3^{(2)-}- v)\ov {S}_2(u_3^{(1)-}- v)}
+
\sum_{M=1}^\infty\log {{S}_{M}(r_{M}^-- v)\ov {S}_{M}(r_{M}^+- v) }\,.
\end{align}

\subsection*{Canonical equations for  $Q$-particles}

The excited state canonical TBA equation for $Y_Q$ can be written in the form
\begin{align}\notag
\log Y_Q =&\, - L_{\rm TBA}\, \tH_{Q} + \log\left(1+Y_{M} \right) \star_{C_M} K_{\sl(2)}^{MQ}
 + 2 \log\left(1+ \frac{1}{Y_{M|vw}} \right) \star_{C_{M|w}}  K^{MQ}_{vwx}
\\[1mm] 
&\quad +  \log {1- \frac{1}{Y_-} \ov 1-\frac{1}{Y_+} } \star_{C_\pm}  K_{Q}  +  \log \big(1-\frac{1}{Y_-}\big)\big( 1 - \frac{1}{Y_+} \big) \star_{C_\pm}  K_{yQ} \, .
\end{align}

Taking the integration contours back to the real line of the mirror $u$-plane and using \eqref{YQKQ}, we obtain
\begin{align}\notag
\log Y_Q =&\, - L_{\rm TBA}\, \tH_{Q} + \log\left(1+Y_{Q'} \right) \star K_{\sl(2)}^{Q'Q}
+ 2 \log\left(1+ \frac{1}{Y_{M'|vw}} \right) \star_{p.v} K^{M'Q}_{vwx}
\\[1mm]\la{YQcanTBA}
&\quad +  \log {1- \frac{1}{Y_-} \ov 1-\frac{1}{Y_+} } \hstar K_{Q}  +  \log \big(1-\frac{1}{Y_-}\big)\big( 1 - \frac{1}{Y_+} \big) \hstar K_{yQ} \\[1mm] \notag
&\quad -\log S_{\sl(2)}^{1_*Q}(u_1,v){S_{\sl(2)}^{1Q}(u_2^{(1)},v)\ov
S_{\sl(2)}^{1Q}(u_3^{(1)},v)}+\log \frac{S_{\sl(2)}^{2Q}(u_2^{(1)+},v)}{S_{\sl(2)}^{2Q}(u_2^{(2)+},v)}\frac{S_{\sl(2)}^{2Q}(u_3^{(2)-},v)}{S_{\sl(2)}^{2Q}(u_3^{(1)-},v)}
\\[1mm]\notag
&\quad-\log S^{1Q}_{vwx}(u_1,v)+\log S^{1Q}_{vwx}(r_0,v) - \log{S_Q}(r_0^--v) - \log{S_{yQ}}(r_0^-,v)\notag\,.
\notag
\end{align}
The driving terms dependent on $S_{\sl(2)}^{1Q}$ of the mirror-mirror region can be rewritten in terms of $S_{\sl(2)}^{1_*Q}$ of the string-mirror region by noting that $u_2^{(1)}$ lies in overlap of the string and mirror regions, meaning that $S_{\sl(2)}^{1Q}(u_2^{(1)},v)=S_{\sl(2)}^{1_*Q}(u_2^{(1)},v)$, and that
$u_3^{(1)}$ lies in overlap of anti-string and mirror regions, meaning that we can use crossing relations \cite{Janik} to replace
$S_{\sl(2)}^{1Q}(u_3^{(1)},v)$ with  $1/S_{\sl(2)}^{1_*Q}(u_3^{(1)},v)$.
\begin{equation}
\label{Cross}
S_{\sl(2)}^{1Q}(u_3^{(1)},v)S_{\sl(2)}^{1_* Q}(u_3^{(1)},v) = \frac{1}{h_Q(u_3^{(1)},v)^2},
\end{equation}
where $h_Q(u,v)$ is defined as  \cite{AF06}
\bea
h_Q(u,v) = {x_s(u-{i\ov g})-x(v+{i\ov g}Q)\ov x_s(u-{i\ov g})- x(v-{i\ov g}Q) }\,  {1 - {1\ov x_s(u+{i\ov g})x(v+{i\ov g}Q)}\ov  1-{1\ov x_s(u+{i\ov g})x(v-{i\ov g}Q)}}\,.
\eea
We would also like to point out that
 since $u_3^{(1)}$  is in the second strip we can rewrite     $h_Q(u_3^{(1)},v)$ as
\begin{align}
\label{h2identity}
&h_Q^2(u_3^{(1)},v)=\frac{S_{yQ}(u_3^{(1)+},v)}{S_{yQ}(u_3^{(1)-},v)}S_Q(u_3^{(1)+}-v)S_Q(u_3^{(1)-}-v)\,.
\end{align}

\section{Simplified and hybrid TBA equations}\la{sect:simpTBA}

The canonical TBA equations can be used to derive the simplified and hybrid TBA equations following the consideration in  \cite{AF09b,AFS09}.
To this end we apply the operator $(K+1)^{-1}_{NM}$ to both sides of the canonical TBA equations, sum over $N$ and use identities listed in appendix \ref{identY}.

\subsection*{Simplified equations for  $Y_{M|w}$}

The simplified TBA equations for $Y_{M|w}$ are found to be
\bea\la{TBAsimyw}
\log Y_{M|w} &=&  \log(1 +  Y_{M-1|w})(1 +
Y_{M+1|w})\star s\\\nonumber
 &+& \delta_{M1}\, \log{1-{1\ov Y_-}\ov 1-{1\ov Y_+} }\hstar s - \log S(r_{M-1}^- - v)S(r_{M+1}^- - v)\, .~~~~~
\eea
The above driving terms appear due to the zero of $1-Y_{-}$ at $r_0^-$ and the zeroes of $1+Y_{M|w}$ at $r_M^-$.

\subsection*{Simplified equations for  $Y_{M|vw}$}

The simplified equations for $Y_{M|vw}$ are found by applying $(K+1)^{-1}$ and subsequently rewriting them by using the simplified equation for $Y_+/Y_-$, convoluted with $s$. This gives the following equations for $vw$ strings
\begin{align}
\log Y_{M|vw} = & - \log(1 +  Y_{M+1})\star s +
\log(1 +  Y_{M-1|vw} )(1 +  Y_{M+1|vw})\star
s\nonumber\\\la{TBAsimyvw}
&  + \delta_{M1} \log{1-Y_-\ov 1-Y_+}\hstar s \\
& + \delta_{M1}\Big(
\log\frac{S(u_2^{(2)+} - v)}{S(u_3^{(2)-} - v)} - \log S(u_1^- - v)S(r_{0}^- - v)\Big)\,\nonumber.
\end{align}
The contour deformation trick explains the driving terms for $M=1$ as follows.
From $1+Y_2$ we get
$$
\log S(u_2^{(2)+} - v) - \log S(u_2^{(1)+} - v) + \log S(u_3^{(2)---} - v) - \log S(u_3^{(3)---} - v)\,.~~~~
$$
Next, $1-Y_-$ contributes $
+ \log S(u_2^{(1)+} - v)- \log S(r_{0}^- - v)
$,
while $1-Y_+$ contributes
$
- \log S(u_1^- - v)
$.
Finally, $1+Y_{2|vw}$ gives
$
+\log S(u_3^{(3)---} - v)
$.
Summing this up we get the desired driving terms.

\smallskip

The contributions from the poles of $1+Y_{M+1}$ for higher $M$ cancel the contribution from the poles of $1+Y_{M-1|vw}$ and $1+Y_{M+1|vw}$. Note that to explain the driving terms in the simplified equations we would have to take into account the poles of $Y_{M|vw}$ outside the analyticity strip.

\subsection*{Simplified equations for  $Y_\pm$}

The simplified TBA equation for the ratio $Y_+/Y_-$ coincides with \eqref{ypovym}.

To derive the equation for $Y_+Y_-$, we need to compute the infinite sums involving the $Y_{M|w}$ and $Y_{M|vw}$-functions which is done in appendix \ref{identY}.
Using these formulae, the TBA equation \eqref{ypovym} for $Y_+/Y_-$, and the identities from appendix \ref{identY} , the TBA equation for $Y_+Y_-$ can be transformed to the simplified form
\begin{align}\nonumber
 &\log {Y_+ Y_-} =\  2\log {1+Y_{1|vw}
 \ov 1+Y_{1|w}}\star s- \log\left(1+Y_Q
\right)\star K_Q
+2\log(1 + Y_{Q}) \star K_{xv}^{Q1}\star s
 \\
 \nonumber
&\qquad\qquad\quad+2\log S(r_{1}^- - v)-2 \log S_{xv}^{1_*1}(u_1,v)\star s+\log {S_2(u_1- v)} \star s
 \\\la{TBAsimypym}
&\qquad\qquad\quad -2 \log {S_{xv}^{11}(u_2^{(1)},v)\ov S_{xv}^{11}(u_3^{(1)},v)}\star s+\log {{S}_1(u_2^{(1)}- v)\ov {S}_{1}(u_3^{(1)}- v)}
  \\\nonumber
 &\qquad\qquad\quad
-\log {{S}_2(u_2^{(1)+}- v)\ov {S}_2(u_2^{(2)+}-v)}{{S}_2(u_3^{(2)-}- v)\ov {S}_2(u_3^{(1)-}- v)}
+2 \log {S^{21}_{xv}(u_2^{(1)+}, v)S^{21}_{xv}(u_3^{(2)-}, v)\ov S^{21}_{xv}(u_2^{(2)+}, v)S^{21}_{xv}(u_3^{(1)-}, v)} \star s \,.
\end{align}
We recall that the integration contours run a bit above the real line.
The driving terms in this equation can obviously be explained by our choice of integration contours. To be sure that no other driving terms appear the kernel $K_{xv}^{Q1}\star s$ and its S-matrix should be analytically continued to complex points in the mirror and string $u$-planes. This is non-trivial because
$K_{xv}^{Q1}$ has poles, and we have not attempted to derive \eqref{TBAsimypym} starting with the simplified equation with deformed contours.
We have however checked that $Y_-$ satisfies its Y-system equation \cite{GKV09} which requires a very delicate balance of the driving terms in \eqref{TBAsimypym} and \eqref{ypovym}.

\smallskip

Let us finally present yet another form of the simplified TBA equation
\begin{align}\nonumber
 &\log {Y_+ Y_-} =\  2\log {1+Y_{1|vw}
 \ov 1+Y_{1|w}}\star s- \log\left(1+Y_Q
\right)\star K_Q
+2\log(1 + Y_{Q}) \star K_{xv}^{Q1}\star s
 \\
\la{TBAsimypym2}
&\qquad\qquad\quad+ 2\log{S(r_1^- - v)}-  \sum_i \log {\big(S_{xv}^{1_*1}\big)^2\ov S_2}\star s(u_i^{(1)},v) + \log {S(u_2^{(1)}-v)\ov S(u_3^{(1)}-v)}
  \\\nonumber
 &\qquad\qquad\quad
-\log {{S}_2(u_2^{(1)+}- v)\ov {S}_2(u_2^{(2)+}-v)}{{S}_2(u_3^{(2)-}- v)\ov {S}_2(u_3^{(1)-}- v)}
+2 \log {S^{21}_{xv}(u_2^{(1)+}, v)S^{21}_{xv}(u_3^{(2)-}, v)\ov S^{21}_{xv}(u_2^{(2)+}, v)S^{21}_{xv}(u_3^{(1)-}, v)}\star s  \,,
\end{align}
where we used identities from appendix \ref{identY} to
replace the mirror-mirror S-matrices $S_{xv}^{11}$ with the string-mirror $S_{xv}^{1_*1}$. As before this form indicates that it might be possible to choose the integration contours for $Y_Q$ so that they would only
enclose  the zeroes of $1+Y_1$ in the string region, and all zeroes and poles of $Y_2$ in the analyticity strip of the mirror $u$-plane. Such a choice, however, would require very intricate integration contours for the auxiliary Y-functions which we will not attempt to describe.

\subsection*{Simplified TBA equations for $Y_Q$}

Applying $(K+1)^{-1}$
to the canonical equations,
the terms which depend on the Y-functions (and involve only the kernels) produce the usual contributions \cite{AF09b,AF09d}. The contribution of the driving terms can be found through the identities from appendix \ref{identY}.
This yields the following simplified TBA equations for $Q$-particles

\bigskip
 \noindent
$\bullet$ $\ Q\ge 4\ $
\bea
\log Y_{Q}&=&\log{\left(1 +  {1\ov Y_{Q-1|vw}} \right)^2\ov (1 +  {1\ov Y_{Q-1} })(1 +  {1\ov Y_{Q+1} }) }\star s\la{YforQ4}
\,~~~~~~~
\eea
The contributions of the zeroes of $1+Y_{Q}$ cancel each
other for $Q\ge 3$, and therefore no driving term appears.

\bigskip
 \noindent
$\bullet$ $\ Q=3\ $

\bea
\log Y_{3}&=&\log S(u_2^{(2)+}-v)-\log S(u_3^{(2)-}-v)+\log{\left(1 +  {1\ov Y_{2|vw}} \right)^2\ov (1 +  {1\ov Y_{2} })(1 +  {1\ov Y_{4} }) }\star s\la{YforQ3}
\,.~~~~~~~
\eea
The driving terms are explained by the zeroes of $1+Y_2$ at  $u_2^{(2)+}$ and $u_3^{(2)---}$. The contributions of the two zeroes of $1+Y_4$ cancel each
other.

\bigskip
 \noindent
$\bullet$ $\ Q=2\ $
\bea\la{YforQ2}
\log Y_{2}&=&\log S(u_2^{(1)}-v)-\log S(u_3^{(1)}-v)+\log{\left(1 +  {1\ov Y_{1|vw}} \right)^2\ov (1 +  {1\ov Y_{1} })(1 +  {1\ov Y_{3} }) }\star_{p.v} s
\,,~~~~~~~
\eea
The contribution due to the zero of $1+Y_1$ at $u_1$  in the string region is canceled because of the zero of $Y_{1|vw}$ at $u_1$. Next, both $Y_{1|vw}$ and $Y_{1}$ have zeroes at $u=r_0$ which contributions cancel each other. The contributions of the two zeroes of $1+Y_3$ cancel each
other, and we are left with the two driving terms produced by the zeroes of
$1+Y_1$ at $u_2^{(1)}$ and $u_3^{(1)--}$.

\bigskip
 \noindent
$\bullet$ $\ Q=1\ $
\begin{align}
\log Y_1 &=  \log \left(1-\frac{1}{Y_-}\right)^2\hstar s -  \log \left(1+\frac{1}{Y_2}\right) \star s+ \log \frac{S(u_2^{(2)+} - v)}{S(u_3^{(2)-} - v)} -2 \log S(r_0^--v)
\nonumber\\
&- \check{\Delta} _v{\cstar} s +\log \check{S_1}{\cstar} s (u_1,v) - \log \check{S_1}{\cstar} s (r_0,v)+ 2\log \check{S}{\cstar} s (r_0^-,v)  \label{STbaQ1}
\\
\nonumber
&+\log \check{\Sigma}_{1_*}^2(u_1,v){\check{\Sigma}_{1}^2 (u_2^{(1)},v) \ov \check{\Sigma}_{1}^2(u_3^{(1)},v)  }{\cstar} s
 -  \log {\check{\Sigma}_2^2  (u_2^{(1)+},v) \check{\Sigma}_2^2 (u_3^{(2)-},v) \ov \check{\Sigma}_2^2  (u_3^{(1)-},v) \check{\Sigma}_2^2 (u_2^{(2)+},v) } {\cstar} s \nonumber
\end{align}
where
\bea \nonumber
\check{\Delta}_v&=&L\check{{\cal E}} +\log\left(1-\frac{1}{Y_-}\right)^2\left(1-\frac{1}{Y_+}\right)^2{\hstar}\check{K}+\log\left(1+\frac{1}{Y_{M|vw}}\right)^2 \star_{p.v.}\check{K}_M\\\la{delv}
&+&2\log(1+Y_Q)\, \star\, \check{K}^{\Sigma}_Q\, . \eea
Note that the terms $+\log \check{S_1}{\cstar} s (u_1,v) - \log \check{S_1}{\cstar} s (r_0,v)$  on the second line of \eqref{STbaQ1} combine with
$\log\left(1+\frac{1}{Y_{M|vw}}\right)^2 \star_{p.v.}\check{K}_M$ and remove the principal value prescription in the integral. The driving terms in this equation can be explained by the zero of $1-Y_-$ at $r_0^-$, the zeroes of  $Y_{1|vw}$ at $r_0$ and $u_1$, the zeroes and poles of $1+Y_Q$, and our choice of the integration contours. The infinite sum involving $Y_{M|vw}$-functions in \eqref{delv}  can be computed in the same way as it was done in \cite{AF09d}, producing additional driving terms.
\smallskip

Using the identities in appendix  \ref{identY}, the TBA equation for $Y_1$ can be rewritten to contain $\check{\Sigma}_{1_*}$-terms similar
to the ones for the Konishi state, namely
\begin{align}
\log Y_1 &=  \log \left(1-\frac{1}{Y_-}\right)^2\hstar s -  \log \left(1+\frac{1}{Y_2}\right) \star s+ \log \frac{S(u_2^{(2)+} - v)}{S(u_3^{(2)-} - v)} {S(r_0^+-v) \ov S(r_0^--v)}
\nonumber\\
&- \check{\Delta} _v{\cstar} s +\sum_i \log \check{\Sigma}_{1_*}^2 \check{S_1} (u_i^{(1)},v)\cstar s
 + \sum_{j=2,3}\log\frac{\check{S}(u_j^{(1)+},v)}{\check{S}(u_j^{(1)-},v)}\cstar s
  \label{STbaQ1b}
\\
\nonumber
& +\log {\check{S} (r_0^-,v)\ov \check{S} (r_0^+,v)}{\cstar} s -  \log {\check{\Sigma}_2^2  (u_2^{(1)+},v) \check{\Sigma}_2^2 (u_3^{(2)-},v) \ov \check{\Sigma}_2^2  (u_3^{(1)-},v) \check{\Sigma}_2^2 (u_2^{(2)+},v) } {\cstar} s \,.\nonumber
\end{align}
The simplified equations for the Y-functions can be used to prove their real analyticity.

\subsection*{Hybrid TBA equations for $Y_Q$}

The hybrid form of the TBA equations for $Y_Q$  is derived from the corresponding canonical equations
and the simplified equations for $Y_{M|vw}$ in the same way as was done in \cite{AFS09}.
To make the presentation transparent, we introduce a function which combines the terms on the right hand side of the hybrid ground state TBA equation
\bea\la{GQ}
G_Q(v)&=& - L_{\rm TBA}\, \tH_{Q} +\log \left(1+Y_{Q'} \right)
\star (K_{\sl(2)}^{Q'Q}+2 s\star K_{vwx}^{Q'-1,Q})\\
&+&  2 \log \(1 + Y_{1|vw}\) \star s \hstar K_{yQ} +2\log(1+Y_{Q-1|vw})\star s\nonumber \\
&  -&  2  \log{1-Y_-\ov 1-Y_+} \hstar s \star K^{1Q}_{vwx} +  \log
{1- \frac{1}{Y_-} \ov 1-\frac{1}{Y_+} } \hstar K_{Q}  +  \log
\big(1-\frac{1}{Y_-}\big)\big( 1 - \frac{1}{Y_+} \big) \hstar
K_{yQ} \, . \nonumber  \eea
With the help of $G_Q$, the hybrid TBA equations for $Y_Q$ read as
\begin{align}
\log Y_Q(v) &= G_Q(v) -\log \frac{S_{\sl(2)}^{1Q}(u_2^{(1)},v)}{S_{\sl(2)}^{1Q}(u_3^{(1)},v)}S_{\sl(2)}^{1_*Q}(u_1,v)\nonumber+\log \frac{S_{\sl(2)}^{2Q}(u_3^{(2)-},v)}{S_{\sl(2)}^{2Q}(u_3^{(1)-},v)}\frac{S_{\sl(2)}^{2Q}(u_2^{(1)+},v)}{S_{\sl(2)}^{2Q}(u_2^{(2)+},v)} \\ \nonumber
& -\log
S^{1Q}_{vwx}(u_1,v) +\log S^{1Q}_{vwx}(r_0,v) -\log S_Q(r_0^--v)S_{yQ}(r_0^-,v)
 \\
&+ 2\log S(u_1^-,v)S(r_0^-,v)\star_{p.v.} K_{vwx}^{1Q}
-2\log {S(u_2^{(2)+},v)\ov S(u_3^{(2)-},v)}\star K_{vwx}^{1Q} \label{HybridQ}\, . \end{align}
It is worth mentioning that the first two terms on the second line combine nicely with the first term on the third line and give the term $2\log S(u_1^-,v)S(r_0^-,v)\star K_{vwx}^{1Q}$ with the usual integration contour, {\it i.e.} running above $u_1$ but below $r_0$. Finally, we point out that equation (\ref{Cross}) allows us to rewrite  (\ref{HybridQ}) in terms of  the S-matrices $S_{\sl(2)}^{1_*Q}$ which is  useful for analyzing the exact Bethe equations and for numerics.

\section{Exact Bethe equations}\label{sect:EBandQ}
In this section we   discuss the exact Bethe equations (quantization conditions)
for the roots $u_1$ and $u_i^{(1,2)}$ where
$i=2,3$.
Let us recall that according to the discussion in section 3 we can choose the following equations as our quantization conditions
\bea\la{EBEall1}
Y_{1_*}(u_1)=-1\,,\quad Y_1(u_2^{(1)++})=-1\ &\Leftrightarrow&\ Y_1(u_3^{(1)--})=-1\,,\\
\la{EBEall2}
 Y_2(u_2^{(2)+})=-1\ &\Leftrightarrow&\ Y_2(u_3^{(2)-})=-1\,.
\eea
This is the simplest set of exact Bethe equations because the complex roots
$u_2^{(Q)++}$ and $u_2^{(Q)+}$ are inside the analyticity strip of the mirror $u$-plane and the analytic continuation of the TBA equations for $Y_Q$ functions to these points is straightforward - all we need to do is to set in \eqref{HybridQ} the variable $v$ to $u_2^{(1)++}$ in the equation for $Y_1$ and to $u_2^{(2)+}$  in the equation for $Y_2$, and then equate the result to $2\pi i n$.
Note that since $u_2^{(2)}\approx u_2^{(1)}$ for small $g$ the mode number appearing in the equation for $u_2^{(2)}$  depends on the one for $u_2^{(1)}$. The exact Bethe equations for $u_3$ are equivalent to those for $u_2$ due to the real analyticity of Y-functions.
For the real rapidity $u_1$ the quantization condition is unique and we must
analytically continue the hybrid equation for $Y_1$ to the string region.
Following the derivation in \cite{AFS09} and using the identities from appendix \ref{app:EBE1}, we get
\begin{align}\nonumber
2\pi i n_1 &= G_{1_*}(u_1) -\log \frac{S_{\sl(2)}^{11_*}(u_2^{(1)},u_1)}{S_{\sl(2)}^{11_*}(u_3^{(1)},u_1)} +\log \frac{S_{\sl(2)}^{21_*}(u_3^{(2)-},u_1)}{S_{\sl(2)}^{21_*}(u_3^{(1)-},u_1)}\frac{S_{\sl(2)}^{21_*}(u_2^{(1)+},u_1)}{S_{\sl(2)}^{21_*}(u_2^{(2)+},u_1)} \\ \nonumber
&-2\log {S(u_2^{(2)+},u_1)\ov S(u_3^{(2)-},u_1)}\star K_{vwx}^{11_*}-\log S_1(r_0^--u_1)S_{y1_*}(r_0^-,u_1)
 \\ \label{EBE1}
&+ 2\log {\rm Res}\, S\star K_{vwx}^{11_*}(r_0^-,u_1)-2\log\left(u_1-r_0-\frac{2i}{g}\right)
\frac{x_s^+(r_0)-x_s^+(u_1)}{x_s^+(r_0)-x_s^-(u_1)}
\\
&
 + 2  \log {\rm Res}\, S\star K^{11_*}_{vwx} (u_1^-,u_1) -2\log\Big(-{2i\ov g}\,
\frac{x_s^-(u_1)-\frac{1}{x_s^-(u_1)}}{x_s^-(u_1)-\frac{1}{x_s^+(u_1)}}\Big)
\notag
\, . \end{align}
where $G_{1_*}(u_1)$ is obtained by analytically continuing \eqref{GQ}
\bea
G_{1_*}(u_1) &=& i L_{\rm TBA}\, p_1 +\log \left(1+Y_{Q} \right)
\star (K_{\sl(2)}^{Q1_*}+2 s\star K_{vwx}^{Q-1,1_*})\\
&+&  2 \log \(1 + Y_{1|vw}\) \star (s \hstar K_{y1_*}+\ts ) \nonumber \\
&  -&  2  \log{1-Y_-\ov 1-Y_+} \hstar s \star K^{11_*}_{vwx} +  \log
{1- \frac{1}{Y_-} \ov 1-\frac{1}{Y_+} } \hstar K_{1}  +  \log
\big(1-\frac{1}{Y_-}\big)\big( 1 - \frac{1}{Y_+} \big) \hstar
K_{y1_*} \, , \nonumber
\eea
and $\ts(u)=s(u^-)$. Since the root $u_1$ is real while $u_2$ and $u_3$ are complex conjugate to each other the real part of equation \eqref{EBE1} must vanish. We show that this is indeed the case in appendix \ref{app:EBE1}.
\smallskip

We further notice that we can express all $S_{\sl(2)}^{11_*}$ S-matrices via $S_{\sl(2)}^{1_*1_*}$ by using
\begin{equation}
\la{ssid}
\begin{array}{l}
S_{\sl(2)}^{11_*}(u_2^{(1)},u_1)=S_{\sl(2)}^{1_*1_*}(u_2^{(1)},u_1)\, ,\\[1mm]
{S_{\sl(2)}^{11_*}(u_3^{(1)},u_1)}{S_{\sl(2)}^{1_*1_*}(u_3^{(1)},u_1)}=1/h_{1_*}(u_3^{(1)},u_1)^2= h_1(u_1,u_3^{(1)})^2\, ,
\end{array}
\end{equation}
where the last formula  is the analytic continuation of the identity \eqref{Cross}.
The representation of the exact Bethe equations
via  $S_{\sl(2)}^{1_*1_*}$ is useful in proving the vanishing of the real part of equation \eqref{EBE1} and in checking the Bethe-Yang equations in the limit $g\to 0$ as  discussed below.

\subsection*{Equivalence of quantization conditions}
An important fact to emphasize is that the equations \eqref{EBEall1} and  \eqref{EBEall2} are not the only quantization conditions. In addition we should have
\bea\la{EBEall1b}
Y_1(u_2^{(1)})=-1\ &\Leftrightarrow&\ Y_1(u_3^{(1)})=-1\,,\\
\la{EBEall2b}
 Y_2(u_2^{(2)+++})=-1\ &\Leftrightarrow&\ Y_2(u_3^{(2)---})=-1\,,
\eea
since these conditions have also been used to derive the TBA equations. These extra quantization conditions obviously have to be equivalent to (\ref{EBEall1}) and (\ref{EBEall2})
respectively, {\it i.e.} we want to verify
\bea
\label{QC1}
Y_1(u_2^{(1)})=-1 ~~~~~&&\stackrel{?}{\Longleftrightarrow}~~~~~~Y_1(u_2^{(1)++})=-1\, ,
\\
\label{QC2}
Y_2(u_2^{(2)+})=-1 ~~~~~&&\stackrel{?}{\Longleftrightarrow}~~~~~~Y_2(u_2^{(2)+++})=-1\, ,
\eea
and we will do so by making use of the Y-system.
As can be checked, the Y-functions which solve the TBA equations also solve the corresponding Y-system equations.
In particular $Y_1$ and $Y_2$ satisfy the following equations
\bea\la{Ys1}
Y_{1}(v^-)Y_{1}(v^{+}) =\frac{\big(1-\frac{1}{Y_-}\big)^2}{1+{1\ov Y_2}}(v)\,,\\
\la{Ys2}
Y_{2}(v^-)Y_{2}(v^{+}) =\frac{\big(1+\frac{1}{Y_{1|vw}}\big)^2}{(1+{1\ov Y_1})(1+{1\ov Y_3})}(v)\, ,\eea
which are valid for any $v$ on the mirror $u$-plane (excluding points on its cuts).

\smallskip

Now let us consider the equation for $Y_1$ at $v=u_2^{(1)+}$ and the equation for $Y_2$ at $v=u_2^{(2)++}$. Then  taking into account that
\bea\nonumber
Y_2(u_2^{(1)+})=Y_-(u_2^{(1)+})=\infty\, , \quad Y_1(u_2^{(2)++})=Y_3(u_2^{(2)++})=Y_{1|vw}(u_2^{(2)++})=\infty\, ,
\eea
we find
\bea
Y_{1}(u_2^{(1)})Y_{1}(u_2^{(1)++})=1\,,\quad Y_{2}(u_2^{(2)+})Y_{2}(u_2^{(2)+++})=1\,  \eea
which clearly implies the equivalence of the quantization conditions.

\subsubsection*{Mirror and string quantization conditions}
In addition to the quantization conditions discussed above, we could also expect to have the exact Bethe equation
$Y_{1_*}(u_3^{(1)})=-1$, where $Y_{1_*}$ is the analytic continuation of $Y_1$ to the string region. In other words,
we would then have
\bea\la{EBEall1c}
Y_1(u_3^{(1)--})=-1\ \Leftrightarrow\ Y_1(u_3^{(1)})=-1\ \Leftrightarrow\ Y_{1_*}(u_3^{(1)})=-1\,.
\eea
The last condition in  \eqref{EBEall1c} is not necessary for our derivation of the TBA equations because the point $u_3^{(1)}$ of the string $u$-plane is not enclosed by the integration contours. Nevertheless, we will show that this condition holds and therefore the exact Bethe equations can be written in precisely the same form as for real momenta
\bea\la{EBEallr}
Y_{1_*}(u_i^{(1)})=-1\,,\quad i=1,2,3\,,
\eea
where we have also taken into account that $u_2^{(1)}$ lies in the overlap of the mirror and string regions, so that $Y_1(u_2^{(1)})=Y_{1_*}(u_2^{(1)})$.

\smallskip

To show that the quantization condition $Y_1(u_3^{(1)}) = -1$ in the mirror region implies the usual exact Bethe equation $Y_{1_*}(u_3^{(1)}) = -1$ in the string region, we will analytically continue the TBA equation for $Y_1$ to a point $v$ close to $u_3^{(1)}$ in the mirror $u$-plane, and to the same point in the string $u$-plane. The resulting two equations are then added up and used to show that $Y_1(u_3^{(1)}) Y_{1_*}(u_3^{(1)}) =1$. The considerations below require the use of crossing relations for various kernels and S-matrices because the point $u_3^{(1)}$ lies in the overlap of the mirror and anti-string regions.
We find it easier to handle the canonical TBA equation \eqref{YQcanTBA} for $Y_1$ because its kernels and S-matrices have simpler properties under the crossing transformation.

\smallskip

The analytic continuation of the canonical TBA equation \eqref{YQcanTBA} to $v\approx u_3^{(1)}$ of the mirror $u$-plane is straightforward and gives

\begin{align}\notag
\log Y_1(v) =&\, - L_{\rm TBA}\, \tH_{1} + \log\left(1+Y_{Q} \right) \star K_{\sl(2)}^{Q1}
+ 2 \log\left(1+ \frac{1}{Y_{Q|vw}} \right) \star_{p.v} K^{Q1}_{vwx}
\\[1mm]\notag
&\quad +  \log {1- \frac{1}{Y_-} \ov 1-\frac{1}{Y_+} } \hstar K_{1}  +  \log \big(1-\frac{1}{Y_-}\big)\big( 1 - \frac{1}{Y_+} \big) \hstar K_{y1} \\[1mm] \notag
&\quad -\log S_{\sl(2)}^{1_*1}(u_1,v){S_{\sl(2)}^{11}(u_2^{(1)},v)\ov
S_{\sl(2)}^{11}(u_3^{(1)},v)}+\log \frac{S_{\sl(2)}^{21}(u_2^{(1)+},v)}{S_{\sl(2)}^{2Q}(u_2^{(2)+},v)}\frac{S_{\sl(2)}^{21}(u_3^{(2)-},v)}{S_{\sl(2)}^{21}(u_3^{(1)-},v)}
\\[1mm]\notag
&\quad-\log S^{11}_{vwx}(u_1,v)+\log S^{11}_{vwx}(r_0,v) - \log{S_1}(r_0^--v){S_{y1}}(r_0^-,v)\notag\\[1mm]
&\quad-\log(1+Y_2(v^{-}))+2\log\left(1-\frac{1}{Y_-(v^{-})}\right)\label{eq:EBu3mirror}\,,
\end{align}
where the terms on the last line of \eqref{eq:EBu3mirror} appear because of the poles of $K_{\sl(2)}^{21}(t,v)$ and $(K_1 + K_{y1})(t,v)$ at $v = t +i/g$.

\medskip

The analytic continuation of the canonical TBA equation for $Y_1$ to $v\approx u_3^{(1)}$ in the string region is discussed in detail in appendix \ref{app:EBE1}, and the resulting TBA equation for $v\approx u_3^{(1)}$ is
\begin{align}\notag
&\log Y_{1_*}(v)=\, - L_{\rm TBA}\, \tH_{1_*} + \log\left(1+Y_{Q} \right) \star K_{\sl(2)}^{Q1_*}+
2 \log\big(1+ \frac{1}{Y_{Q|vw}} \big) \star_{p.v} K^{Q1_*}_{vwx} \nonumber \\
& +  \log {1- \frac{1}{Y_-} \ov 1-\frac{1}{Y_+} } \hstar K_{1}+  \log \big(1-\frac{1}{Y_-}\big)\big( 1 - \frac{1}{Y_+} \big) \hstar K_{y1_*} -\log(1+Y_2(v^-))\nonumber \\
&+2\log\big( 1-\frac{1}{Y_{+_{\hat{*}}}(v^+)}\big) +2\log\left(1+\frac{1}{Y_{1_{\hat{*}}|vw}(v)}\right)
+2\log\left(1+\frac{1}{Y_{2|vw}(v^-)}\right)\nonumber
\\
&-\log \frac{S_{\sl(2)}^{11_*}(u_2^{(1)},v)}{S_{\sl(2)}^{11_*}(u_3^{(1)},v)}S_{\sl(2)}^{1_*1_*}(u_1,v)+\log \frac{S_{\sl(2)}^{21_*}(u_3^{(2)-},v)}{S_{\sl(2)}^{21_*}(u_3^{(1)-},v)}\frac{S_{\sl(2)}^{21_*}(u_2^{(1)+},v)}{S_{\sl(2)}^{21_*}(u_2^{(2)+},v)} \nonumber\\
&-\log S^{11_*}_{vwx}(u_1,v)+\log \frac{S^{11_*}_{vwx}(r_0,v)}{{S_1}(r_0^-v){S_{y1_*}}(r_0^-,v)} \label{eq:EBu3string}\,.
\end{align}
In the above, $Y_{1_{\hat{*}}|vw}$ and $Y_{+_{\hat{*}}}$ are the analytic continuations of $Y_{1|vw}$ and $Y_+$ through their cuts at $i/g$ and $2i/g$ respectively, {\it cf.} appendix \ref{app:EBE1}.

To proceed further we add the right hand sides of equations (\ref{eq:EBu3mirror}) and (\ref{eq:EBu3string}). Then, by using the crossing relations \eqref{crossQ1} for the bound-state dressing factors and other identities from appendix \ref{app:EBE1}, we find for $v\approx u_3^{(1)}$
\begin{align}\notag
\log &Y_1(v)Y_{1_*}(v)= -2\log\left(1+Y_{Q} \right)\star
(K_{xv}^{Q1}(v)-K_{Qy} (v^-)) \nonumber \\
&
+2 \log\left(1+ \frac{1}{Y_{Q|vw}} \right) \star_{p.v} K_{Q1}(v)
+  2\log {1- \frac{1}{Y_-} \ov 1-\frac{1}{Y_+} } \hstar K_{1}(v) \nonumber  \\[1mm]\notag
&-2\log(1+Y_2(v^-))+2\log\left(1-\frac{1}{Y_-(v^-)}\right)\left(1-\frac{1}{Y_{+_{\hat{*}}}(v^+)}\right) \nonumber
\\
&+2\log\left(1+\frac{1}{Y_{1_{\hat{*}}|vw}(v)}\right)+2\log\left(1+\frac{1}{Y_{2|vw}(v^-)}\right)\nonumber \\
&-\log \frac{S_{\sl(2)}^{11}(u_2^{(1)},v)S_{\sl(2)}^{11_*}(u_2^{(1)},v)}{S_{\sl(2)}^{11}(u_3^{(1)},v)S_{\sl(2)}^{11_*}(u_3^{(1)},v)}S_{\sl(2)}^{1_*1}(u_1,v)S_{\sl(2)}^{1_*1_*}(u_1,v) \nonumber \\
&+\log \frac{S_{\sl(2)}^{21}(u_3^{(2)-},v)S_{\sl(2)}^{21_*}(u_3^{(2)-},v)}{
S_{\sl(2)}^{21}(u_3^{(1)-},v)S_{\sl(2)}^{21_*}(u_3^{(1)-},v)}
+\log \frac{S_{\sl(2)}^{21}(u_2^{(1)+},v)S_{\sl(2)}^{21_*}(u_2^{(1)+},v)}{
S_{\sl(2)}^{21}(u_2^{(2)+},v)S_{\sl(2)}^{21_*}(u_2^{(2)+},v)} \la{YQcanTBA2add}\\
&-\log S_2(u_1-v)+\log \frac{S_{2}(r_0-v)}{S_1(r_0^- -v)^2} \notag\,.
\end{align}
To show that the right hand side of this equation in fact vanishes at $v= u_3^{(1)}$, we use the canonical TBA equations for $vw$-strings continued to $v\approx u_3^{(1)}$ through the cut at $i/g$. Noting that $K_{xv}^{Q1_{\hat{*}}}(u,v) = K_{xv}^{Q1}(u,v)-K_{Qy}(u,v^-)$, it reads
\bea
\nonumber
Y_{1_{\hat{*}}|vw}(v)&=&\log\left(1+\frac{1}{Y_{Q|vw}}\right)\star K_{Q1}(v)+\log\left(1+\frac{1}{Y_{2|vw}(v^-)}\right)+\\
&+&\log\frac{1-\frac{1}{Y_-}}{1-\frac{1}{Y_+}}\hat{\star}K_1-\log(1+Y_Q)\star (K_{xv}^{Q1}(v)-K_{Qy} (v^-)) \label{eq:Y1vwcont}\\
&-&\log(1+Y_2(v^-)) + \frac{1}{2}\log\frac{S_2(r_0-v)}{S_2(u_1-v)}-\log S_1(r_0^--v)\nonumber \\
&-&\log\frac{S_{xv}^{11}(u_3^{(1)},v)}{S_{xv}^{11}(u_2^{(1)}, v)} + \log\frac{S_{1y}(u_3^{(1)},v^-)}{S_{1y}(u_2^{(1)}, v^-)}  + S_{xv}^{1*1}(u_1,v) -\log S_{1_*y}(u_1^{(1)},v^-)\nonumber \\
&-&\log\frac{S^{21}_{xv}(u_2^{(1)+},v)S^{21}_{xv}(u_3^{(2)-},v)}{S^{21}_{xv}(u_2^{(2)+},v)S^{21}_{xv}(u_3^{(1)-},v)} + \log {S_{2y}(u_2^{(1)+}, v^-)\ov S_{2y}(u_2^{(2)+}, v^-)}{S_{2y}(u_3^{(2)-}, v^-)\ov S_{2y}(u_3^{(1)-}, v^-)}\nonumber
\eea
Using this equation and crossing relations \eqref{crossQ1}, all driving terms and convolution terms cancel and we find a simple result
 \begin{align}\notag
Y_1(v)Y_{1_*}(v)= (1+Y_{1_{\hat{*}}|vw}(v))^2\left(1-\frac{1}{Y_-(v^-)}\right)^2\left(1-\frac{1}{Y_{+_{\hat{*}}}(v^+)}\right)^2\, .
\end{align}
It is now straightforward to show that  $Y_1(u_3^{(1)})Y_{1_*}(u_3^{(1)}) =1$.
Firstly, considering the equation for $Y_{1_{\hat{*}}|vw}$ at $u_3^{(1)}$, it is clear that we have
\begin{equation}
Y_{1_{\hat{*}}|vw}(u_3^{(1)})=0 \, ,
\end{equation}
because $S_{1y}(u_3^{(1)},u_3^{(1)-})$ is zero, while the poles of $Y_2$ at $u_3^{(1)-}$ and $S^{21}_{xv}(u_3^{(1)-},v)$ at $u_3^{(1)}$ cancel each other  and all other terms in \eqref{eq:Y1vwcont} are finite. Then, analytically continuing the canonical equations for $y$-particles, we find that after crossing the cut at $2i/g$
\begin{equation}
\log{Y_{+_{\hat{*}}}(u_3^{(1)+})} \sim \log { \left(1+\frac{1}{Y_{1_{\hat{*}}|vw}}\right)} + \mbox{reg.} \, \, \Rightarrow \, \, Y_{+_{\hat{*}}}(u_3^{(1)+}) = \infty \, ,
\end{equation}
so that we obtain the desired result
\begin{align}
Y_1(u_3^{(1)})Y_{1_*}(u_3^{(1)}) = 1\, .
\end{align}

\subsection*{Exact Bethe equations for roots $r_M$}

The TBA equations also depend on additional roots $r_M$. The exact Bethe equations for the roots are just obtained by analytically continuing the equations for $-Y_-$ and $Y_{M|w}$ to $r_0^-$ and $r_M^-$ respectively, and setting the values of these functions to $-1$.

\subsection*{Relation to the asymptotic Bethe Ansatz}
In the asymptotic limits $g\to 0$ with $J$ fixed or $J\to \infty$ with $g$ fixed the exact quantization conditions for the Bethe roots should reduce to the Bethe-Yang equations
{\small
\bea \la{ebe0}
\pi i (2n_k+1)=ip_kJ - \sum_{j=1}^3\log
S_{\su(2)}^{1_*1_*}(u_j,u_k) \, ,~~~~~~n_k\in {\mathbb Z}\, ,
\eea }
where $S_{\su(2)}^{1_*1_*}$ is the S-matrix in the $\su(2)$-sector related to
$S_{\sl(2)}^{1_*1_*}$ as
\bea\la{su2S}
S_{\su(2)}^{1_*1_*}(u_j,u_k)=S_{\sl(2)}^{1_*1_*}(u_j,u_k)\, \prod_{j=1}^3\left(\frac{x_k^+ -x_j^-
}{x_k^--x_j^+}\sqrt{\frac{x_j^+ x_k^-}{x_j^- x_k^+}} \right)^2 \,.
\eea
Since in these equations  the S-matrix has both arguments in the string region it is convenient to  express all $S_{\sl(2)}^{11_*}$ S-matrices in the exact Bethe equations via $S_{\sl(2)}^{1_*1_*}$ at the final stage of deriving the Bethe-Yang equations from them.

\medskip

According to the discussion in section \ref{exactYQ} in the asymptotic limit $u_2^{(2)}\to u_{2}^{(1)}$ and  $u_3^{(2)}\to u_{3}^{(1)}$, and by using \eqref{u1u0} we find
\bea\la{u2u1}
u_{2}^{(2)}-u_{2}^{(1)}\approx -{\rm Res}\, Y_2(u_2^{(1)+})\, ,\quad
u_{2}^{(1)}-u_{2}^{(2)}\approx -{\rm Res}\, Y_1(u_2^{(2)++})\, ,
\eea
where we have taken into account that $1+Y_1$ has a zero at $u_{2}^{(1)++}$
and a pole at $u_{2}^{(2)++}$ while $1+Y_2$ has a zero at $u_{2}^{(2)+}$
and a pole at $u_{2}^{(1)+}$. Comparing these two expressions we immediately conclude that in the asymptotic limit the residues of $Y_1$ and $Y_2$ must obey the relation
\bea
{\rm Res}\, Y_2(u_2^{+}) + {\rm Res}\, Y_1(u_2^{++}) = 0\,,
\eea
where we have equated $u_2^{(2)}= u_{2}^{(1)}\equiv u_2$.
This is indeed satisfied, as can be readily verified through the Bajnok-Janik formula \eqref{YQexact} for $Y_Q$ functions.

\smallskip

Restricting ourselves for definiteness to the limit $g\to 0$ with $J$ fixed and rescaling the rapidities $u\to u/g$ so that the rescaled Bethe roots have a finite limit as $g\to 0$,  we  find that the leading term of ${\rm Res}\, Y_2(u_2^{(1)+})$ scales as $g^{2L}$.
Hence, we arrive at the following asymptotic relation for the rescaled rapidities 
\bea
u_2^{(2)}-u_2^{(1)}=g^{2L}a+{\cal O}(g^{2L+1})\, ,
\eea
where the constant $a$ can be found either from the TBA equation for $Y_2$ or from the Bajnok-Janik formula \eqref{YQexact}. This formula shows that as expected at weak coupling the corrections to the asymptotic Bethe ansatz start at $L$-loop order.

\medskip

Taking the limit  $u_2^{(2)}\to u_{2}^{(1)}\equiv u_2$ and  $u_3^{(2)}\to u_{3}^{(1)}\equiv u_3$ and dropping  the subleading terms $\log(1+Y_Q)$
in the exact Bethe equation \eqref{EBE1} for $u_1$ is straightforward, and it is easy to verify numerically that the resulting equation coincides with \eqref{ebe0}.

\medskip

Considering the asymptotic limit of the exact quantization condition for the complex root $u_2^{(1)}$ is more involved and it is convenient to do this by using the equation $Y_1(u_{2}^{(1)})=-1$ because there the $S_{\sl(2)}^{11}$ S-matrices depend on $u_i^{(1)}$ only.
To write down the exact Bethe equation for $u_2^{(1)}$, we need to analytically continue the
hybrid TBA equation\footnote{Of course we can perform the analytic continuation at the level of the canonical or simplified
equation for $Y_1$ as well. The hybrid form is preferred because it is the simplest one.}  for $Y_1$ to this point. This is done in appendix \ref{app:EBE1} and the resulting exact Bethe equation at $u_2^{(1)}$ is
{\small
\begin{align}
&\log(-1)=\log Y_{1}(u_2^{(1)}) = G_1(u_2^{(1)})
+  2 \log \(1 + Y_{1|vw}\) \star \tilde{s}  -\log \frac{S_{\sl(2)}^{11}(u_2^{(1)},u_2^{(1)})}{S_{\sl(2)}^{11}(u_3^{(1)},u_2^{(1)})}
S_{\sl(2)}^{1_*1}(u_1,u_2^{(1)}) \nonumber \\
&-2\log {S(u_2^{(2)+}, u_2^{(1)})\ov S(u_3^{(2)-}, u_2^{(1)})}\star
K_{vwx}^{11} -\log S_1(r_0^--u_2^{(1)})S_{y1}(r_0^-,u_2^{(1)})
\nonumber \\
&+\log \frac{
S_{\sl(2)}^{21}(u_3^{(2)-},u_2^{(1)})
}
{     S_{\sl(2)}^{21}(u_3^{(1)-},u_2^{(1)})  }
+\log \frac{{\rm Res}\, S_{\sl(2)}^{21}(u_2^{(1)+},u_2^{(1)})}
{S_{\sl(2)}^{21}(u_2^{(2)+},u_2^{(1)}) \,{\rm Res}\, Y_2(u_2^{(1)+})} \label{ExactBYE1} \\
& +2\log{\rm Res}\, S\star K_{vwx}^{11}(u_1^-,u_2^{(1)})- \log\Big(u_1-u_2^{(1)}-\frac{2i}{g}\Big)^2
\left(\frac{x_s^-(u_1)-\frac{1}{x^-(u_2^{(1)})}}{x_s^-(u_1)-\frac{1}{x^+(u_2^{(1)})}}\right)^2\nonumber\\
& +2\log {\rm Res}\, S\star K_{vwx}^{11}(r_0^-,u_2^{(1)})-\log\Big(r_0-u_2^{(1)}+\frac{2i}{g}\Big)^2
\left(\frac{x_s^+(r_0)-x^+(u_2^{(1)})}{x_s^+(r_0)-x^-(u_2^{(1)})}\right)^2 \, . \nonumber\end{align}
}

Taking the limit  $u_2^{(2)}\to u_{2}^{(1)}$
in this equation is not straightforward because
the S-matrix $S_{\sl(2)}^{21}(u_2^{(2)+},u_2^{(1)}) $ develops a singularity. For $u_2^{(2)}\sim u_{2}^{(1)}$
we have
\bea
\log S_{\sl(2)}^{21}(u_2^{(2)+},u_2^{(1)}) =\log\frac{{\rm Res}\, S_{\sl(2)}^{21}(u_2^{(1)+},u_2^{(1)})}{u_{2}^{(1)}-u_{2}^{(2)}}+
o(\delta u)\, ,
\eea
where $\delta u=u_2^{(1)}-u_2^{(2)}$. Taking into account \eqref{u2u1}, we get that in the limit  $u_2^{(2)}\to u_{2}^{(1)}$ the
terms on the third line of
equation (\ref{ExactBYE1}) vanish, and therefore equation (\ref{ExactBYE1}) acquires the form
\begin{align}
\nonumber
&\log(-1)= G_1^{\rm asympt}(u_2)+  2 \log \(1 + Y_{1|vw}\) \star \tilde{s} -\log \frac{S_{\sl(2)}^{11}(u_2,u_2)}{S_{\sl(2)}^{11}(u_3,u_2)}
S_{\sl(2)}^{1_*1}(u_1,u_2)\nonumber \\
&-2\log {S(u_2^+, u_2)\ov S(u_3^-, u_2)}\star
K_{vwx}^{11} -\log S_1(r_0^--u_2)S_{y1}(r_0^-,u_2)
\nonumber  \\
& +2\log {\rm Res}\, S\star K_{vwx}^{11}(u_1^-,u_2)- \log\Big(u_1-u_2-\frac{2i}{g}\Big)^2
\left(\frac{x_s^-(u_1)-\frac{1}{x^-(u_2)}}{x_s^-(u_1)-\frac{1}{x^+(u_2)}}\right)^2\label{ExactBYE1asympt}\\
& +2\log {\rm Res}\, S\star K_{vwx}^{11}(r_0^-,u_2)-\log\Big(r_0-u_2+\frac{2i}{g}\Big)^2
\left(\frac{x_s^+(r_0)-x^+(u_2)}{x_s^+(r_0)-x^-(u_2)}\right)^2\, , \nonumber\end{align}
where $G_1^{\rm asympt}$ is $G_1$ with the subleading terms $\log(1+Y_Q)$ neglected.

\smallskip

It is worth mentioning that our consideration is valid for both the asymptotic limit $g\to 0$ with $J$ fixed, and  $J\to \infty$ with $g$ fixed.
Thus, this formula should coincide with the expression for the
asymptotic Bethe ansatz for {\it any} value of $g$! In other words, if we substitute the asymptotic expressions for the Y-functions in equation (\ref{ExactBYE1asympt}) it should turn into the BY equation \eqref{ebe0} for $u_2$. This is indeed the case as we have verified numerically.

\begin{figure}[t]
\begin{center}
\includegraphics[width=.48\textwidth]{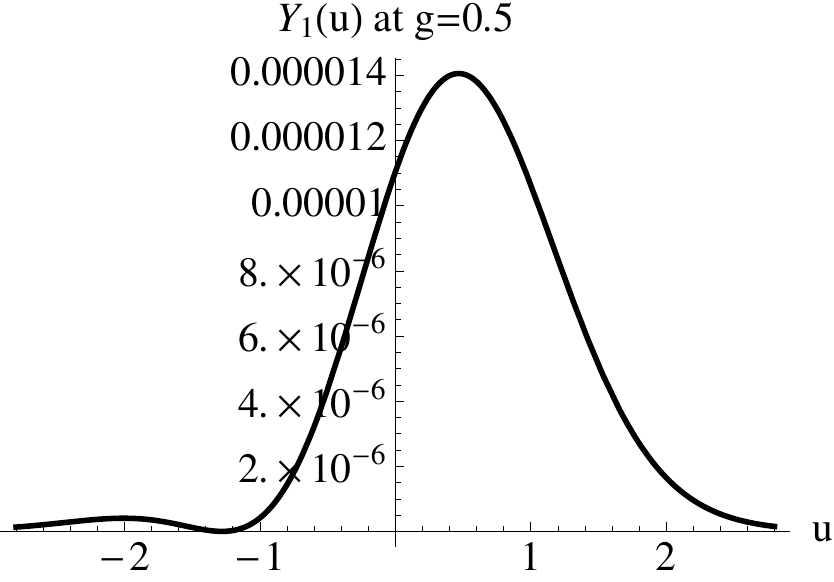}\quad \includegraphics[width=.48\textwidth]{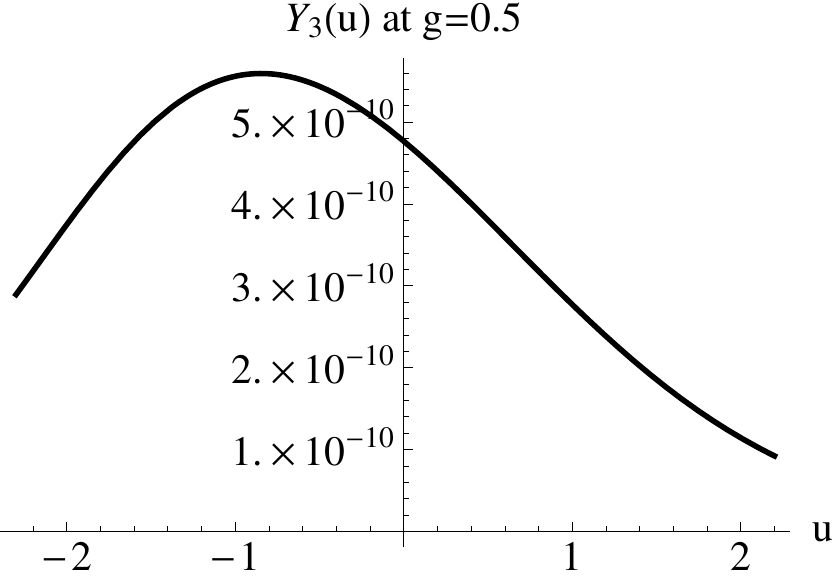}
\end{center}
\caption{The asymptotic $Y_1$- and $Y_3$-functions on the real mirror line at $g=0.5$.  }
\la{figY1Y3}
\end{figure}

\begin{figure}[t]
\begin{center}
\includegraphics[width=.44\textwidth]{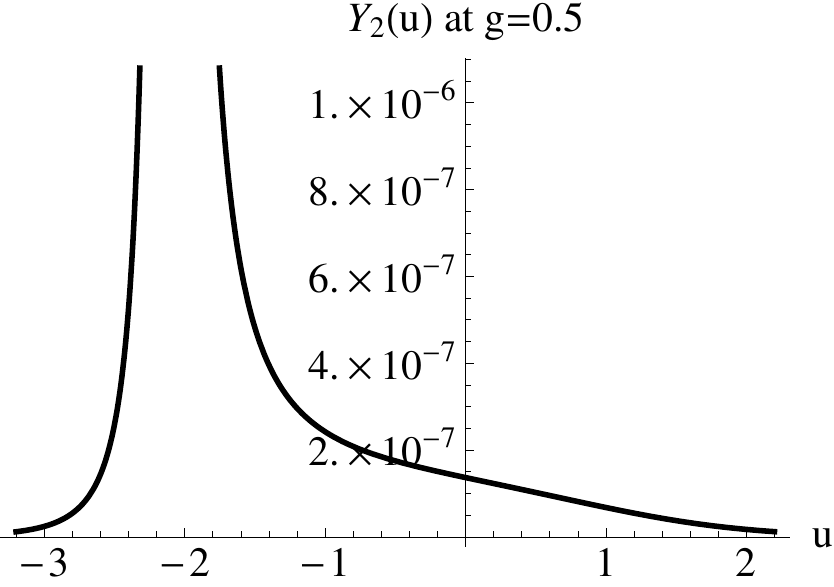}\quad \includegraphics[width=.52\textwidth]{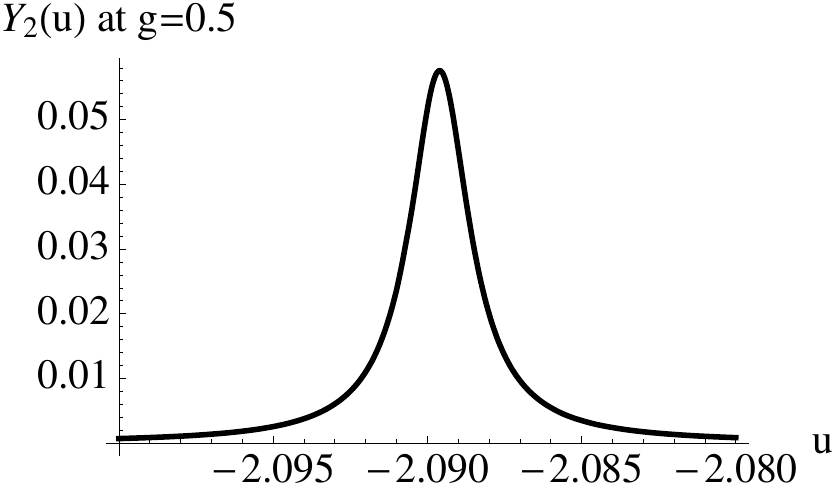}
\end{center}
\caption{The asymptotic $Y_2$-function on the real mirror line at $g=0.5$.  }
\la{figY2}
\end{figure}

\section{Conclusions}
 In this paper we have developed a description of string excited states with complex momenta in the framework of the mirror Thermodynamic Bethe Ansatz.
 For suitably small $g$ the asymptotic solution is reliable and the corresponding TBA equations can be constructed by applying the contour deformation trick. However, as soon as $g$
exceeds a certain critical value, the description of a state through the BY equations breaks down
as its energy becomes complex.
In our main example of the $L=7$ three-particle state this happens for $g\gtrsim 0.53$.
Therefore, it is important to understand how the TBA equations may cure this problem, and what happens to the state at large values of coupling. The answers to these questions do not appear to be straightforward, requiring analysis of the coupled system of TBA and exact Bethe equations.
However, the following scenario seems quite plausible; due to the TBA corrections to the BY equations the motion of the complex Bethe roots towards the boundaries of the analyticity strip slows down so that they actually freeze as $g\to \infty$. Indeed, for $g=0.5$ which is close to  the problematic value of $0.53$
the asymptotic $Y_Q$-functions are very small, see figure \ref{figY1Y3}  and \ref{figY2}, and they approximate the exact
Y-functions with very high precision.  At the same time the exact positions of the Bethe roots $u_k$ can change much more noticeably because the roots $u_2$ and $u_3$ are close to the lines Im$(u)=\pm 1/g$, and some of the kernels appearing in the exact Bethe equations develop singularities as Im$(u_{2,3})\to\pm 1/g$ and  give large contributions to the r.h.s. of the equations.
It is less clear what might happen to complex roots of states which fall in the $k$th strip at weak
coupling. The BY equations allow for these roots to move
towards the $k=1$ analyticity strip as $g$ increases. For the exact Bethe equations, various scenarios can take place, for instance, the roots always stay in the $k$th strip or, just as in the asymptotic case, they move towards the boundaries of the first strip and get frozen there.  Clearly, understanding of these issues will shed further light on how the string spectrum
is organized.

\smallskip

Recently a way of obtaining excited state TBA equations alternative to
the contour deformation trick has been discussed in the literature.
It has been argued in \cite{CFT10,Cavaglia:2011kd}  and shown for a large class
of states from the $\sl(2)$ sector \cite{BH11a} that the Y-system functional relations  \cite{GKV09} supplied with the jump discontinuity
conditions and with some analyticity assumptions on the distribution of zeroes and poles of the Y- functions are sufficient to transform the Y-system to TBA integral equations. It is not difficult to see that the TBA lemmas
of \cite{BH11a}  allow us to also reconstruct the TBA equations for $Y_{M|w}$- and $Y_{M|vw}$-functions
for the cases we study here. How this method is applied for $Y_{\pm}$ and $Y_1$ requires more careful considerations which we have not attempted. In general, it would be of interest to understand how the Y-system can be transformed into
TBA integral equations for states from the $\su(2)$ sector. This is undoubtedly possible because all TBA equations
we constructed are compatible with the Y-system functional relations, as we have checked. Also,
the driving terms in the simplified TBA equations can be rewritten to depend on the positions of zeros and poles of
Y-functions inside the analyticity strip.
We should stress however that the Y-system does not provide an intrinsic definition of the energy formula,
and for this reason the integration contour is still essential in determining the energy.

\smallskip

 Finally, we would like to mention that there has been recent interesting progress
\cite{Gromov,Suzuki:2011dj,Gromov:2011cx} towards obtaining a finite set of non-linear integral equations (NLIE) as a complementary approach
to the TBA description of the spectrum of the $\AdS$ superstring. It would be important to see how
states with complex momenta from the $\su(2)$ sector whose TBA equations we have proposed can be accommodated
within the NLIE approach.  Another interesting direction is to understand implications of our results to the spectral problem
in $\beta$-deformed and orbifold theories  \cite{Ahn:2010yv}-\cite{Ahn:2011xq}.

\section*{Acknowledgements}
We are grateful to J. Balog for sharing with us his unpublished note \cite{Balog} and we thank
J. Balog, Z. Bajnok, N. Beisert, R. Janik, A. Sfondrini and R. Suzuki for useful discussions.
G.A. acknowledges support by the Netherlands Organization for Scientific Research (NWO) under the VICI grant 680-47-602.
The work of S.F. was supported in part by the Science Foundation Ireland under Grant 09/RFP/PHY2142.
The work by S.T. is a part of the ERC Advanced grant research programme No. 246974,  {\it ``Supersymmetry: a
 window to non-perturbative physics"}.

\appendix

\renewcommand{\theequation}{\thesubsection.\arabic{equation}}

\section{Appendices}

\subsection{Contribution from $\log(1+Y_Q)\star_{C_Q} {\cal K}_Q$}
\label{sec:contcontrib}

In this appendix we consider the contribution of the terms of the form
$\log(1+Y_Q)\star_{C_Q} {\cal K}_Q$ where ${\cal K}_Q$ is an arbitrary kernel and
${\cal S}_Q$ is the corresponding S-matrix. First we will discuss the contribution for the $L=7$ state considered in the main text. Below we also discuss the contribution for a three-particle state with rapidities in the $k$th strip.

\subsubsection*{$\mathbf{L=7}$}

Taking the contour described in the main text back to the real line, we obtain the following contributions from the different $Y_Q$ functions\\

 \noindent
$\bullet$\ $Q=1$
\begin{align}
\log {\cal S}_1(u_3^{(2)--},v)-\log {\cal S}_1(u_3^{(1)--}, v)-\log {\cal S}_1(u_2^{(1)}, v)-\log {\cal S}_{1_*}(u_1, v)
\end{align}

 \noindent
$\bullet$\ $Q=2$
\begin{align}
&+\log {\cal S}_2(u_2^{(1)+}, v)+\log {\cal S}_2(u_3^{(3)---}, v)\nonumber\\
&-\log {\cal S}_2(u_2^{(2)+}, v)-\log {\cal S}_2(u_3^{(2)---}, v)
\end{align}

 \noindent
$\bullet$\ $Q\ge 3$
\begin{align}\la{YQKQ3}
&+\log {\cal S}_Q(u_3^{(Q-1)}-{i\ov g}(Q-1), v)+\log {\cal S}_Q(u_3^{(Q+1)}-{i\ov g}(Q+1), v)\nonumber\\
&-\log {\cal S}_Q(u_3^{(Q)}-{i\ov g}(Q-1), v)-\log {\cal S}_Q(u_3^{(Q)}-{i\ov g}(Q+1), v)
\end{align}

\noindent Now let us assume that ${\cal S}_Q$ satisfies the discrete Laplace equation
$$
{\cal S}_{Q-1} (u,v){\cal S}_{Q+1} (u,v)={\cal S}_{Q} (u^-,v){\cal S}_{Q} (u^+,v) \,.
$$
Then we take a sum over $Q\ge3$ of the terms in \eqref{YQKQ3} and get
\begin{align}
&\sum_{Q=3}^\infty \log {{\cal S}_Q(u_3^{(Q-1)}-{i\ov g}(Q-1), v){\cal S}_Q(u_3^{(Q+1)}-{i\ov g}(Q+1), v)\ov {\cal S}_Q(u_3^{(Q)}-{i\ov g}(Q-1), v){\cal S}_Q(u_3^{(Q)}-{i\ov g}(Q+1), v)}
=\log {{\cal S}_3(u_3^{(2)--}, v)\ov {\cal S}_2(u_3^{(3)---}, v) }\,.
\end{align}
Adding the contributions from $Q=1,2$,  we finally get the driving terms originating from
$\log(1+Y_Q)\star_{C_Q} {\cal K}_Q$
 \begin{align}\la{YQKQ}
&-\log {\cal S}_{1_*}(u_1, v)-\log{ {\cal S}_{1}(u_2^{(1)}, v)\ov  {\cal S}_1(u_3^{(1)}, v)}
+\log {{\cal S}_2(u_2^{(1)+}, v)\ov {\cal S}_2(u_2^{(2)+}, v)}{{\cal S}_2(u_3^{(2)-}, v)\ov {\cal S}_2(u_3^{(1)-}, v)}\,.
\end{align}
In the asymptotic limits $g\to 0$ with $J$ fixed or $J\to \infty$ with $g$ fixed the  last term in \eqref{YQKQ} goes to 0.
Using the discrete Laplace equation, equation (\ref{YQKQ}) can be also written in the form
\bea
-\log {\cal S}_{1_*}(u_1, v)+\log {{\cal S}_2(u_2^{(1)++}, v)\ov {\cal S}_2(u_3^{(1)--}, v)}{{\cal S}_2(u_3^{(2)-}, v)\ov {\cal S}_2(u_2^{(2)+}, v)}\,.\eea

It is worth mentioning that all the driving terms in \eqref{YQKQ}
depend only on the singularities of $Y_1$ and $Y_2$ functions, and in fact they
can also be explained by the integration contours which pick up the contribution of the real zero of $1+Y_1$ in the string $u$-plane, the zeroes $u_2^{(1)}$ and $u_3^{(1)}$  of $1+Y_1$ in the mirror $u$-plane, and all zeroes and poles of $1+Y_2$ in the analyticity strip of the mirror $u$-plane, but avoid all the other zeroes and poles of $1+Y_Q$ even those which are inside the analyticity strip of the mirror $u$-plane.

\subsubsection*{General three-particle states}

Here we give the generalization of the above contribution for the $L=7$ state to three-particle states with rapidities in the $k$th strip. Let us discuss the $u_3$ contribution in some detail. Since the poles and zeroes associated to $u_3$ are always shifted down, we simply need to determine when they start to contribute. If $u_3$ is in the $k$th strip, it needs to be shifted down $k$ times to lie below the real line and contribute. This means we get all contributions from $Y_{Q\geq k+1}$
\begin{align}
\sum_{Q=k+1}^\infty \log {{\cal S}_Q(u_3^{(Q-1)}-{i\ov g}(Q-1), v){\cal S}_Q(u_3^{(Q+1)}-{i\ov g}(Q+1), v)\ov {\cal S}_Q(u_3^{(Q)}-{i\ov g}(Q-1), v){\cal S}_Q(u_3^{(Q)}-{i\ov g}(Q+1), v)}
=& \\
& \hspace{-90pt} =\log {{\cal S}_{k+1}(u_3^{(k)}-{i\ov g}k, v)\ov {\cal S}_k(u_3^{(k+1)}-{i\ov g}(k+1), v) }\,. \nonumber
\end{align}
Next, from $Y_{k}$ we get a contribution
\begin{equation}
\log {{\cal S}_{k}(u_3^{(k+1)}-{i\ov g}(k+1), v)\ov {\cal S}_k(u_3^{(k)}-{i\ov g}(k+1), v) } \, ,\nonumber
\end{equation}
while from $Y_{k-1}$ we get
\begin{equation}
\log {{\cal S}_{k-1}(u_3^{(k)}-{i\ov g}k, v)\ov {\cal S}_{k-1}(u_3^{(k-1)}-{i\ov g}k, v) } \, .\nonumber
\end{equation}
Summing these up immediately yields
\begin{align}
\log \frac{{\cal S}_{k+1}(u_3^{(k)}-{i\ov g}k, v){\cal S}_{k-1}(u_3^{(k)}-{i\ov g}k, v)}{{\cal S}_{k-1}(u_3^{(k-1)}-{i\ov g}k, v){\cal S}_k(u_3^{(k)}-{i\ov g}(k+1), v)} = & \nonumber \\
& \hspace{-110pt}=
 \log \frac{{\cal S}_{k}(u_3^{(k)}-{i\ov g}(k-1), v){\cal S}_{k}(u_3^{(k)}-{i\ov g}(k+1), v)}{{\cal S}_{k-1}(u_3^{(k-1)}-{i\ov g}k, v){\cal S}_k(u_3^{(k)}-{i\ov g}(k+1), v)} \nonumber \\
&\hspace{-110pt}= \log \frac{{\cal S}_{k}(u_3^{(k)}-{i\ov g}(k-1), v)}{{\cal S}_{k-1}(u_3^{(k-1)}-{i\ov g}k, v)} \nonumber\\
&\hspace{-110pt}= \log \frac{{\cal S}_{k}(u_3^{(k)}-{i\ov g}(k-1), v)}{{\cal S}_{k}(u_3^{(k-1)}-{i\ov g}(k-1), v)}\frac{{\cal S}_{k-1}(u_3^{(k-1)}-{i\ov g}(k-2), v)}{{\cal S}_{k-2}(u_3^{(k-1)}-{i\ov g}(k-1), v)}\nonumber\\
&\hspace{-110pt}= \log \frac{{\cal S}_{k}(u_3^{(k)}-{i\ov g}(k-1), v)}{{\cal S}_{k}(u_3^{(k-1)}-{i\ov g}(k-1), v)} {\cal S}_{1}(u_3^{(k-1)}, v)
\end{align}
All identities follow from the discrete Laplace equations. The last identity in particular is immediately clear by rewriting the discrete Laplace equation as
\begin{equation}
\frac{S_Q(u-{i\ov g}(Q-1),v)}{S_{Q-1}(u-{i\ov g}Q,v)} = \frac{S_{Q-1}(u-{i\ov g}(Q-2),v)}{S_{Q-2}(u-{i\ov g}(Q-1),v)} \, .
\end{equation}
Similarly summing up the $u_2$ contributions, the total contribution is then
\begin{align}
& -\log {\cal S}_{1_*} (u_1,v) +
\log \frac{{\cal S}_{1}(u_3^{(k-1)}, v)}{{\cal S}_{1}(u_2^{(k-1)}, v)} \nonumber \\
& \quad + \log \frac{{\cal S}_{k}(u_2^{(k-1)}+{i \ov g}(k-1), v)}{{\cal S}_{k}(u_2^{(k)}+{i \ov g}(k-1), v)} \frac{{\cal S}_{k}(u_3^{(k)}-{i \ov g}(k-1), v)}{{\cal S}_{k}(u_3^{(k-1)}-{i \ov g}(k-1), v)} \, , \label{eq:YQkQgeneral}
\end{align}
where we also added the string contribution for $u_1$. Note that in generalizing the $L=7$ case, the generic contribution \eqref{eq:YQkQgeneral} has lost its seemingly obvious connection to the singularities of $Y_1$.

\subsection{Identities to simplify the TBA equations}\la{identY}

Here we collect the identities necessary to derive the simplified TBA equations from the canonical ones. For brevity we have unified the discussion of the identities used for the $L=7$ state, with rapidities in the second strip, and the $L=40$ state, with rapidities in the third strip respectively. The basic identities hold for rapidities $u_2$ and $u_3$ in the second strip; additional terms which arise upon changing the location of $u_2$ and $u_3$ to the third strip are indicated in {\color{blue}\underline{blue and are underlined}}.

\smallskip

Before listing the specific identities, let us discuss a frequently encountered situation; integrating $\log f$ of a complex function $f$ over an integration contour which runs either a bit above or below the real line. For any function $f(t)$ which has real zeroes at $u_i^o$ and real poles at $u_j^\infty$ we define
$\log f$ as
\bea\la{logf}
\log f(t) \equiv \log \Big(f(t)\, {\prod_j (t-u_j^\infty)\ov \prod_i (t-u_i^o)}\Big) +\sum_i \log(t-u_i^o) - \sum_j \log(t-u_j^\infty)\,.
\eea
Since $\tilde f(t)\equiv f(t)\, {\prod_j (t-u_j^\infty)\ov \prod_i (t-u_i^o)} $ has no real zeroes or real poles, the cuts of $\log \tilde f$ can and must be chosen so that they would not intersect the real line. With such a choice of the cuts of $\log \tilde f$ the imaginary part of $\log f^{p.v}$ is continuous on the real line where the function $f^{p.v.}$ is defined as $f^{p.v.}(t)\equiv  f(t)\, {\prod_j {\rm sign}(t-u_j^\infty)\,\prod_i {\rm sign}(t-u_i^o)}$. If $f(t)$ is real for real $t$ and $f(\infty)>0$ then  $f^{p.v.}(t)= |f(t)|$. The function $f^{p.v.}$ is used to define the principal value prescription by the formula
\bea\la{logf2}
\log f\star_{p.v.}K \equiv \log f^{p.v.}\star K\,,
\eea
where  on the right hand side the Cauchy principal value of the integral is computed over the real line. This definition is a generalization of the one used in  \cite{AFS09} to complex functions $f$.

The formulae \eqref{logf} and \eqref{logf2} are also used if some of the zeroes or poles coincide, {\it e.g.} if $f$ has a real double pole at $u^\infty$ then $\log f$ is understood as
\bea
\log f(t)  \equiv \log \big(f(t)(t-u^\infty)^2\big) - 2\log (t-u^\infty )\, ,
\eea
and a similar expression if $f$ has a real double zero.
\medskip

In all the formulae below
we define two actions of the operator $(K+1)^{-1}_{NM}$ on any set of functions $\log f_N$. The first one  is defined as
\bea
\log f_N\, (K+1)^{-1}_{NM} \equiv \log f_M - \log f_{M-1}\star s -\log f_{M+1}\star s\,,
\eea
where the integration contour for the $\star$-convolution runs a bit above the real line to deal with zeroes and poles of $f_{M-1}$ and $f_{M+1}$ on the real line.

The second action explicitly takes into account the real zeroes and  poles  by using the principal value prescription defined above
\bea
\log f_N\star_{p.v.}  (K+1)^{-1}_{NM} \equiv \log f_M^{p.v.} - \log f_{M-1}\star_{p.v.} s -\log f_{M+1}\star_{p.v.} s\,.
\eea

To simplify the notations, in this paper we often use the conventions
\bea
f(u-v)\star {\rm K} \equiv \int \, dt \, f(u-t){\rm K}(t,v)\,,\quad g(u,v)\star {\rm K} \equiv \int \, dt \, g(u,t){\rm K}(t,v)\,, \nonumber
\eea
where $f$, $g$, and ${\rm K}$ are arbitrary kernels or functions. Notice that according to our other conventions
\bea
f(u-v)\star {\rm K} \equiv f\star {\rm K}(u,v) \,,\quad g(u,v)\star {\rm K} \equiv g\star {\rm K}(u,v)\,.
\eea

\subsection*{Identities  for $Y_{M|w}$ }
Firstly we have
\bea
\sum_{N=1}^{\infty}\log S_{1N} (r_0-v) (K+1)^{-1}_{NM}=\delta_{2,M}\log S(r_0-v)\, ,\eea
and
\bea
\nonumber
&\tfrac{1}{2}\sum_{P,N=1}^{\infty} \log \frac{S_{PN} (r_{N}^+-v)}{S_{PN} (r_{N}^--v)}\star_{p.v.} (K+1)^{-1}_{NM}=
\left\{\begin{array}{l} -\log |S(r_{M-1}^--v)S(r_{M+1}^--v)|~~{\rm if}~~M\neq 1 \, ,\\
-\log |S(r_{2}^--v)|~~~~~~~~~~~~~~~~{\rm if}~~M= 1 \, ,\\\end{array}\right.\eea
where the p.v. prescription has been used to deal with zeroes and poles of
$S_{N-1,N}$ and $S_{N+1,N}$ at $v=r_{N}$.

Moreover, we need the sum
\bea
\sum_{N=1}^{\infty}\log S_{N} (r_0^--v)\star_{p.v.} (K+1)^{-1}_{NM}=\left\{\begin{array}{cl} 0 &~~ {\rm for}~~M\geq 3 \,,\\
\tfrac{1}{2}\log S(r_0-v)&~~ {\rm for}~~M= 2 \,,\\
\log |S(r_{0}^--v)| &~~ {\rm for}~~M= 1 \, .\\
\end{array}\right.\,~~~~~
\eea

\subsection*{Identities for $Y_{M|vw}$ }

In addition to the identities for $Y_w$ we also need

\begin{align}\nonumber
\log S^{1_*Q}_{xv}\star_{p.v} (K+1)^{-1}_{QM} (u_1,v) = &\delta_{M,1}\big( \log S_{1_*y}\star s (u_1,v) - \log S(u_1^- - v)\big)+ {1\ov 2}\delta_{M,2}\log S(u_1 - v)\,,\\
\nonumber
\log \frac{S^{1Q}_{xv} (u_2,v)}{S^{1Q}_{xv} (u_3,v)} \star (K+1)^{-1}_{QM} =&  \delta_{M,1} \Big(\log S_{1_*y}\hstar s(u_2,v)+\log S_{1_*y}\hstar s(u_3,v) +\log {S(u_2^+ -v)\ov S(u_3^- -v)}\Big)\, , \nonumber \\
& {\color{blue}\underline{ + \delta_{M,2} \left(\log \frac{S(u_3-v)}{S(u_2-v)}\right)}}
\end{align}
where we have rewritten $S_{1y}$ in terms of $S_{1_*y}$. Next we have the identities
\begin{align}
\log \frac{S^{2Q}_{xv} (u_2^+,v)}{S^{2Q}_{xv} (u_3^-,v)}(K+1)^{-1}_{QM} = & \delta_{M,1} \Big( \log \frac{S_{2y}(u_2^+,v)}{S_{2y}(u_3^-,v)}\hstar s +\log {S(u_2^+ -v)\ov S(u_3^- -v)}\Big) \nonumber \\
& {\color{blue}\underline{ + \delta_{M,2} \log \frac{S(u_3-v)}{S(u_2-v)}}}\, ,
\end{align}
and
\begin{align}\nonumber
{\color{blue}\underline{
\log \frac{S^{3Q}_{xv} (u_2^{++},v)}{S^{3Q}_{xv} (u_3^{--},v)}(K+1)^{-1}_{QM} =}} &\underline{\color{blue} \delta_{M,1} \Big( \log \frac{S_{3y}(u_2^{++},v)}{S_{3y}(u_3^{--},v)}\hstar s\Big)  + \delta_{M,2} \log \frac{S(u_3-v)}{S(u_2-v)}} \, .
\end{align}

\subsection*{Identities  for $Y_+Y_-$ }
To simplify the canonical TBA equation for $Y_+Y_-$ we need to compute the infinite sums involving $Y_{w}$ and $Y_{vw}$ functions.
Using
the method from section 8.4 of \cite{AFS09}  which we modify slightly due to the presence of zeroes of Y-functions and driving terms, we get the following two formulae
\begin{align}\label{identityYMvw}
\sum_{M=1}^\infty& \log \(1 + {1 \over Y_{M|vw}}\) \star_{p.v.} {K}_M =
\log \(1 + Y_{1|vw}\) \star s
\\[1mm]
&\qquad\qquad\qquad -  \log{1-Y_-\ov 1-Y_+} \hstar s\star {K}_1+ \sum_{M=1}^\infty \log(1 + Y_{M+1}) \star s \star {K}_M
\notag
\\[1mm]
&\qquad\qquad\qquad-\Big(
\log\frac{S(u_2^{(2)+} - v)}{S(u_3^{(2)-} - v)} - \log S(u_1^- - v)S(r_{0}^- - v)\Big)\star_{p.v.} {K}_1\,,
\notag
\end{align}
and
\begin{align}\label{identityYMw}
\sum_{M=1}^\infty \log \(1 + {1 \over Y_{M|w}}\) \star_{p.v.} {K}_M &=
\log \(1 + Y_{1|w}\) \star s
-  \log{1-{1\ov Y_-}\ov 1-{1\ov Y_+}} \hstar s\star {K}_1
\\[1mm]
&+\sum_{M=1}^\infty  \log S(r_{M-1}^- - v)S(r_{M+1}^- - v) \star_{p.v.} {K}_M\,.
\notag
\end{align}
The sum on the second line of equation \eqref{identityYMw} can be transformed to the form
\begin{align}\notag
\sum_{M=1}^\infty  &\log S(r_{M-1}^- - v)S(r_{M+1}^- - v) \star_{p.v.} {K}_M =
 \log S(r_{0}^- - v) \star_{p.v.} {K}_1\\[1mm]
&\label{sumYMw}
\qquad\qquad\qquad\qquad\qquad -\log S(r_{1}^- - v) +{1\ov 2}
\sum_{M=1}^\infty\log {{S}_{M}(r_{M}^-- v)\ov {S}_{M}(r_{M}^+- v) }
\,.
\end{align}
Then we also use
\begin{align}
& \log S(u_1^- - v)\star_{p.v.} {K}_1= \log |S_1(u_1^- - v)|\star s\,,
 \\[1mm]
& \log S_{1_*y}(u_1,v)
\hstar {K}_1 = \log {S^{1_*1}_{xv}(u_1, v)} - \log {S_1(u_1^{+}- v)}
\,, \\[1mm]
&
\log |S_1(u_1^- - v)|\star s
+\log {S_1(u_1^{+}- v)} \star s = \frac{1}{2}\log {S_2(u_1- v)} \star s
\,, \\[1mm]
& \log S_{xv}^{1_*1}(u_1,v)\star s-\frac{1}{2}\log {S_2(u_1- v)} \star s = \frac{1}{2}\log {S_{xv}^{1_*1}(u_1,v)^2\ov S_2(u_1- v)} \star s
  \,,\\[1mm]\nonumber
&\nonumber
\log\frac{S(u_2^{(2)+} - v)}{S(u_3^{(2)-} - v)}\star {K}_1 = \log\frac{S_1(u_2^{(2)+} - v)}{S_1(u_3^{(2)-} - v)}\star s + {\color{blue} \underline{ \log \frac{S_1(u_3^{--}-v)}{S_1(u_2^{++}-v)}\frac{S(u_3-v)}{S(u_2-v)} }}\,,
 \\[1mm]\nonumber
 &\nonumber{\color{blue}\underline{
\log\frac{S(u_3 - v)}{S(u_2 - v)}\star {K}_2 = \log\frac{S_1(u_2^{++} - v)}{S_1(u_3^{--} - v)}- \log \frac{S(u_3 - v)}{S(u_2 - v)}}}\,,
\end{align}
\begin{align}
 & \log {S_{1y}(u_2^{(1)},v)\ov  S_{1y}(u_3^{(1)},v)}
\hstar {K}_1= \log {S^{11}_{xv}(u_2^{(1)}, v)\ov S^{11}_{xv}(u_3^{(1)}, v)} - \log {S_1(u_2^{(1)+}- v)\ov S_1(u_3^{(1)-}- v)}
\,,\\[1mm]
&\log {S_{2y}(u_2^{(1)+}, v)\ov S_{2y}(u_3^{(1)-}, v)}\hstar {K}_1 = \log {S^{21}_{xv}(u_2^{(1)+}, v)\ov S^{21}_{xv}(u_3^{(1)-}, v)} - \log {S_1(u_2^{(1)+}- v)\ov S_1(u_3^{(1)-}- v)}
\,, \\[1mm]
&{\color{blue}  \underline{ \log {S_{3y}(u_2^{++}, v)\ov S_{3y}(u_3^{--}, v)}\hstar {K}_1 = \log {S^{31}_{xv}(u_2^{++}, v)\ov S^{31}_{xv}(u_3^{--}, v)} - \log {S_2(u_2^{++}- v)\ov S_2(u_3^{--}- v)}}}
\,,\\[1mm]
&
 \log {S^{11}_{xv}(u_3^{(1)}, v)}
 =- \log {S^{1_*1}_{xv}(u_3^{(1)}, v)} + \log {S_2(u_3^{(1)}- v)}\,,
\\[1mm]
& \nonumber \log {{S}_1(u_2^{(1)}- v)\ov {S}_{1}(u_3^{(1)}- v)}-2 \log {S_{xv}^{11}(u_2^{(1)},v)\ov S_{xv}^{11}(u_3^{(1)},v)}\star s
 =-\log {S_{xv}^{1_*1}(u_2^{(1)},v)^2\ov S_2(u_2^{(1)}- v)} {S_{xv}^{1_*1}(u_3^{(1)},v)^2\ov S_2(u_3^{(1)}- v)} \star s  \\
 \nonumber
 & \hspace{7cm} \hspace{-2pt} + \log {S(u_2^{(1)}- v)\ov S(u_3^{(1)}- v)} -{\color{blue}  \underline{\log {S(u_2^{(1)}- v)\ov S(u_3^{(1)}- v)}}} \nonumber
  \,.
\end{align}

\subsection*{Identities  for $Y_Q$ }
Let us start by recalling that the $S_{\sl(2)}^{1_*Q}$ S-matrix has the following structure
\bea
\log S^{1_*Q}_{sl(2)}(u,v)&=&-2\log \Sigma_{1_*Q}(u,v) -\log S_{1Q}(u-v)\\\nonumber
&=&-2\log \Sigma_{1_*Q}(u,v) -\log S_{Q-1}(u-v)-\log S_{Q+1}(u-v)\, .
\eea
Thus identities involving $S_{\sl(2)}^{1_*M}$ follow from the corresponding identities for $\Sigma_{1_*Q}$ and $S_{Q}$. Firstly for $S_Q$  with the both arguments in the analyticity strip we have the following identity
\bea\nonumber
\log S_M\star (K+1)^{-1}_{MQ}
&=&\log S_Q -\log S_{Q-1}\star s-\log S_{Q+1}\star s=\delta_{Q1}\log S(u-v)\,,\quad u\in {\bf R}\,.~~~~
\eea
Analytically continuing this identity in the variable $u$ outside the analyticity strip to the locations of $u_2$ and $u_3$, we get for $u_2$ and $u_3$ in the second strip ($L=7$)
\bea\nonumber
\log S_M\star (K+1)^{-1}_{MQ} (u_2,v)
&=&\delta_{Q1}\log S(u_2-v)+\delta_{Q2}\log S(u_2^+-v)\,,~~~~\\\nonumber
\log S_M\star (K+1)^{-1}_{MQ} (u_3,v)
&=&\delta_{Q1}\log S(u_3-v)+\delta_{Q2}\log S(u_3^--v)\,,
\eea
while for rapidities in the third strip ($L=40$) the relevant identities are
\begin{align}\nonumber
\color{blue} \underline{\log S_{1M}\star (K+1)^{-1}_{MQ} (u_2,v)=}& \color{blue}\underline{\delta_{Q1}\log S(u_2^+-v)-\delta_{Q2}\log S(u_2-v) }\\
& \color{blue} \underline{+\delta_{Q3}\log S(u_2^+-v)-\delta_{Q4}\log S(u_2-v)}\,,~~~~\\\nonumber
\color{blue}\underline{\log S_{1M}\star (K+1)^{-1}_{MQ} (u_3,v)
=}&\color{blue}\underline{\delta_{Q1}\log S(u_3^--v)-\delta_{Q2}\log S(u_3-v)} \\
& \color{blue}\underline{+\delta_{Q3}\log S(u_3^--v)-\delta_{Q4}\log S(u_3-v)}\,,
\end{align}
The next identity is for the dressing factor $\Sigma_{1_*Q}$ with the first argument on the real line of the string $u$-plane, {\it e.g.} equal to $u_1$
\bea
\log \Sigma_{1_*M}\star (K+1)^{-1}_{MQ} (u_1,v)
&=&\delta_{Q1}\log \check{\Sigma}_{1_*}\cstar s(u_1,v)\,.~~~~
\eea
It was derived in \cite{AFS09} where the precise definition of $\check{\Sigma}_{1_*}$ can be found, see equation (8.24) there. It is worth mentioning that the expression (8.24) in  \cite{AFS09}  for  $\check{\Sigma}_{1_*}(u,v)$ is valid for
any real $v$ and any complex $u$ on the string $u$-plane because the cuts  there are inside the vertical strip $-2\le {\rm Re}\,  u\le 2$.

\smallskip

It is convenient to use the canonical TBA equations in the form \eqref{YQcanTBA}. This means we need identities for $\Sigma_{1Q}$ with the both arguments in the mirror $u$-plane. Since $\Sigma_{1Q}$ is a holomorphic function in the mirror region they have the same form for any value of the first argument, {\it i.e.} also for $u_2$ and $u_3$
\bea
\log \Sigma_{1M}\star (K+1)^{-1}_{MQ} (u_i,v)
&=&\delta_{Q1}\log \check{\Sigma}_{1}\cstar s(u_i,v)\,.~~~~
\eea
We also need the standard identities
\begin{equation}
\log \Sigma_{2M}(K+1)^{-1}_{MQ} = \delta_{Q,1} \log \check{\Sigma}_2 {\cstar} s \, ,
\end{equation}
and
\begin{align}
\log S_{2M}(K+1)^{-1}_{MQ} =& (\delta_{Q,1}+\delta_{Q,3}) \log S \, ,\\
\color{blue} \underline{\log S_{3M}(K+1)^{-1}_{MQ} =}&\color{blue}\underline{ (\delta_{Q,2}+\delta_{Q,4}) \log S }\, ,
\end{align}
which can be applied directly since the relevant arguments lie within the analyticity strip in both the $L=7$ and $L=40$ cases.

Finally we need the following identities for the auxiliary S-matrices
\bea\nonumber
\log S^{1M}_{vwx}\star (K+1)^{-1}_{MQ} (u,v)
=\delta_{Q2}\log S(u-v)-\delta_{Q1}\log \check S_1\cstar s(u,v)\,,\quad u,v\in {\bf R}\,,~~~~~~
\eea
\begin{align}
\log S_{yM}\star_{p.v.} (K+1)^{-1}_{MQ}(r_0^-,v) =&  \delta_{Q,1} (\log S(r_0^--v)-2\log \check{S}\cstar s(r_0^-,v))  \nonumber\\
& +{1\ov2} \delta_{Q,2} \log S(r_0 -v) \,.
\end{align}

For the $L=7$ state, to replace $ \check{\Sigma}_{1}$ by $ \check{\Sigma}_{1_*}$ in the simplified TBA equation for $Y_1$ we also need
\bea
 \check{\Sigma}_{1}(u_2,v) =  \check{\Sigma}_{1_*}(u_2,v)\,,\quad
1/\check{\Sigma}_{1}(u_3,v) =  \check{\Sigma}_{1_*}\check{S}_{1}(u_3,v)\,.
\eea
The last identity uses
\bea
\check{S}_{1}(u_3,v)=\check{S}(u_3^+,v)/\check{S}(u_3^-,v)\,,
\eea
Other useful identities are
\bea
\check{S}_{1}(u_1,v)=\check{S}(u_1^+,v)\check{S}(u_1^-,v)\,,\quad \check{S}_{1}(u_2,v)=\check{S}(u_2^-,v)/\check{S}(u_2^+,v)\,.
\eea
Finally let us also note this identity for rapidities $u_2$ and $u_3$ in the third strip ($L=40$)
\begin{align}
\label{eq:hL40id1}
\color{blue} \underline{\frac{S^{2Q}_{vwx} (u_2^+,v)}{S^{1Q}_{vwx} (u_2^{++},v)}  \color{blue}= h_Q (u_2,v) \, , }\quad\underline{
\color{blue}\frac{S^{2Q}_{vwx} (u_3^-,v)}{S^{1Q}_{vwx} (u_3^{--},v)}  \color{blue}= h_Q (u_3,v) \, .}
\end{align}

\subsection{Identities for the exact Bethe equations}\la{app:EBE1}
\subsection*{Identities for the exact Bethe equation for $u_1$}\la{app:EBE1}
The derivation of the exact Bethe equation for the real root $u_1$requires the following  identities
\begin{align}
\label{Resu1}
& 2\log S\star_{v.p.} K_{vwx}^{11}(u_1^-,v)-\log
S^{11}_{vwx}(u_1,v)= \nonumber \\
&~~~~~~~~~~~2\log{\rm Res}\, S\star K_{vwx}^{11}(u_1^-,v)- 2\log\left(u_1-v-\frac{2i}{g}\right)
\frac{x_s^-(u_1)-\frac{1}{x^-(v)}}{x_s^-(u_1)-\frac{1}{x^+(v)}}\, ,
\\
&2\log S\star_{v.p.}  K_{vwx}^{11}(r_0,v)+\log
S^{11}_{vwx}(r_0,v) =\nonumber\\
&~~~~~~~~~~~2\log{\rm Res}\, S\star K_{vwx}^{11}(r_0^-,v)-2\log\left(v-r_0-\frac{2i}{g}\right)
\frac{x_s^+(r_0)-x^+(v)}{x_s^+(r_0)-x^-(v)}\, ,
\label{Resro}\end{align}
where we use the notation
\bea
\log{\rm Res}\, S\star K_{vwx}^{11_*}(u^-,v)=\int_{-\infty}^{\infty}{\rm d}t\, \log \Big[S(u^--t)(t-u)\Big]K^{11_*}_{vwx}(t+i0,v)\, .~~~~~\eea

\medskip

To show that the real part of equation \eqref{EBE1} vanishes we use the following identities valid for real $t$ and $v$
\bea
&&{\rm Re}\Big(2 s\star K_{vwx}^{Q,1_*}(t,v) \Big)= K_Q(t-v)\\
&&{\rm Re}\Big(K_{\sl(2)}^{Q1_*}(t,v)+2 s\star K_{vwx}^{Q-1,1_*}(t,v)\Big) = -K_{yQ}\hstar K_1(t,v)\eea
which allow us to prove that
\bea
{\rm Re}\Big(G_{1_*}(u_1) \Big)&=& -\log \left(1+Y_{Q} \right)\star K_{yQ}\hstar K_1(u_1)+ \log {Y_+\ov Y_-} \hstar K_1(u_1)  \\\nonumber
&=&- \sum_i \log S_{1_*y}(u_i^{(1)},u_1)\hstar K_1  +\log {S_{2y}(u_2^{(1)+}, u_1)\ov S_{2y}(u_3^{(1)-}, u_1)}{S_{2y}(u_3^{(2)-}, u_1)\ov S_{2y}(u_2^{(2)+}, u_1)}\hstar K_1\,.
\eea
To handle the driving terms in \eqref{EBE1} we further use \eqref{ssid} to write
\bea
\log\frac{S_{\sl(2)}^{11_*}(u_2^{(1)},u_1)}{S_{\sl(2)}^{11_*}(u_3^{(1)},u_1)} = \log{S_{\sl(2)}^{1_*1_*}(u_2^{(1)},u_1)S_{\sl(2)}^{1_*1_*}(u_3^{(1)},u_1)\ov h_1(u_1,u_2)h_1(u_1,u_3)}+ \log {h_1(u_1,u_2)\ov h_1(u_1,u_3)}\,.~~~~~~
\eea
It can be shown that the first term on the r.h.s. is imaginary while the second one is real.  Then we need the identities
\bea
{\rm Re}{S_{\sl(2)}^{21_*}(u_2^{+},u_1)\ov S_{\sl(2)}^{21_*}(u_3^{-},u_1)}= {S_{xv}^{21}(u_3^{-},u_1)\ov S_{xv}^{21}(u_2^{+},u_1)} \, ,\
{\rm Re}\Big(2\log {S(u_2^{+},u_1)\ov S(u_3^{-},u_1)}\star K_{vwx}^{11_*}\Big) = \log {S_1(u_2^{+}-u_1)\ov S_1(u_3^{-}-u_1)}\,.~~~~~~ \nonumber
\eea
By using these identities it is then straightforward to check numerically that the real part of \eqref{EBE1} vanishes.

\subsection*{Exact Bethe equation for $u_3$ in the string region}

\begin{figure}[t]
\begin{center}
\includegraphics[width=.45\textwidth]{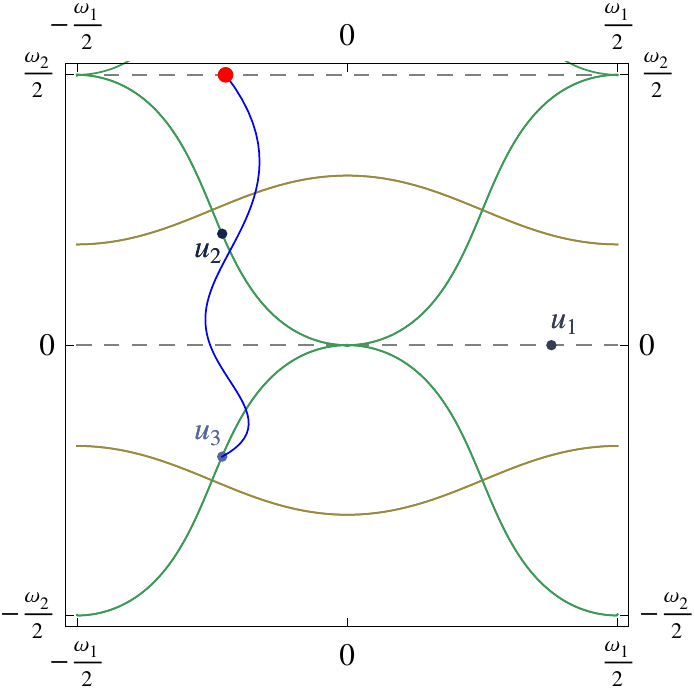}\qquad\includegraphics[width=.45\textwidth]{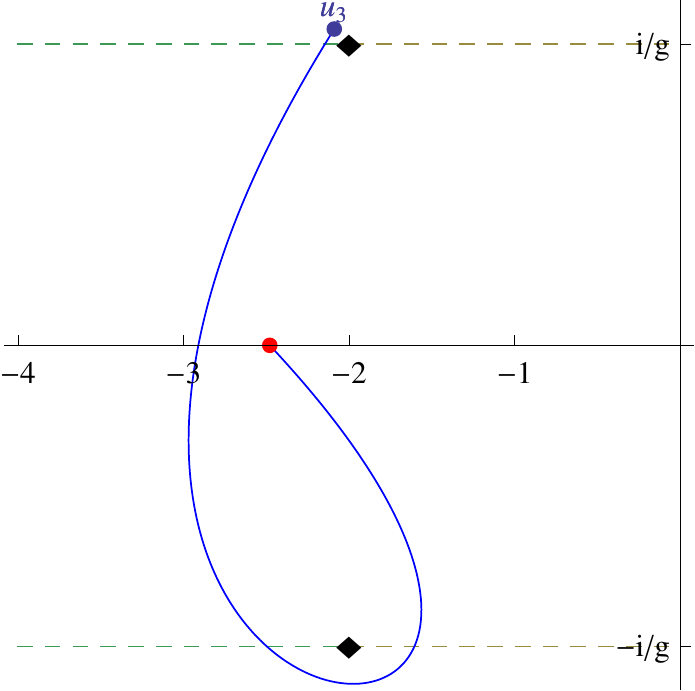}
\caption{The left picture depicts a continuation path connecting
a point on the real line of the mirror theory to the point $u_3^{(1)}$
in the string region. The right picture shows the same (homotopic) path on the $u$-plane.}
\label{fig:contpicture}
\end{center}
\end{figure}

Here we  discuss the continuation of the canonical TBA equation for $Y_1$ to $v\approx u_3^{(1)}$ in the string region. Since Im$\,u_3>1/g$ once we are in the string region we need to cross the real line and the line  ${\rm Im}\,  v=1/g$ from below and \emph{outside} of $(-2+i/g,2+i/g)$, as illustrated in figure \ref{fig:contpicture}. Let us consider the continuation of the relevant terms individually.
\begin{enumerate}[label=\it{\Roman{*}})~]
\item
$\log\left(1+Y_{Q} \right) \star K_{\sl(2)}^{Q1}(v)$. Continuation of this term gives
$$\log\left(1+Y_{Q} \right) \star K_{\sl(2)}^{Q1}(v)\to \log\left(1+Y_{Q} \right) \star K_{\sl(2)}^{Q1*}(v)
-\log(1+Y_2(v^-))\, .$$
Note that the line ${\rm Im}\,  v=-1/g$ is crossed twice during the continuation, giving vanishing net contribution, while the line  ${\rm Im}\,  v=1/g$ is crossed once.

\item $2 \log\left(1+ \frac{1}{Y_{Q|vw}} \right) \star_{p.v} K^{Q1}_{vwx} (v)$. Nothing happen to the kernels
upon crossing the line ${\rm Im}\,  v=-1/g$. However entering the string region we have to change
\bea
2 \log\left(1+ \frac{1}{Y_{Q|vw}} \right) \star_{p.v} K^{Q1}_{vwx} (v)\to
2 \log\left(1+ \frac{1}{Y_{Q|vw}} \right) \star_{p.v} K^{Q1_*}_{vwx} (v)\,.
\nonumber
\eea
Continuing further we cross the real line where $K^{11*}_{vwx}$ exhibits a pole, and  crossing the line  ${\rm Im}\,  v=1/g$ we encounter a pole of  $K^{21*}_{vwx}$
\bea\nonumber
K^{11^*}_{vwx}(u,v)&=&-\frac{1}{2\pi i}\frac{1}{u-v} + \mbox{reg.} \, ,\quad
K^{21*}_{vwx}(u,v)=-\frac{1}{2\pi i}\frac{1}{u-v+i/g} + \mbox{reg.}
\eea
Resolving these singularities gives
\bea
\nonumber
&&2 \log\left(1+ \frac{1}{Y_{Q|vw}} \right) \star_{p.v} K^{Q1}_{vwx} (v)\to
2 \log\left(1+ \frac{1}{Y_{Q|vw}} \right) \star_{p.v} K^{Q1_*}_{vwx} (v)\\
&& \hspace{1cm }+2\log\left(1+ \frac{1}{Y_{1_{\hat{*}}|vw} (v)}\right) +2\log\left(1+ \frac{1}{Y_{2|vw} (v^-)}\right)\,,
\eea
where $Y_{1_{\hat{*}}|vw}$ is the analytic continuation of $Y_{1|vw}$ across its cut at ${\rm Im}\,  v=1/g$.

\item $\log\big(1- \frac{1}{Y_-} \big) \hstar (K_{1}  +  K_{y1}).$ Taking into account the pole of $(K_1 + K_{y1})(t,v)$ at $v = t -i/g$ we get
\bea\nonumber
&& \hspace{-10pt} \log\big(1- \frac{1}{Y_-} \big) \hstar (K_{1}  +  K_{y1})\to \log\big(1- \frac{1}{Y_-} \big) \hstar (K_{1}  +  K_{y1_*})
+ 2\log\big(1-\frac{1}{Y_{+_{\hat{*}}}(v^+)} \big)
\eea
Here $Y_+(v^+)$ appears because we continue $Y_-$ across its cut on the real line; $Y_{+_{\hat{*}}}$ denotes  $Y_+$ analytically continued  across  its cut at $2i/g$.

\item $\log\big(1- \frac{1}{Y_+} \big) \hstar (-K_{1}  +  K_{y1}).$  This term does not produce any extra term.
\end{enumerate}
The resulting analytic continuation of the TBA equation to the string region for $v\approx u_3^{(1)}$  is given by \eqref{eq:EBu3string}.

\smallskip

To proceed further, we recall that the  crossing relations for the bound-state dressing factors imply  the following identity for the $S_{\sl(2)}^{Q1}$ S-matrix
\cite{AF08}
\bea\la{crossQ1}
S_{\sl(2)}^{Q1}(v_1,v_2)S_{\sl(2)}^{Q1_*}(v_1,v_2)=\left(\frac{x_1^+-x_2^+}{x_1^--x_2^+}\frac{1-\frac{1}{x_1^+x_2^-}}{1-\frac{1}{x_1^-x_2^-}}\right)^2=\left(\frac{S_{Qy}(v_1,v_2^-)}{S^{Q1}_{xv}(v_1,v_2)}\right)^2\, .
\eea
This identity in its turn leads to the following crossing relations
\begin{align}
K_{\sl(2)}^{Q1}(t,v) & +K_{\sl(2)}^{Q1_*}(t,v) =2 K_{Qy}(t,v^-)- 2 K^{Q1}_{xv}(t,v) \, .
\end{align}
Then we can easily check that for $v\approx u_3^{(1)}$
\begin{align}
K_{y1}(t,v) & + K_{y1_*} (t,v)=0 \, ,\quad
K^{Q1}_{vwx}(t,v)   + K^{Q1_*}_{vwx}(t,v)  = K_{Q1}(t-v) \, .\nonumber
\end{align}
Thus, adding the right hand sides of equations (\ref{eq:EBu3mirror}) and (\ref{eq:EBu3string}) and using the above crossing relations, we find equation \eqref{YQcanTBA2add} for $v\approx u_3^{(1)}$.

\subsection*{Asymptotic limit of the exact Bethe equation for $u_2$}
Here we analytically continue the
hybrid TBA equation for $Y_1$ to the point $u_2^{(1)}$.
Recall that
$u_2^{(1)}$ is in the intersection of the string and mirror regions and it lies below the line $-\frac{i}{g}$ in the mirror theory. We have
$$\log\left(1+Y_{Q} \right) \star
K_{\sl(2)}^{Q1} = -\log\left(1+Y_{Q} \right) \star
K_{Q1} -2\log\left(1+Y_{Q} \right) \star
K^{\Sigma}_{Q1} \,.$$
Since the dressing kernel is holomorphic in the region containing the continuation path \cite{AFdp}, it is sufficient to consider
$-\log\left(1+Y_{Q} \right) \star K_{Q1}$. Since $K_{n1}=K_{n+1}+K_{n-1}$
and at $u - v-{i\ov g}Q\sim 0$
\bea
K_Q(u-v) = {1\ov 2\pi i}{1\ov u - v-{i\ov g}Q} +\ldots\,,
\eea
we conclude that only the term with $K_{21}$ containing $K_1$ plays  a role for analytic continuation to $u_2^{(1)}$.
Continuing beyond the line
Im$(v)=-1/g$ from above produces the term $\log\left(1+Y_{2} \right)(v+i/g) $.
Taking this into account we get for Im$(v)<-1/g$
\bea
-\log\left(1+Y_{2} \right) \star K_{21}(v) \to -\log\left(1+Y_{2}(v+{i\ov g})  \right) -\log\left(1+Y_{2} \right) \star K_{21}(v)\,.~~~~~~
\eea
Thus  the continuation to $v=u_2^{(1)}$ produces an extra term $-\log(1+Y_{2}(u_2^{(1)+}))=-\log \infty $ which is
actually divergent! However, the equation (\ref{HybridQ}) contains the driving term
$
+\log \frac{S_{\sl(2)}^{21}(u_2^{(1)+},v)}{S_{\sl(2)}^{21}(u_2^{(2)+},v)}\, .
$
Since
\bea S_{\sl(2)}^{QQ'}(u,v)=S^{QQ'}(u-v)^{-1}\Sigma_{QQ'}(u,u')^{-2}\, , \eea
upon continuation of this term to
$u_2^{(1)}$ we get another divergent contribution arising due to the S-matrix $S^{21}$
$$
\log {S_{\sl(2)}^{21}(u_2^{(1)+},u_2^{(1)}) }\to -\log S^{21}(u_2^{(1)+}-u_2^{(1)})=-\log S^{21}(i/g)=-\log 0=\log\infty\, ,$$
which precisely cancels  the infinity coming from $-\log\left(1+Y_{2} \right) \star K_{21}$. Therefore, it makes
sense to combine these divergent terms into a regular expression
\bea
\lim_{v\to u_2^{(1)}}\log \frac{S_{\sl(2)}^{21}(u_2^{(1)+},v)}
{1+Y_2(v^+)} =\log \frac{{\rm Res}\, S_{\sl(2)}^{21}(u_2^{(1)+},u_2^{(1)})}
{{\rm Res}\, Y_2(u_2^{(1)+})} \, .\eea
Continuation of all the other terms in equation (\ref{HybridQ}) goes without any difficulty and as a result we find the exact Bethe equation \eqref{ExactBYE1} .

\subsection{The $L=40$, $n=2$, $k=3$ three-particle state}\la{L40}

\begin{figure}
\begin{center}
\includegraphics[width=7cm]{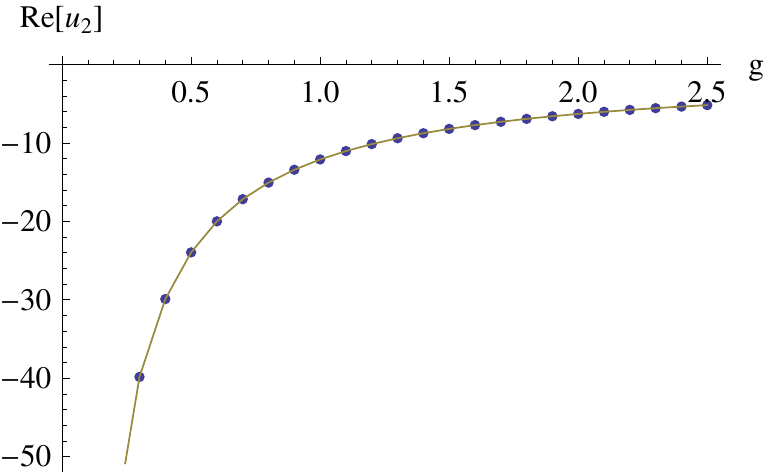}\quad \includegraphics[width=7cm]{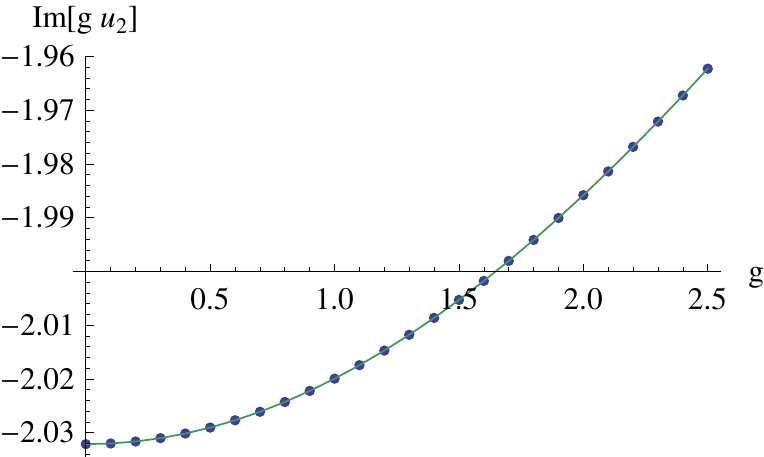}
\caption{The solution to the Bethe Yang equation for $u_2$. The imaginary part of the rapidity has been rescaled by a factor of $g$.}
\label{fig:BYsolL40}
\end{center}
\end{figure}

In this appendix we discuss a three-particle state with one real rapidity and two complex conjugate rapidities with $2/g< |\mbox{Im}(u)|<3/g$. The state we are considering is a solution of the BY equations at $L=40$ with $n=2$. The numerical solution of the BY equation for $u_2$ has been plotted in figure \ref{fig:BYsolL40}.
The asymptotic solution for this state has similar analytic properties to the $L=7$ state discussed in the main text, namely the poles of $Y_Q$ functions are at the same locations relative to the rapidities. However, this immediately implies that some of these poles lie in different regions with respect to the universal integration contour, as well as with respect to the string and mirror regions. We refer the reader back to figure \ref{fig:contour} for a qualitative picture on the torus; the roots $u_2$ and $u_3$ can qualitatively be identified with $u_2^-$ and $u_3^+$ respectively in the $L=7$ picture\footnote{The actual solution to the BY equation places the rapidities much closer to the real mirror and string lines however (of course within the same analyticity region), so that a quantitatively accurate picture would place all details on top of each other.}. The upshot of this change is that the contribution of the contour deformation trick for the $Y_Q$ functions is now given by \eqref{eq:YQkQgeneral} for $k=3$
\begin{align} \nonumber
\mathcal{D}(u_{123},v) \equiv
&-\log {\cal S}_{1_*}(u_1, v) -\log \frac{{\cal S}_{1}(u_2^{(2)}, v)}{{\cal S}_{1}(u_3^{(2)}, v)}+\log \frac{{\cal S}_{3}(u_2^{(2)++}, v)}{{\cal S}_{3}(u_2^{(3)++}, v)}\frac{{\cal S}_{3}(u_3^{(3)--}, v)}{{\cal S}_{3}(u_3^{(2)--}, v)} \, .
\end{align}
The corresponding TBA equations confirm the general discussion in the main text, fitting nicely into the picture painted there. Let us mention explicitly that exactly as for the $L=7$ state, most of the simplified equations immediately agree with their construction through the TBA lemmas of \cite{BH11a}.

Since the above combination of driving terms enters frequently, we will extensively use the shorthand $\mathcal{D}$ below. Any additional labels the $S$-matrices have will label this shorthand notation in the same way, for example
\begin{equation} \nonumber
\mathcal{D}_{\sl(2)}^{Q}(u_{123},v) = - \log S_{\sl(2)}^{1_*Q}(u_1,v)\frac{S_{\sl(2)}^{1Q}(u_2^{(2)},v)}{S_{\sl(2)}^{1Q}(u_3^{(2)},v)} \frac{S_{\sl(2)}^{3Q}(u_2^{(3)++},v)}{S_{\sl(2)}^{3Q}(u_2^{(2)++},v)}\frac{S_{\sl(2)}^{3Q}(u_3^{(2)++},v)}{S_{\sl(2)}^{3Q}(u_3^{(3)++},v)}\, .
\end{equation}

\subsection*{Further analytic properties}

Apart from the contributions from the contour deformation trick involving the rapidities indicated above, we also need to take into account exactly the same type of roots for $Y_w$ functions as we observed for the $L=7$ state. In addition however, for this state $Y_{Q|vw}$ has four real roots for $Q=1,\ldots4$ and two for $Q=5,6$, at $g=\frac{1}{2}$\footnote{These roots are also present at $g=\frac{1}{10}$, hinting that these roots are not associated to critical behavior.}. Concretely we have
\begin{align}
Y_{M|vw}(\mathbf{r}_{M\pm 1})=Y_{M|vw}(\tilde{\mathbf{r}}_{M\pm 1}) = 0 \, ,
\end{align}
where $\mathbf{r}_M$ is real and relevant to the equations for $M=0,\ldots 5$. Due to the asymmetric configuration of the state, $\mathbf{r} \neq - \tilde{\mathbf{r}}$, but the roots are of opposite sign; we denote the negative root by $\mathbf{r}$. The $Y_Q$ functions also have roots at these points in the usual fashion, removing the need for a principal value prescription in the simplified equations for $Q$ particles. As usual, these roots give
\begin{align}
Y_{M|vw}(\mathbf{r}_{M}^\pm)=Y_{M|vw}(\tilde{\mathbf{r}}_{M}^\pm) = -1 \, .
\end{align}
Since this gives some less than pleasant looking driving terms in the canonical TBA equations, we will use the following shorthand for contributions of  $\log(1+\frac{1}{Y_{Q|vw}})\star_C \mathcal{K}_{Q} \rightarrow \log(1+\frac{1}{Y_{Q|vw}})\star_{p.v.} \mathcal{K}_{Q}$
\begin{align}
D(\mathbf{r},v) = &  \sum_{M=1}^5 \log
\frac{\mathcal{S}^{(M+1)}(\mathbf{r}_M,v)}{\mathcal{S}^{(M+1)}(\tilde{\mathbf{r}}_M,v)}+  \sum_{M=2}^5 \log \frac{\mathcal{S}^{(M-1)}(\mathbf{r}_M,v)}{\mathcal{S}^{(M-1)}(\tilde{\mathbf{r}}_M,v)} \nonumber \\
&  -2 \sum_{M=1}^5 \log \mathcal{S}^{M}(\mathbf{r}_M^-,v)\mathcal{S}^{M}(\tilde{\mathbf{r}}_M^-,v) \, .
\end{align}
which is to be labeled analogously to $\mathcal{D}$ for the rapidities just above.

\subsection*{Canonical TBA equations}

Given the above discussion, we immediately derive the following TBA equations

\subsubsection*{$w$ strings}

\begin{align}
\log Y_{M|w} = & \log( 1+\frac{1}{Y_{N|w}})\star_{p.v.} K_{NM} + \log \frac{1-\frac{1}{Y_-}}{1-\frac{1}{Y_+}}\hstar K_M \\
& +\tfrac{1}{2}\sum_{N=1}^{\infty} \log \frac{S_{NM} (r_{N}^+-v)}{S_{NM} (r_{N}^--v)}
+ \tfrac{1}{2}\log S_{1M} (r_0-v)-\log S_{M} (r_0^--v) \, , \nonumber
\end{align}

\subsubsection*{$vw$ strings}

\begin{align}
\log Y_{M|vw} & =\log( 1+\frac{1}{Y_{N|vw}})\star_{p.v.} K_{NM} + \log \frac{1-\frac{1}{Y_-}}{1-\frac{1}{Y_+}}\hstar K_M - \log (1+Y_Q)\star K^{QM}_{xv} \nonumber \\
&+ \frac{1}{2} \log \frac{S_{1M} (r_0-v)}{S_{1M} (u_1-v)} - \log S_{M} (r_0^--v) + \mathcal{D}^M_{xv}(u_{123},v) + D_M(\mathbf{r},v)\, ,\nonumber
\end{align}

\subsubsection*{$y$ particles}

\begin{align}
\log {Y_+\ov Y_-} =  & \log(1 +  Y_{Q})\star K_{Qy} + \mathcal{D}_{y}(u_{123} ,v) \,,\\
\log {Y_+ Y_-} = & - \log\left(1+Y_Q\right)\star K_Q+2\log {1+{1\ov Y_{M|vw}}
 \ov 1+{1\ov Y_{M|w}}}\star_{p.v.} K_M \\\nonumber
& \quad+\sum_{M=1}^\infty\log {{S}_{M}(r_{M}^-- v)\ov {S}_{M}(r_{M}^+- v) } +\mathcal{D}(u_{123} ,v) +D(\mathbf{r},v)\,,
\end{align}

\subsubsection*{$Q$ particles}

\begin{align}
&\log Y_Q = - L\, \tH_{Q} + \log\left(1+Y_{Q'} \right) \star K_{\sl(2)}^{Q'Q}
+ 2 \log\left(1+ \frac{1}{Y_{M|vw}} \right) \star_{p.v} K^{MQ}_{vwx}\nonumber \\
&\quad +  \log {1- \frac{1}{Y_-} \ov 1-\frac{1}{Y_+} } \hstar K_{Q}  +  \log \big(1-\frac{1}{Y_-}\big)\big( 1 - \frac{1}{Y_+} \big) \hstar K_{yQ} \\
&\quad - \log \frac{S_Q (r_0^-,v) S_{yQ} (r_0^-,v)}{S^{1Q}_{vwx} (r_0,v)} - \log S^{1Q}_{vwx} (u_1,v)+ \mathcal{D}_{\sl(2)}^{Q}(u_{123},v)  +D^{Q}_{vwx}(\mathbf{r},v)\, . \nonumber
\end{align}

\subsection*{Simplified TBA equations}

Using the identities in appendix \ref{identY} we find the following simplified TBA equations\footnote{For brevity we omit presenting the simplified equation for $Y_1$, instead presenting the hybrid equations for $Q$ particles below.}

\subsubsection*{$w$ strings}

\begin{align}
\log Y_{M|w} = &  \log(1 +  Y_{M-1|w})(1 +
Y_{M+1|w})\star s
 + \delta_{M1}\, \log{1-{1\ov Y_-}\ov 1-{1\ov Y_+} }\hstar s \nonumber  \\
& - \log S(r_{M-1}^- - v)S(r_{M+1}^- - v)\,,~~~~~
\end{align}

\subsubsection*{$vw$ strings}

\begin{align}
\log Y_{M|vw}(v) = & - \log(1 +  Y_{M+1})\star s +
\log(1 +  Y_{M-1|vw} )(1 +  Y_{M+1|vw})\star
s\\
&  + \delta_{M1} \log{1-Y_-\ov 1-Y_+}\hstar s \nonumber \\
& + \delta_{M1}( \log\frac{S(u_2^{(2)+} - v)}{S(u_3^{(2)-} - v)} - \log{S(u_1^- - v)S(r_{0}^- - v)}) +\delta_{M2}( \log \frac{S(u_3^{(3)}-v)}{S(u_2^{(3)}-v)} )\,\nonumber\\
& - \log S(\mathbf{r}_{M-1}^- - v)S(\tilde{\mathbf{r}}_{M-1}^- - v)S(\mathbf{r}_{M+1}^- - v)S(\tilde{\mathbf{r}}_{M+1}^- - v)\,\nonumber .
\end{align}
Here the terms involving $\mathbf{r}$ and $\tilde{\mathbf{r}}$ roots should naturally be interpreted in accordance with their existence.

\subsubsection*{$y$ particles}

\begin{align}
\log {Y_+\ov Y_-} = \,  & \log(1 +  Y_{Q})\star K_{Qy} + \mathcal{D}_{y}(u_{123} ,v) \,,\nonumber\\[2mm]
\log {Y_- Y_+}(v) = \, &  2 \log{1 +  Y_{1|vw} \ov 1 +  Y_{1|w} }\star s - \log\left(1+Y_Q \right)\star K_Q + 2 \log(1 +Y_{Q})\star K_{xv}^{Q1} \star s \nonumber \\
& + 2\log{\frac{S(r_1^- - v)}{S(\mathbf{r}_1^- - v)S(\tilde{\mathbf{r}}_1^- - v)}}-  \sum_i \log {\big(S_{xv}^{1_*1}\big)^2\ov S_2}\star s(u_i^{(2)},v) \nonumber \\
&  + 2\log \frac{S(u_2^{(2)}-v)}{S(u_2^{(3)}-v)}\frac{S(u_3^{(3)}-v)}{S(u_3^{(2)}-v)}
-\log {{S}_3(u_2^{(2)++}- v)\ov {S}_3(u_2^{(3)++}-v)}{{S}_3(u_3^{(3)--}- v)\ov {S}_3(u_3^{(2)--}- v)} \nonumber \\
&
+2 \log {S^{31}_{xv}(u_2^{(2)++}, v)S^{31}_{xv}(u_3^{(3)--}, v)\ov S^{31}_{xv}(u_2^{(3)++}, v)S^{31}_{xv}(u_3^{(2)--}, v)}\star s  \,,
\end{align}

\subsubsection*{$Q$ particles}

$\bullet$\ $\ Q=2\ $\\
\bea
\log Y_{2}&=&\log{\left(1 +  {1\ov Y_{1|vw}} \right)^2\ov (1 +  {1\ov Y_{1} })(1 +  {1\ov Y_{3} }) }\star_{p.v} s + \log \frac{S(u_3^{(3)}-v)}{S(u_2^{(3)}-v)}+ 2 \log S(\mathbf{r}_1^- - v)S(\tilde{\mathbf{r}}_1^- - v) \, , \nonumber
\,~~~~~~~
\eea
\\
\bigskip
 \noindent
$\bullet$\ $\ Q=3\ $
\bea
\log Y_{3}&=&\log{\left(1 +  {1\ov Y_{2|vw}} \right)^2\ov (1 +  {1\ov Y_{2} })(1 +  {1\ov Y_{4} }) }\star s + 2 \log S(\mathbf{r}_2^- - v)S(\tilde{\mathbf{r}}_2^- - v) + \log \frac{S(u_2^{(2)+} - v)}{S(u_3^{(2)-} -v)} \, ,\nonumber
\,~~~~~~~
\eea
\vspace{10pt}\\
\bigskip
\noindent
$\bullet$\  $\ Q=4\ $
\bea
\log Y_{4}&=&\log{\left(1 +  {1\ov Y_{3|vw}} \right)^2\ov (1 +  {1\ov Y_{3} })(1 +  {1\ov Y_{5} }) }\star s + 2 \log S(\mathbf{r}_3^- - v)S(\tilde{\mathbf{r}}_3^- - v) + \log \frac{S(u_3^{(3)} - v)}{S(u_2^{(3)} -v)} \, , \nonumber
\,~~~~~~~
\eea
\vspace{10pt}\\
\bigskip
 \noindent
$\bullet$\  $\ Q\ge 5\ $
\bea
\log Y_{Q}&=&\log{\left(1 +  {1\ov Y_{Q-1|vw}} \right)^2\ov (1 +  {1\ov Y_{Q-1} })(1 +  {1\ov Y_{Q+1} }) }\star s + 2 \log S(\mathbf{r}_{Q-1}^- - v)S(\tilde{\mathbf{r}}_{Q-1}^- - v)\, . \nonumber
\,~~~~~~~
\eea
Again $\mathbf{r}$ contributions are to be taken in accordance with their existence.\vspace{10pt}\\
\bigskip
\noindent
$\bullet$\  Hybrid equations
\begin{align}
\log Y_Q =& - L\, \tH_{Q} + \log\left(1+Y_{Q'} \right) \star (K_{\sl(2)}^{Q'Q}+ 2 \, s \star K^{Q'-1,Q}_{vwx} )
\nonumber \\[1mm]
&  +  2 \log \(1 + Y_{1|vw}\) \star s \hstar K_{yQ} +2\log(1+Y_{Q-1|vw})\star s
\nonumber\, \\[1mm]
& +  \log {1- \frac{1}{Y_-} \ov 1-\frac{1}{Y_+} } \hstar K_{Q}  +  \log \big(1-\frac{1}{Y_-}\big)\big( 1 - \frac{1}{Y_+} \big) \hstar K_{yQ} \nonumber \\[1mm]
& - 2  \log{1-Y_-\ov 1-Y_+} \hstar s \star K^{1Q}_{vwx} \label{eq:hybridL40}\\[1mm]
\notag
&+2\sum_i \log S(u_i^{(2)-})\star K^{1Q}_{vwx}+2\log S(r_0^-)\star_{p.v.}K^{1Q}_{vwx}\\[1mm]
\nonumber
&  -2\log \frac{S(u_2^{(3)++})}{S(u_3^{(3)--})}\star K_{vwx}^{2Q} - 2 \log S(\mathbf{r}^{-}_{1})S(\tilde{\mathbf{r}}^{-}_{1})\hat{\star} K_{yQ} \\[1mm] \nonumber
 \notag
& + \mathcal{D}_{\sl(2)}^{Q}(u_{123},v) - \log \frac{S_Q (r_0^-,v) S_{yQ} (r_0^-,v)}{S^{1Q}_{vwx} (r_0,v)} - \log S^{1Q}_{vwx} (u_1,v)\, .
\notag
\end{align}
We would like to note here that similarly to the $L=7$ case, some of the driving terms can be rewritten by using identities such as \eqref{eq:hL40id1}.

\subsection*{Energy formula}

As discussed in the introduction, the energy formula for the $L=40$ state is given by
\begin{align}
E&=\mathcal{E}(u_{123})
-{1\ov 2\pi}\sum_{Q=1}^{\infty}\int_{-\infty}^\infty\, du {d\tilde{p}_Q\ov du}\log(1+Y_Q)
\nonumber\\
&=\
\sum_{i=1}^3\E(u_i^{(2)})
-{1\ov 2\pi}\sum_{Q=1}^{\infty}\int_{-\infty}^\infty\, du {d\tilde{p}_Q\ov du}\log(1+Y_Q)
\\
&\quad -i\tilde{p}_3(u_2^{(2)++})+i\tilde{p}_3(u_2^{(3)++})
-i\tilde{p}_2(u_3^{(3)--})+i\tilde{p}_3(u_3^{(2)--})\,.\nonumber
\end{align}
We would like to point out once again that this expression does not explicitly depend on the Bethe roots $u_{2,3}^{(1)}$.

\subsection{A four-particle state of two bound states}\la{4pt}

In this section we discuss a four-particle state given by a scattering state of two identical bound-like states with opposite momenta. In other words, the momenta of the four particles are arranged as $\{p_i\} = \{p,p^*,-p,-p^*\}$. Such configurations exist at the level of the asymptotic Bethe ansatz, but the region on the z torus where such momenta exist depends on the length of the state. For the three-particle states described in the main text, solutions with rapidities inside the analyticity strip of the $u$ plane do not exist. As such, here we are most interested in potential states with complex rapidities inside this first strip. Such solutions in fact exist for the configuration we are considering here, at least as long as the length of the operator is ten or greater.

\smallskip

\begin{figure}
\begin{center}
\includegraphics[width=7cm]{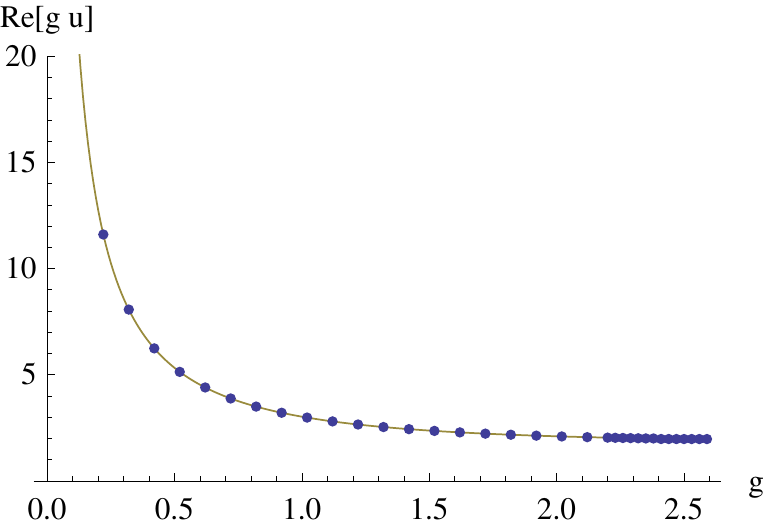}\quad \includegraphics[width=7cm]{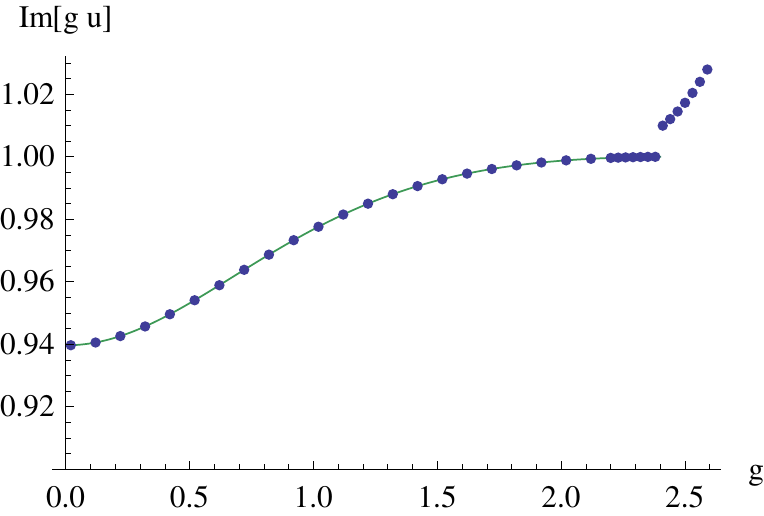}
\caption{The solution to the Bethe Yang equation for $u_1$. For the imaginary part, the rapidity has been rescaled by a factor of $g$. Note that the rapidity asymptotes to $2+i/g$ before breakdown of the BY equations.}
\label{fig:BYsol2bs}
\end{center}
\end{figure}

\begin{figure}
\begin{center}
\includegraphics[width=\textwidth]{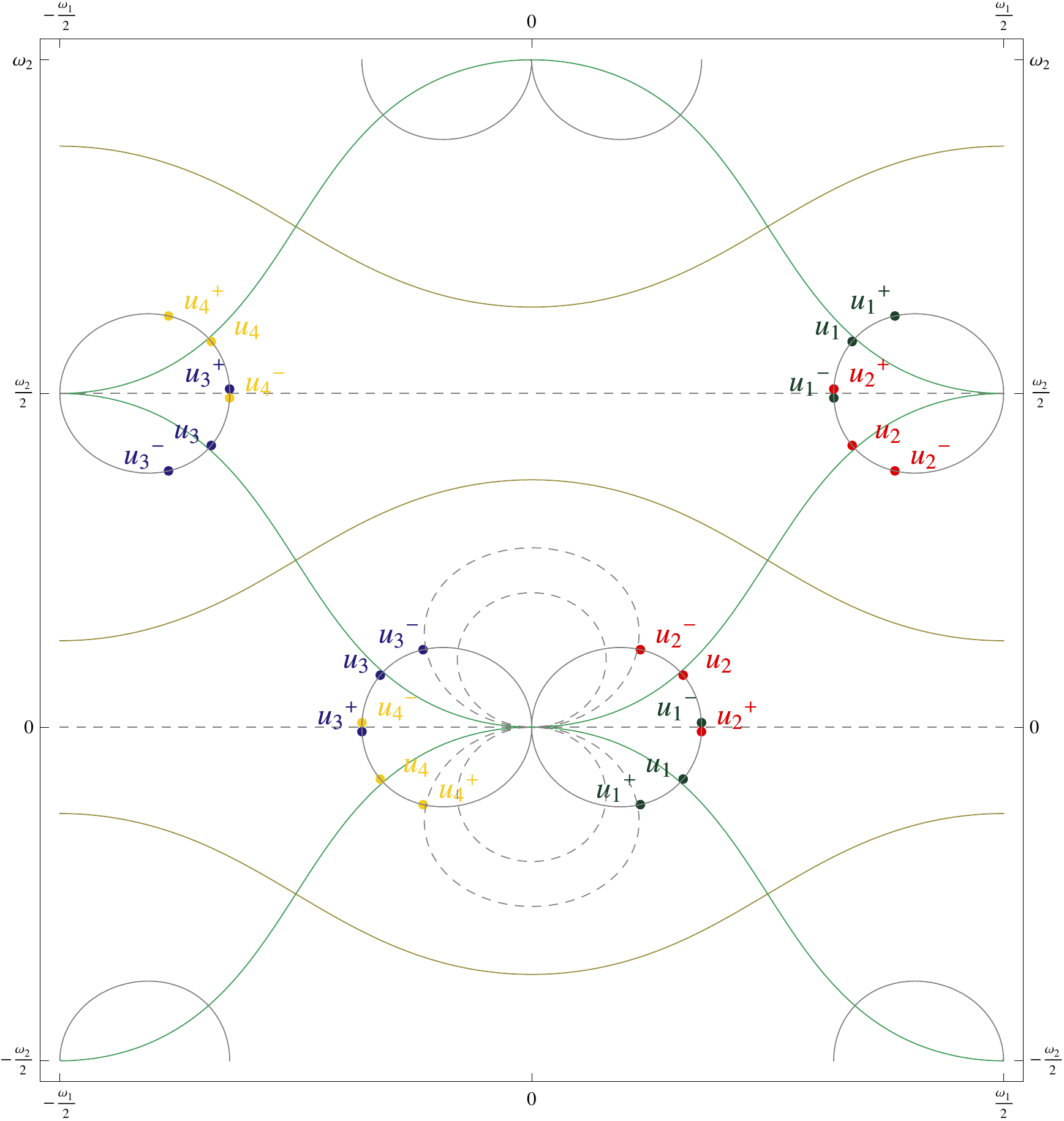}
\caption{The $z$-torus with the (shifted) rapidities $u_i$ at $g=1$. The gray lines again correspond to the contours $\mbox{Re}(u(z)) = \mbox{Re}(u_i)$, while the curved dashed lines indicate the lines $-2 i/g$ and $-3 i/g$.}
\label{fig:torus2bs}
\end{center}
\end{figure}

For numerical reasons we prefer to study a state of moderate length since the complex solution of the Bethe-Yang equation which lies inside the analyticity strip appears to move closer to the real line as the length is increased. The numerical solution of the Bethe-Yang equations at length $16$ for $n=2$ is plotted in figure \ref{fig:BYsol2bs}. We see that around $g=2.4$ we run into trouble similar to the length seven three-particle state discussed in the main text, and from this point the solution of the Bethe-Yang equations can no longer be trusted. Up to this point however, we can use the solution of the Bethe-Yang equations to study the analytic properties of the asymptotic solution and use them to engineer the TBA equations in the usual fashion. The main difference with the three-particle state naturally lies in the fact that the rapidities are inside the physical strip leading to drastic simplifications in the story. In fact this appears to remove the need for any explicit higher quantization conditions. This leaves us with the simplest possible situation which is as close as possible to previously studied states \cite{AFS09,Sfondrini:2011rr}.
The analytic properties of the asymptotic $Y$-functions have been summarized in table \ref{tab:analyticproperties2bs}.
\begin{table}
\begin{center}
\begin{tabular}{|c|c|c|}
\hline Y${}^o$-function  & Zeroes   & Poles\\
\hline $Y_{M|w}$   & $\pm r_{M\pm1}$ &  \\
\hline $1+Y_{M|w}$ & $(\pm r_M)^- \,,\ (\pm r_M)^+$ &   \\
\hline $Y_{1|vw}$ &$u_i \, , \ \pm r_0$ & \\
\hline $Y_-$      &  & $\pm r_1$ \\
\hline $Y_+$       & & $\pm r_1\,,\ u_i^\pm$ \\
\hline $1-Y_-$   & $(\pm r_0)^-  \,,\ (\pm r_0)^+$ & \\
\hline $Y_1$   & &$ u_i^{\pm\pm}$ \\
\hline $Y_2$   & &$ u_i^{\pm} \,,\ u_i^{\pm\pm\pm}$\\
\hline $Y_{Q\geq2}$  & &$ u_i\pm{i\ov g}(Q-1)  \,,\  u_i\pm{i\ov g}(Q+1)$ \\
\hline
\end{tabular}
\end{center}
\caption{Relevant roots and poles of asymptotic Y-functions in the mirror region. Note that only $Y_2$ has poles within the analyticity strip.}
\label{tab:analyticproperties2bs}
\end{table}
To make the differences with the three-particle state apparent, we have also illustrated the location of the rapidities in the mirror and string regions on the
$z$-torus in figure \ref{fig:torus2bs}.

\subsection*{The TBA equations}

By means of the contour deformation trick with considerations entirely analogous to those for Konishi-like states \cite{AFS09} we can derive a set of consistent TBA equations for our state. We would like to emphasize that there appears to be no direct need to introduce a sum over zeroes and poles of $1+Y_Q$. Analogously to the Konishi case, we take a contour for $Y_Q$ functions that encloses all rapidities in the string plane, but such that any other potential contributions vanish. Next, the relevant contours should of course enclose the points $(\pm r_M)^-$ below the real mirror line. Finally we take a natural extension of the principal value prescription normally taken for $Y_{1|vw}$ functions with roots at real rapidities $u_i$; the contour encloses the roots of $Y_{1|vw}$ at the rapidities \emph{above} the real line, {\it i.e.} $u_1$ and $u_4$. This contour gives both the canonical and simplified TBA equations. The resulting equations are compatible with the asymptotic solution. For brevity, except for the case of $Q$ particles, in this appendix we only present the simplified equations.
Finally, let us note that once again the equations appear to be naturally compatible with the TBA lemmas of \cite{BH11a}.

\subsection*{Simplified, hybrid and exact Bethe equations}

Below we present the simplified TBA equations for $w$ and $vw$ strings, $y$ particles and $Q$ particles for $Q>1$, the hybrid TBA equations for $Q$ particles, and the exact Bethe equations.

\subsubsection*{$w$ strings}

\bea\la{TBAsimyw4b}
\log Y_{M|w} &=&  \log(1 +  Y_{M-1|w})(1 +
Y_{M+1|w})\star s + \delta_{M1}\, \log{1-{1\ov Y_-}\ov 1-{1\ov Y_+} }\hstar s \\\nonumber
 &&-\log S((\pm r_{M-1})^- - v)S((\pm r_{M+1})^- - v)\, .~~~~~
\eea

\subsubsection*{$vw$ strings}

\begin{align}
\log Y_{M|vw}(v) = & - \log(1 +  Y_{M+1})\star s +
\log(1 +  Y_{M-1|vw} )(1 +  Y_{M+1|vw})\star
s\la{TBAsimyvw4b} \\
&  + \delta_{M1} \left( \log{1-Y_-\ov 1-Y_+}\hstar s -\log  S((\pm r_{0})^- - v)\prod_i S(u_i^- - v)\right) \nonumber \, .
\end{align}

\subsubsection*{$y$ particles}

\begin{align}\la{ypovym4b}
\log {Y_+\ov Y_-} = &\,   \log(1 +  Y_{Q})\star K_{Qy} - \sum_i \log S_{1_*y}(u_i,v)  \, , \\
\log {Y_+ Y_-} =& \  2\log {1+Y_{1|vw} \ov 1+Y_{1|w}}\star s- \log\left(1+Y_Q\right)\star K_Q
+2\log(1 + Y_{Q}) \star K_{xv}^{Q1}\star s \la{TBAsimypym24b} \\
&+ 2\log{S((\pm r_1)^- - v)}-  \sum_i \log {\big(S_{xv}^{1_*1}\big)^2\ov S_2}\star s(u_i,v) + \log {S(u_1-v)S(u_4-v)\ov
S(u_2-v)S(u_3-v)}\,.\nonumber
\end{align}

\subsubsection*{$Q$ particles, $Q>1$}

\noindent
$\bullet \ Q=2\ $
\bea
\log Y_{2}&=&\log{\left(1 +  {1\ov Y_{1|vw}} \right)^2\ov (1 +  {1\ov Y_{1} })(1 +  {1\ov Y_{3} }) }\star s
- \log {S(u_1-v)S(u_4-v)\ov S(u_2-v)S(u_3-v)}\,.
\,~~~~~~~
\eea

\bigskip
 \noindent
$\bullet \ Q>2\ $
\bea
\log Y_{Q}&=&\log{\left(1 +  {1\ov Y_{Q-1|vw}} \right)^2\ov (1 +  {1\ov Y_{Q-1} })(1 +  {1\ov Y_{Q+1} }) }\star s \, .
\,~~~~~~~
\eea

\subsubsection*{Hybrid equations for $Q$ particles}

\begin{align}
\label{eq:hybrid2bs}
\log Y_Q(v) = & - L_{\rm TBA}\, \tH_{Q} + \log \left(1+Y_{M} \right)
\star \(K_{\sl(2)}^{MQ} + 2 \, s \star K^{M-1,Q}_{vwx} \)   \\
& - \sum_i \log S_{\sl(2)}^{1_*Q}(u_i,v)  +  2 \log \(1 + Y_{1|vw}\) \star s \hstar K_{yQ} +2 \log \left(1 + Y_{Q-1|vw} \right)\star s\nonumber\\
&  - 2  \log{1-Y_-\ov 1-Y_+} \hstar s \star K^{1Q}_{vwx} +  \log
{1- \frac{1}{Y_-} \ov 1-\frac{1}{Y_+} } \hstar K_{Q}  +  \log \big(1-\frac{1}{Y_-}\big)\big( 1 - \frac{1}{Y_+} \big) \hstar K_{yQ}  \nonumber\\
&+2\sum_i \log S\star K_{vwx}^{1Q}(u_i^-,v)\nonumber - 2\log S_{vwx}^{1Q}(u_1,v) S_{vwx}^{1Q}(u_4,v) \\
&+2\log S\star K_{vwx}^{1Q}((\pm r_0)^-,v) - \log S_Q((\pm r_0)^- - v)S_{yQ}((\pm r_0)^-,v) \, .
\nonumber
\end{align}

\subsubsection*{Exact Bethe equations}

Continuation of the hybrid equation for $Q=1$ to the string region is straightforward, and immediately gives the exact Bethe equations for $u_2$ and $u_3$
\begin{align}
\label{eq:hybrid2bs}
\log(-1) = & - L_{\rm TBA}\, \tH_{1_*} + \log \left(1+Y_{M} \right)
\star \(K_{\sl(2)}^{M1_*} + 2 \, s \star K^{M-1,1_*}_{vwx} \)   \\
& - \sum_i \log S_{\sl(2)}^{1_*1_*}(u_i,u_k)  +  2 \log \(1 + Y_{1|vw}\) \star (s \hstar K_{y1_*}+\tilde{s})\nonumber\\
&  - 2  \log{1-Y_-\ov 1-Y_+} \hstar s \star K^{11_*}_{vwx} +  \log
{1- \frac{1}{Y_-} \ov 1-\frac{1}{Y_+} } \hstar K_{1}  +  \log \big(1-\frac{1}{Y_-}\big)\big( 1 - \frac{1}{Y_+} \big) \hstar K_{y1_*}  \nonumber\\
&+2 \sum_i\log S\star K_{vwx}^{11_*}(u_i^-,u_k)\nonumber - 2\log S_{vwx}^{11_*}(u_1,u_k) S_{vwx}^{11_*}(u_4,u_k) \\
&+2\log S\star K_{vwx}^{11_*}((\pm r_0)^-,u_k) - \log S_1((\pm r_0)^- - v)S_{y1_*}((\pm r_0)^-,u_k)\, .
\nonumber
\end{align}
As discussed in \cite{AFS09} there should in general be a $\log\left(1+\frac{1}{Y_+(u_k^-)}\right)$ term in the above. However, due to the pole of $Y_+$ at $u_k^-$ it does not contribute in the exact Bethe equations.
\smallskip

Upon continuation to $u_1$ and $u_4$ we necessarily cross the cut of $f \star K^{1,1_*}_{vwx}$ on the real line. Taking this into account we obtain
\begin{align}
\label{eq:hybrid2bs}
\log(-1) &=  - L_{\rm TBA}\, \tH_{1_*} + \log \left(1+Y_{M} \right)
\star \(K_{\sl(2)}^{M1_*} + 2 \, s \star K^{M-1,1_*}_{vwx} \)   \\
 -& \sum_i \log S_{\sl(2)}^{1_*1_*}(u_i,u_k)  +  2 \log \(1 + Y_{1|vw}\) \star (s \hstar K_{y1_*}+\tilde{s})\nonumber\\
  -& 2  \log{1-Y_-\ov 1-Y_+} \hstar s \star K^{11_*}_{vwx} +  \log{1- \frac{1}{Y_-} \ov 1-\frac{1}{Y_+} } \hstar K_{1}  +  \log \big(1-\frac{1}{Y_-}\big)\big( 1 - \frac{1}{Y_+} \big) \hstar K_{y1_*}  \nonumber\\
 +&2 \sum_i\log S\star K_{vwx}^{11_*}(u_i^-,u_k)\nonumber - 2\log S_{vwx}^{11_*}(u_1,u_k) S_{vwx}^{11_*}(u_4,u_k) \nonumber\\
+&2\log S\star K_{vwx}^{11_*}((\pm r_0)^-,u_k) - \log S_1((\pm r_0)^- - v)S_{y1_*}((\pm r_0)^-,u_k) \nonumber\\
+&2 \left(\log \left(1+Y_{2}\right)\star s -  \log{1-Y_-\ov 1-Y_+} \hstar s + \log S((\pm r_0)^- - u_k) \prod_i S(u_i^- - u_k)\right)\, . \nonumber
\end{align}
As we show below, these equations are compatible with the complex conjugate nature of the momenta.

\subsection*{Conjugation of the exact Bethe equations}

The exact Bethe equation for $u_1$, respectively $u_3$, should be anti-conjugate to the one for $u_2$, respectively $u_4$, however this is not manifest from their derivation. Analogously to how crossing relations and the equations for $vw$ strings were used to show equivalence of string and anti-string\footnote{Again, for the three-particle states we consider $u_2$ lies within the overlap of the string and mirror regions.} quantization conditions for the three-particle state, here we will use conjugation relations together with the equations for $vw$ strings to show that the exact Bethe equations for $u_1$ and $u_2$ are anti-conjugate, meaning that the resulting momenta are conjugate. The discussion is most elegant at the level of canonical equations, which for $Q$ particles are given by
\begin{align}
\log Y_Q =&\, - L_{\rm TBA}\, \tH_{Q} + \log\left(1+Y_{Q'} \right) \star K_{\sl(2)}^{Q'Q}
+ 2 \log\left(1+ \frac{1}{Y_{M'|vw}} \right) \star_{p.v} K^{M'Q}_{vwx}\nonumber \\
&\quad +  \log {1- \frac{1}{Y_-} \ov 1-\frac{1}{Y_+} } \hstar K_{Q}  +  \log \big(1-\frac{1}{Y_-}\big)\big( 1 - \frac{1}{Y_+} \big) \hstar K_{yQ}\nonumber \\
&\quad -\log \prod_i S_{\sl(2)}^{1_*Q}(u_{i},v)+\log \frac{S^{1Q}_{vwx}((\pm r_0),v)}{S_Q((\pm r_0)^--v) S_{yQ}((\pm r_0)^-,v)} \nonumber \\
&\quad - 2\log S_{vwx}^{1Q}(u_1,v) S_{vwx}^{1Q}(u_4,v) \, .
\end{align}
Note the $S_{vwx}^{1Q}(u_{1,4},v)$ terms arising from the roots of $Y_{1|vw}$. The continuation of the canonical equation to $u_2$ (equivalently $u_3$) is trivial apart from a vanishing contribution of the form $\log(1+\frac{1}{Y_+})$ and we directly obtain
\begin{align}
\log (-1) =&\, i L_{\rm TBA}\,p_2 + \log\left(1+Y_{Q'} \right) \star K_{\sl(2)}^{Q'1_*}
+ 2 \log\left(1+ \frac{1}{Y_{M'|vw}} \right) \star_{p.v} K^{M'1_*}_{vwx}\nonumber \\
&\quad +  \log {1- \frac{1}{Y_-} \ov 1-\frac{1}{Y_+} } \hstar K_{1}  +  \log \big(1-\frac{1}{Y_-}\big)\big( 1 - \frac{1}{Y_+} \big) \hstar K_{y1_*}\nonumber \\
&\quad -\log \prod_i S_{\sl(2)}^{1_*1_*}(u_{i},u_2)+\log \frac{S^{11_*}_{vwx}((\pm r_0),u_2)}{S_1((\pm r_0)^--u_2) S_{y1_*}((\pm r_0)^-,u_2)} \nonumber \\
&\quad - 2\log S_{vwx}^{11_*}(u_1,u_2) S_{vwx}^{11_*}(u_4,u_2) \, .\label{eq:EBp2can}
\end{align}
Next, continuation to the point $u_1$ (equivalently $u_4$) requires intersection of the cut of {\mbox{$\log\left(1+ \frac{1}{Y_{1|vw}} \right) \star_{p.v} K^{11_*}_{vwx}$}}, yielding a divergent contribution $\log\left(1+ \frac{1}{Y_{1|vw}(u_1)} \right)$ which naturally cancels the divergence of $\log S_{vwx}^{11_*}(u_1,u_1)$, leaving behind
\begin{align}
\log (-1) =&\, i L_{\rm TBA}\,p_1 + \log\left(1+Y_{Q'} \right) \star K_{\sl(2)}^{Q'1_*}
+ 2 \log\left(1+ \frac{1}{Y_{M'|vw}} \right) \star_{p.v} K^{M'1_*}_{vwx}\nonumber \\
&\quad +  \log {1- \frac{1}{Y_-} \ov 1-\frac{1}{Y_+} } \hstar K_{1}  +  \log \big(1-\frac{1}{Y_-}\big)\big( 1 - \frac{1}{Y_+} \big) \hstar K_{y1_*}\nonumber \\
&\quad -\log \prod_i S_{\sl(2)}^{1_*1_*}(u_{i},u_1)+\log \frac{S^{11_*}_{vwx}((\pm r_0),u_1)}{S_1((\pm r_0)^--u_1) S_{y1_*}((\pm r_0)^-,u_1)} \nonumber \\
&\quad - 2\log S_{vwx}^{11_*}(u_4,u_1) + 2 \log \frac{1+ \frac{1}{Y_{1|vw}(u_1)}}{S_{vwx}^{11_*}(u_1,u_1)}\, . \label{eq:EBp1can}
\end{align}
In order to relate these two equations we will need certain conjugation relations. For real $t$ and $u$ in the analyticity strip we have
\begin{align}
&(K_{\sl(2)}^{Q1_*})^*(t,u) = - K_{\sl(2)}^{Q1_*}(t,u^*) - 2 K^{Q1}_{xv}(t,u^*)\, ,\nonumber\\
&(K^{M1_*}_{vwx})^*(t,u) =  - K^{M1_*}_{vwx}(t,u^*) + K_{M,1}(t,u^*) \, ,\label{eq:Kstringconj}\\
&(K_{y1_*})^*(t,u) = -K_{y1_*}(t,u^*)\, ,\nonumber\\
&(K_1)^*(t,u) = K_1(t,u^*)\nonumber \, .
\end{align}
Also, we have the following identities for the driving terms
\begin{align}
\left(S^{11_*}_{vwx}(u,v)\right)^*= S_{11} (u-v^*)\left(S^{11_*}_{vwx}(u,v^*) S^{11_*}_{xv}(u,v^*)S^{11_*}_{xv}(u^*,v^*)\right)^{-1} \, ,
\end{align}
where both $u$ and $v$ are in the analyticity strip, and
\begin{align}
\left(\frac{S^{11_*}_{vwx}(u,v)}{S_1(u^- -v)S_{y1_*}(u^-,v)}\right)^*= \frac{S_{11}(u-v^*)}{S_1 (u^- - v^*)^2} \, ,
\end{align}
where $u$ is taken to be real. Finally, from the canonical exact Bethe equations (\ref{eq:EBp2can}) and (\ref{eq:EBp1can}), and the above conjugation relations we find
\begin{align}
(\log& (-1) - i L_{\rm TBA}\,p_2)^* =  -(\log(-1) - i L_{\rm TBA}\,p_1) + 2 \log\left(1+\frac{1}{Y_{1|vw}(u_1)}\right) \nonumber\\
 +& 2 \log\left(1+ \frac{1}{Y_{M'|vw}} \right) \star_{p.v}K_{M,1}(u_1) + 2 \log {1- \frac{1}{Y_-} \ov 1-\frac{1}{Y_+} } \hstar K_{1}  -2 \log\left(1+Y_{Q'} \right) \star K^{Q1}_{xv}(u_1)\nonumber \\
 +&2 \log \prod_i S_{xv}^{1_*1}(u_{i},u_1)+\log \frac{S_{11}((\pm r_0),u_1)}{S_1((\pm r_0)^--u_1)^2} - 2\log S_{11}(u_1-u_1) S_{11}(u_4-u_1) \nonumber \, ,\\
= & -(\log(-1) + i L_{\rm TBA}\,p_1) + 2 \log\left(1+\frac{1}{Y_{1|vw}(u_1)}\right) + 2\log Y_{1|vw}(u_1) \nonumber\, ,\\
= & -(\log(-1) + i L_{\rm TBA}\,p_1) \, .
\end{align}
In the first equality we have identified a large part of the conjugate of the exact Bethe equation for $p_2$ as minus the corresponding part of the exact Bethe equation for $p_1$ by the conjugation relations. Subsequently we used the canonical equation for $Y_{1|vw}$, and finally we note that $Y_{1|vw}$ is zero at $u_1$. This shows that the exact Bethe equations are compatible with the reality structure of our state.

\subsection{Transfer matrices}

For the explicit form of eigenvalues of the transfer matrix $T_{a,1}^{\rm sl(2)}$ in the $\sl(2)$-grading, depending on
$K^{\rm I}$ main roots, $K^{\rm II}$
auxiliary roots of $y$-type and $K^{\rm III}$ auxiliary roots of $w$-type, we refer the reader
to  the formula (4.14) from \cite{Arutyunov:2011uz}.
From the point of view of the  $\sl(2)$ grading, the $\su(2)$ sector is described by the following excitation numbers
\bea
\nonumber
K^{\rm I} =K^{\rm II}_{\alpha}\equiv M\, , ~~~~~K^{\rm III}_{\alpha}=0\, ,
\eea
where $\alpha=1,2$ corresponds to the left and right wings of auxiliary Bethe equations.
To construct the asymptotic solution, the auxiliary $y$-roots must be found from their Bethe equations and further
substituted in the expression for $T_{a,1}^{\rm sl(2)}$. It is technically simpler but equivalent to perform a duality transformation on $y$-roots, as in terms of the dual description, the number of dual roots $\tilde{y}$ is
\bea
\nonumber
\widetilde{K}^{\rm II}_{\alpha}=K^{\rm I} -K^{\rm II}_{\alpha}+2K^{\rm III}_{\alpha}=0
\eea
for states from the $\su(2)$ sector.  Performing dualization\footnote{This can be also regarded as switching from
the $\sl(2)$ grading to the $\su(2)$ one. }, we find the following formula, which is a particular case of (4.31) in
\cite{Arutyunov:2011uz}
 \bea
 \label{Ta1}
T_{a,1}^{\su(2)}(v)&=&\left(\frac{x^+}{x^-}\right)^{\frac{M}{2}}\left[(a+1)\prod_{i=1}^{M}\frac{x^--x_i^-}{x^+-x_i^-}
-a\, \prod_{i=1}^{M}\frac{x^--x_i^+}{x^+-x_i^-}\sqrt{\frac{x_i^-}{x_i^+}}\right.
-\\ \nonumber &&-a\, \left.
\prod_{i=1}^{M}\frac{x^--x_i^-}{x^+-x_i^-}\frac{x_i^--\frac{1}{x^+}}{x_i^+-\frac{1}{x^+}}
\sqrt{\frac{x_i^+}{x_i^-}}+(a-1)\prod_{i=1}^{M}\frac{x^--x_i^+}{x^+-x_i^-}\frac{x_i^--\frac{1}{x^+}}{x_i^+-\frac{1}{x^+}}
\right]\, . \eea
Here $M$ is naturally interpreted as a number of excited string theory particles from the $\su(2)$ sector.
Also,
$$
v=x^++\frac{1}{x^+}-\frac{i}{g}a=x^-+\frac{1}{x^-}+\frac{i}{g}a\, .
$$
The variable $v$ takes values in the mirror theory $v$-plane, so that $x^{\pm}=x(v\pm \frac{i}{g}a)$ with $x(v)$ being the mirror theory $x$-function.
Similarly, $x^{\pm}_j=x_s(u_j\pm \frac{i}{g})$, where $x_s$ is the string theory $x$-function.

\smallskip

Specifying the formula (4.34) from \cite{Arutyunov:2011uz}, we also get the eigenvalues of $T_{1,s}^{\rm su(2)}$
\bea
\label{T1s}
T_{1,s}^{\su(2)}(v)&=&\mathscr{M}_s \left[
P_c(x^-)\prod_{i=1}^{M} \frac{ 1}{x^--x_i^+}\sqrt{\frac{x^+_i}{x^-_i}}
-{\textstyle{P_c\left(\frac{1}{x^+}\right)}}
\prod_{i=1}^{M}{\textstyle \frac{x^+-x_i^+}{(x^--x_i^+)(\frac{1}{x^-}-x_i^-)} } \right.   \\
&-&\left. \sum_{k=1}^{s-1}{\textstyle{P_c\Big(\frac{1}{x(v-\frac{i}{g}(s-2k))}\Big)P_c\Big(x(v-\frac{i}{g}(s-2k))\Big)}}
\displaystyle{\prod_{i=1}^{M}}{\textstyle\frac{1}{(x^--x_i^+)(\frac{1}{x^-}-x_i^-)}\sqrt{\frac{x^+_i}{x^-_i}}}\right]
\nonumber
\, .
\eea
Here $\mathscr{M}_s$ is the following normalization prefactor
\bea
\nonumber
\mathscr{M}_{s}={\textstyle{(-1)^s }}
{\textstyle \left(\frac{x^+}{x^-}\right)^{\frac{M}{2}} }
\prod_{i=1}^M {\textstyle{\left(\frac{x^-_i}{x^+_i}\right)^{\sfrac{s}{2}}\frac{x^--x^+_i}{x^+-x^-_i}   }}
\prod_{k=1}^{s-1}
{\textstyle{\frac{x(v+\frac{i}{g}(s-2k))-x^+_i}{x(v-\frac{i}{g}(s-2k))-x^-_i}}}\,
\eea
and $P_c$ is a polynomial
\bea
P_c(y)&=&\prod_{i=1}^M(y-x_i^+)\sqrt{\frac{x^-_i}{x^+_i}}
-\prod_{i=1}^M (y-x_i^-)\, .
\nonumber
\eea
Formulas (\ref{Ta1}) and (\ref{T1s}) obey the Hirota equations and they are used to construct the asymptotic Y-functions
corresponding to an $M$-particle state from the $\su(2)$ sector.

\smallskip



\begin{thebibliography}{20}

\bibitem{M}
  J.~M.~Maldacena,
  ``The large N limit of superconformal field theories and supergravity,''
  Adv.\ Theor.\ Math.\ Phys.\  {\bf 2} (1998) 231
  [Int.\ J.\ Theor.\ Phys.\  {\bf 38} (1999) 1113]
  [arXiv:hep-th/9711200].


 \bibitem{Zamolodchikov90}
  A.~B.~Zamolodchikov,
  ``Thermodynamic Bethe Ansatz in Relativistic Models. Scaling Three State Potts and Lee--Yang Models,''
 {\slshape   Nucl.\ Phys.\  B }{\bf 342} (1990) 695.

 \bibitem{Kuniba2}
 A.~Kuniba, T.~Nakanishi, J.~Suzuki,
  ``T-systems and Y-systems in integrable systems,''
J.\ Phys.\ A  {\bf 44} (2011) 103001
  [arXiv:1010.1344 [hep-th]].


\bibitem{Bajnok:2010ke}
  Z.~Bajnok,
  ``Review of AdS/CFT Integrability, Chapter III.6: Thermodynamic Bethe
  Ansatz,''
  arXiv:1012.3995 [hep-th].


\bibitem{Arutyunov:2009ga}
  G.~Arutyunov and S.~Frolov,
  ``Foundations of the $\AdS$ Superstring. Part I,''
  J.\ Phys.\ A  {\bf 42} (2009) 254003
  [arXiv:0901.4937 [hep-th]].

\bibitem{Brev}
  N.~Beisert {\it et al.},
  ``Review of AdS/CFT Integrability: An Overview,''
  arXiv:1012.3982 [hep-th].

\bibitem{Takahashi72}
  M.~Takahashi,
  ``One-Dimensional Hubbard Model at Finite Temperature,''
  Prog.\ Theor.\ Phys.\  {\bf 47} (1972) 69.

\bibitem{AF07}
  G.~Arutyunov and S.~Frolov,
  ``On String S-matrix, Bound States and TBA,''
  JHEP {\bf 0712} (2007) 024, hep-th/0710.1568.

  \bibitem{AF09a}
  G.~Arutyunov and S.~Frolov,
  ``String hypothesis for the $\AdS$ mirror,''
  JHEP {\bf 0903} (2009) 152
  [arXiv:0901.1417 [hep-th]].

  \bibitem{AF09b}
  G.~Arutyunov and S.~Frolov,
  ``Thermodynamic Bethe Ansatz for the $\AdS$ Mirror Model,''
  JHEP {\bf 0905} (2009) 068
  [arXiv:0903.0141 [hep-th]].

\bibitem{BFT}
  D.~Bombardelli, D.~Fioravanti and R.~Tateo,
  ``Thermodynamic Bethe Ansatz for planar AdS/CFT: a proposal,''
  J.\ Phys.\ A  {\bf 42} (2009) 375401
  [arXiv:0902.3930].

\bibitem{GKKV09}
  N.~Gromov, V.~Kazakov, A.~Kozak and P.~Vieira,
  ``Exact Spectrum of Anomalous Dimensions of Planar N = 4 Supersymmetric
  Yang-Mills Theory: TBA and excited states,''
  Lett.\ Math.\ Phys.\  {\bf 91} (2010) 265
  [arXiv:0902.4458 [hep-th]].

\bibitem{AFS09}
  G.~Arutyunov, S.~Frolov and R.~Suzuki,
  ``Exploring the mirror TBA,''
  JHEP {\bf 1005} (2010) 031
  [arXiv:0911.2224 [hep-th]].

\bibitem{BH10b}
  J.~Balog, A.~Hegedus,
  ``The Bajnok-Janik formula and wrapping corrections,''
  JHEP {\bf 1009}, 107 (2010).
  [arXiv:1003.4303 [hep-th]].

\bibitem{Sfondrini:2011rr}
  A.~Sfondrini, S.~J.~van Tongeren,
  ``Lifting asymptotic degeneracies with the Mirror TBA,''
  JHEP {\bf 1109 } (2011)  050.
  [arXiv:1106.3909 [hep-th]].


\bibitem{AF09d}
 G.~Arutyunov and S.~Frolov,
  ``Simplified TBA equations of the $\AdS$ mirror model,''
  JHEP {\bf 0911} (2009) 019
  [arXiv:0907.2647 [hep-th]].

\bibitem{Balog:2011cx}
  J.~Balog, A.~Hegedus,
  ``Quasi-local formulation of the mirror TBA,''
  [arXiv:1106.2100 [hep-th]].


\bibitem{GKV09b}
 N.~Gromov, V.~Kazakov and P.~Vieira,
  ``Exact Spectrum of Planar ${\cal N}=4$ Supersymmetric Yang-Mills Theory:
  Konishi Dimension at Any Coupling,''
  Phys.\ Rev.\ Lett.\  {\bf 104} (2010) 211601
  [arXiv:0906.4240 [hep-th]].

\bibitem{Frolov:2010wt}
  S.~Frolov,
  ``Konishi operator at intermediate coupling,''
  J.\ Phys.\ A  {\bf 44} (2011) 065401
  [arXiv:1006.5032 [hep-th]].

\bibitem{Gromov:2011de}
  N.~Gromov, D.~Serban, I.~Shenderovich, D.~Volin,
  ``Quantum folded string and integrability: From finite size effects to Konishi dimension,''
  JHEP {\bf 1108 } (2011)  046.
  [arXiv:1102.1040 [hep-th]].

\bibitem{Roiban:2011fe}
  R.~Roiban, A.~A.~Tseytlin,
  ``Semiclassical string computation of strong-coupling corrections to dimensions of operators in Konishi multiplet,''
  Nucl.\ Phys.\  {\bf B848 } (2011)  251-267.
  [arXiv:1102.1209 [hep-th]].

\bibitem{Vallilo:2011fj}
  B.~C.~Vallilo, L.~Mazzucato,
  ``The Konishi multiplet at strong coupling,''
  [arXiv:1102.1219 [hep-th]].

\bibitem{Beccaria:2011uz}
  M.~Beccaria, G.~Macorini,
  ``Quantum folded string in $S^5$ and the Konishi multiplet at strong coupling,''
  [arXiv:1108.3480 [hep-th]].

  \bibitem{Gromov09a}
 N.~Gromov,
  ``Y-system and Quasi-Classical Strings,''
  JHEP {\bf 1001} (2010) 112
  [arXiv:0910.3608 [hep-th]].

  \bibitem{AFS10}
   G.~Arutyunov, S.~Frolov and R.~Suzuki,
  ``Five-loop Konishi from the Mirror TBA,''
  JHEP {\bf 1004} (2010) 069
  [arXiv:1002.1711 [hep-th]].

\bibitem{BH10a}
 J.~Balog and A.~Hegedus,
  ``5-loop Konishi from linearized TBA and the XXX magnet,''
  JHEP {\bf 1006} (2010) 080
  [arXiv:1002.4142 [hep-th]].


 \bibitem{BJ08}
  Z.~Bajnok and R.~A.~Janik,
  ``Four-loop perturbative Konishi from strings and finite size effects for multiparticle states,''
 {\slshape   Nucl.\ Phys.\  B }{\bf 807} (2009) 625
  [arXiv:0807.0399 [hep-th]].

  \bibitem{BJ09}
   Z.~Bajnok, A.~Hegedus, R.~A.~Janik and T.~Lukowski,
  ``Five loop Konishi from AdS/CFT,''
  Nucl.\ Phys.\  B {\bf 827} (2010) 426
  [arXiv:0906.4062 [hep-th]].


\bibitem{LRV09}
 T.~Lukowski, A.~Rej and V.~N.~Velizhanin,
  ``Five-Loop Anomalous Dimension of Twist-Two Operators,''
  Nucl.\ Phys.\  B {\bf 831} (2010) 105
  [arXiv:0912.1624 [hep-th]].

\bibitem{Janik:2010kd}
  R.~A.~Janik,
  ``Review of AdS/CFT Integrability, Chapter III.5: Luscher corrections,''
  arXiv:1012.3994 [hep-th].


\bibitem{Sieg}
  F.~Fiamberti, A.~Santambrogio, C.~Sieg and D.~Zanon,
  ``Wrapping at four loops in N=4 SYM,''
  Phys.\ Lett.\  B {\bf 666} (2008) 100
  [arXiv:0712.3522 [hep-th]].

\bibitem{Velizhanin:2008jd}
  V.~N.~Velizhanin,
  ``The four-loop anomalous dimension of the Konishi operator in N=4 supersymmetric Yang-Mills theory,''
  JETP Lett.\  {\bf 89 } (2009)  6-9.
  [arXiv:0808.3832 [hep-th]].


\bibitem{Arutyunov:2011uz}
  G.~Arutyunov, S.~Frolov,
  ``Comments on the Mirror TBA,''
  JHEP {\bf 1105}, 082 (2011).
  [arXiv:1103.2708 [hep-th]].

\bibitem{DT96}
  P.~Dorey and R.~Tateo,
  ``Excited states by analytic continuation of TBA equations,''
 {\slshape   Nucl.\ Phys.\  B }{\bf 482} (1996) 639
  [arXiv:hep-th/9607167].

\bibitem{BLZe}
  V.~V.~Bazhanov, S.~L.~Lukyanov and A.~B.~Zamolodchikov,
  ``Quantum field theories in finite volume: Excited state energies,''
  Nucl.\ Phys.\  B {\bf 489} (1997) 487, hep-th/9607099.

     \bibitem{BS}
  N.~Beisert and M.~Staudacher,
  ``Long-range $PSU(2,2|4)$ Bethe ansaetze for gauge theory and strings,''
 {\slshape   Nucl.\ Phys.\  B }{\bf 727} (2005) 1
  [arXiv:hep-th/0504190].


 \bibitem{BKS}
  N.~Beisert, C.~Kristjansen, M.~Staudacher,
  ``The Dilatation operator of conformal N=4 superYang-Mills theory,''
  Nucl.\ Phys.\  {\bf B664 } (2003)  131-184.
  [hep-th/0303060].

 \bibitem{BDS}
  N.~Beisert, V.~Dippel, M.~Staudacher,
  ``A Novel long range spin chain and planar N=4 super Yang-Mills,''
  JHEP {\bf 0407 } (2004)  075.
  [hep-th/0405001].

\bibitem{Kazakov:2010kf}
  V.~Kazakov, S.~Leurent,
  ``Finite Size Spectrum of SU(N) Principal Chiral Field from Discrete Hirota Dynamics,''
  [arXiv:1007.1770 [hep-th]].

\bibitem{Balog}   J.~Balog, unpublished.



\bibitem{AFS}
  G.~Arutyunov, S.~Frolov, M.~Staudacher,
  ``Bethe ansatz for quantum strings,''
  JHEP {\bf 0410 } (2004)  016.
  [hep-th/0406256].


 \bibitem{BES}
  N.~Beisert, B.~Eden and M.~Staudacher,
  ``Transcendentality and crossing,''
  J.\ Stat.\  Mech.\  {\bf 0701} (2007) P021
  [arXiv:hep-th/0610251].

\bibitem{Kuniba1}
  A.~Kuniba, T.~Nakanishi and J.~Suzuki,
  ``Functional relations in solvable lattice models. 1: Functional relations
  and representation theory,''
  Int.\ J.\ Mod.\ Phys.\  A {\bf 9} (1994) 5215
  [arXiv:hep-th/9309137].


 \bibitem{Hirota}
  R. Hirota, ``Discrete analogue of a generalized Toda equation,'' Journ. of the Phys. Soc. of Japan, {\bf 50} (1981) 3785-3791.

\bibitem{Suzuki:2011dj}
  R.~Suzuki,
  ``Hybrid NLIE for the Mirror $AdS_5 x S^5$,''
  J.\ Phys.\ A {\bf A44 } (2011)  235401.
  [arXiv:1101.5165 [hep-th]].

   \bibitem{CFT10}
 A.~Cavaglia, D.~Fioravanti and R.~Tateo,
  ``Extended Y-system for the $AdS_5/CFT_4$ correspondence,''
  Nucl.\ Phys.\  B {\bf 843} (2011) 302
  [arXiv:1005.3016 [hep-th]].


  \bibitem{Janik}
  R.~A.~Janik,
  ``The $AdS_5\times S^5$ superstring worldsheet S-matrix and crossing symmetry,''
  Phys.\ Rev.\  {\bf D73 } (2006)  086006.
  [hep-th/0603038].

\bibitem{AF06}
  G.~Arutyunov, S.~Frolov,
  ``On $AdS_5\times S^5$ String S-matrix,''
  Phys.\ Lett.\  {\bf B639 } (2006)  378-382.
  [hep-th/0604043].

  \bibitem{GKV09}
 N.~Gromov, V.~Kazakov and P.~Vieira,
  ``Exact Spectrum of Anomalous Dimensions of Planar N=4 Supersymmetric
  Yang-Mills Theory,''
  Phys.\ Rev.\ Lett.\  {\bf 103} (2009) 131601
  [arXiv:0901.3753 [hep-th]].


\bibitem{Cavaglia:2011kd}
  A.~Cavaglia, D.~Fioravanti, M.~Mattelliano and R.~Tateo,
  ``On the $AdS_5/CFT_4$ TBA and its analytic properties,''
  arXiv:1103.0499 [hep-th].

  \bibitem{BH11a}
  J.~Balog and A.~Hegedus,
  ``$AdS_5\times S^5$ mirror TBA equations from Y-system and discontinuity
  relations,''
  arXiv:1104.4054 [hep-th].

 \bibitem{Gromov} N. Gromov (presented by V. Kazakov),
  talk at ÒConference on Integrability in Gauge and String Theory 2010,Ó in Nordita Stockholm, June, 2010.
  http://agenda.albanova.se/contributionDisplay.py?contribId=258\&confId=1561


\bibitem{Gromov:2011cx}
  N.~Gromov, V.~Kazakov, S.~Leurent, D.~Volin,
  ``Solving the AdS/CFT Y-system,''
  [arXiv:1110.0562 [hep-th]].

\bibitem{Ahn:2010yv}
  C.~Ahn, Z.~Bajnok, D.~Bombardelli, R.~I.~Nepomechie,
  ``Finite-size effect for four-loop Konishi of the beta-deformed N=4 SYM,''
  Phys.\ Lett.\  {\bf B693 } (2010)  380-385.
  [arXiv:1006.2209 [hep-th]].

\bibitem{Gromov:2010dy}
  N.~Gromov, F.~Levkovich-Maslyuk,
  ``Y-system and $\beta$-deformed N=4 Super-Yang-Mills,''
  J.\ Phys.\ A {\bf A44 } (2011)  015402.
  [arXiv:1006.5438 [hep-th]].

\bibitem{Arutyunov:2010gu}
  G.~Arutyunov, M.~de Leeuw, S.~J.~van Tongeren,
  ``Twisting the Mirror TBA,''
  JHEP {\bf 1102 } (2011)  025.
  [arXiv:1009.4118 [hep-th]].

\bibitem{deLeeuw:2011rw}
  M.~de Leeuw, S.~J.~van Tongeren,
  ``Orbifolded Konishi from the Mirror TBA,''
  J.\ Phys.\ A {\bf A44 } (2011)  325404.
  [arXiv:1103.5853 [hep-th]].

\bibitem{Beccaria:2011qd}
  M.~Beccaria, G.~Macorini,
  ``Y-system for $Z_S$ Orbifolds of N=4 SYM,''
  JHEP {\bf 1106 } (2011)  004.
  [arXiv:1104.0883 [hep-th]].

\bibitem{Ahn:2011xq}
  C.~Ahn, Z.~Bajnok, D.~Bombardelli, R.~I.~Nepomechie,
  ``TBA, NLO Luscher correction, and double wrapping in twisted AdS/CFT,''
  [arXiv:1108.4914 [hep-th]].

 \bibitem{AFdp}
  G.~Arutyunov and S.~Frolov,
  ``The Dressing Factor and Crossing Equations,''
  J.\ Phys.\ A  {\bf 42} (2009) 425401
  [arXiv:0904.4575 [hep-th]].

\bibitem{AF08}
  G.~Arutyunov, S.~Frolov,
  ``The S-matrix of String Bound States,''
  Nucl.\ Phys.\  {\bf B804 } (2008)  90-143.
  [arXiv:0803.4323 [hep-th]].

 \end{thebibliography}
\end{document}